\documentclass[aps, prd, showpacs, showkeys ]{revtex4}

\usepackage[small]{caption}
\usepackage{amsmath,amsfonts,amscd,amssymb,epsf, epsfig}\usepackage{graphicx}

\renewcommand{\d}{\operatorname{d}}

\newcommand{\pa}{\partial}

\newcommand{\vep}{\varepsilon}

\begin{document}

 \title{Mode Summation Approach to Casimir Effect Between Two Objects}
 \author{L. P. Teo}
 \email{LeePeng.Teo@nottingham.edu.my}
\affiliation{Department of Applied Mathematics, Faculty of Engineering, University of Nottingham Malaysia Campus, Jalan Broga, 43500, Semenyih, Selangor Darul Ehsan, Malaysia.}
\begin{abstract}
In the last few years, several approaches  have been developed  to compute the exact Casimir interaction energy between two nonplanar objects, all lead to the same functional form, which is called the $TGTG$ formula. In this paper, we explore the $TGTG$ formula from the perspective of mode summation approach. Both scalar fields and electromagnetic fields are considered. In this approach, one has to first solve the equation of motion to find a wave basis for each object. The two $T$'s  in the $TGTG$ formula are $\mathbb{T}$-matrices  representing the Lippmann-Schwinger T-operators, one for each of the objects. Each $\mathbb{T}$-matrix can be found by matching the boundary conditions imposed on the object, and it is independent of the other object. However, it depends on whether the object is interacting with an object outside it, or an object inside it.   The two $G$'s in the $TGTG$ formula are the translation matrices, relating the wave basis of an object to the wave basis of the other object. These translation matrices only depend on the wave basis chosen for each object, and they are independent of the boundary conditions on the objects. After discussing the general theory, we apply the prescription to derive the explicit formulas for the Casimir energies for the sphere-sphere, sphere-plane, cylinder-cylinder and cylinder-plane  interactions. First the $\mathbb{T}$-matrices for a plane, a sphere and a cylinder are derived for the following cases: the object is imposed with Dirichlet, Neumann or general Robin boundary conditions; the object is semi-transparent; and the object is a magnetodielectric object  immersed in a magnetodielectric media.  Then the operator approach developed by Wittman [IEEE Trans. Antennas Propag. \textbf{36}, 1078 (1988)] is used to derive the translation matrices. From these,   the explicit $TGTG$ formula for each of the  scenarios can be written down. On the one hand, we have summarized all the $TGTG$ formulas that have been derived so far for the sphere-sphere, cylinder-cylinder, sphere-plane and cylinder-plane configurations. On the other hand, we provide the $TGTG$ formulas for some scenarios that have not been considered before.
\end{abstract}
 \maketitle

\section{Introduction}
Casimir effect is a purely quantum effect that arises from the vacuum fluctuations of a quantum field in the presence of boundaries \cite{20}. It has attracted a lot of attention due to its wide applications in different areas of physics such as quantum field theory, condensed matter physics, atomic and molecular physics, astrophysics,  gravitation and cosmology, nanoscience and mathematical physics \cite{32}. A number of books and reviews have been written on this subject \cite{21,22,23,24,25,26,27,28,29,30,31}.

Naively,  the Casimir energy is a sum of the ground state energies of all the eigenmodes of the quantum field. Various approaches have been developed to compute the exact Casimir energy analytically, such as mode summation method, path integral approach, Green's function method, quantum statistical approach and scattering approach. Nevertheless, before the turn of the century, the exact computations were limited to some simple configurations, such as two parallel plates, a sphere, a cylinder, two concentric spheres, a wedge and a cone. A particularly large amount of works were devoted to two parallel plates.

Since the work of Lamoreaux \cite{33} in 1997, Casimir force have been measured with high precision in various configurations especially the sphere-plane configuration \cite{28,31}. This has called for precise computation of the Casimir force between a sphere and a plate in particular, and between any two objects in general. Several approximation schemes have been developed for this purpose, such as the semiclassical approach \cite{34,38, 35}, the optical path approximation \cite{36, 39, 40} and the multiple reflection approximation \cite{37}.  However, each of these numerical methods have its limitations. Therefore, it is compelling to have an exact formula for the Casimir interaction between two objects, so that one can justify the accuracies of the numerical approximations and the experimental data.

In \cite{41,73,42,43,44}, Gies et al. derived a worldline representation of the Casimir interaction between two objects imposed with Dirichlet boundary conditions. They used the Monte-Carlo technique to develop a numerical scheme to compute the Casimir interactions of various configurations from the worldline representations. This is basically a geometrical approach, and its application has so far been limited to scalar field with Dirichlet boundary conditions.

Another series of breaktrough appeared in 2006. In \cite{10}, Bulgac et al. computed the Casimir interaction energy between Dirichlet spheres or between a Dirichlet sphere and a Dirichlet plate using the multiple scattering approach, whose application in Casimir effect can be dated back to the work of  Balian and   Duplantier \cite{45,46}. In this approach, the Casimir energy is expressed as an integral over the density of states of the fluctuating field, and the Krein's formula is used to relate the density of states to the scattering matrix of the objects. A novel feature about this method is that it is free of ultraviolet divergences.

In \cite{11}, Emig et al. used path integral quantization with partial wave expansion to compute the Casimir interaction energy between a cylinder and a plane when both are imposed with Dirichlet, Neumann or perfectly conducting boundary conditions. This was elaborated and extended to two cylinders in \cite{49}. Their approach can also be put in the framework of multiple scattering theory, as is more obvious in the works \cite{3,5,12,7,4}. In \cite{3}, a general scheme that uses multipole expansions and multiple scattering theory to compute the Casimir interaction energy between two dielectric objects was proposed. This scheme was applied to obtain the Casimir interaction energy between two dielectric spheres. It was further illustrated in \cite{5} for scalar interactions and in \cite{4} for electromagnetic interactions. The Casimir interaction energies between two spheres imposed with Robin boundary conditions, between two perfectly conducting cylinders, between a magnetodielectric sphere and a magnetodielectric half-space, and between a magnetodielectric cylinder and a magnetodielectric half-space were derived.

Back in 2006,  Bordag \cite{9} has also used the path integral and Green's function method to find the Casimir interaction energy in the cylinder-plane and sphere-plane configuration for a scalar field with Dirichlet or Neumann boundary conditions. Later he also obtained the Casimir interaction energy between a semitransparent cylinder and a plane \cite{51}.  The path integral formalism used in these works can be dated back to the works \cite{47,48}.

At the same time, Dalvit et al. \cite{18, 19} used an entirely different approach -- the mode summation method, to compute the Casimir interaction energy of two eccentric cylinders with Dirichlet, Neumann or perfectly conducting boundary conditions. Later in \cite{50}, the case where the cylinders are separated by a dielectric medium was considered.

In \cite{52, 13}, Kenneth and Klich   explored the multiple scattering approach from the point of view of the Lippmann-Schwinger T-operator. In \cite{52}, they have obtained a general formula for the Casimir interaction energy prior to \cite{3} and showed that the Casimir interaction force between two objects related by reflection is always attractive. In \cite{13}, they used their formalism to derive the Casimir interaction energy between two dielectric spheres.

In \cite{14,16}, Milton and Wagner used the multiple scattering approach to compute the exact Casimir interaction between two semitransparent spheres and two semitransparent cylinders, and obtained their weak coupling limits.

Despite the different perspectives, all the methods,  except for the mode summation approach of Dalvit et al., are fundamentally equivalent. In all cases, the formula for the Casimir interaction energy can be cast in the form
\begin{equation*}
E_{\text{Cas}}=\frac{\hbar}{2\pi}\int_0^{\infty} d\xi \text{Tr}\ln\left(\mathbb{I}-\mathbb{T}_1\mathbb{G}_{12}\mathbb{T}_2\mathbb{G}_{21}\right),
\end{equation*}
which is referred to as the $TGTG$ formula. Different approaches lead to different methods for computing the $\mathbb{T}_1, \mathbb{T}_2$ and $ \mathbb{G}_{12}, \mathbb{G}_{21}$ matrices. In multiple scattering theory, $\mathbb{T}_1$ ($\mathbb{T}_2$) is the transition matrix or Lippmann-Schwinger T-operator  of object 1 (object 2), which is related to and can be computed from the scattering matrix of the object. $\mathbb{G}_{12}$ ($\mathbb{G}_{21}$) is the translation matrix that relates the wave basis centered at object 1 (object 2) to the wave basis centered at object 2 (object 1).

As discussed above, the exact Casimir interaction energy between two objects have been mostly computed by the multiple scattering approach. Since mode summation have been proved to be a very powerful method for computing the Casimir energy of a cylinder, a sphere, two concentric cylinders or two concentric spheres, it is timely to extend the mode summation approach to Casimir interactions between two or multiple objects.  This is the task undertaken in this paper, with the hope that one can get some new insight about how to compute the $\mathbb{T}$ and $\mathbb{G}$ matrices.

 In the first part of this paper, we explain how  the mode summation approach can be used to derive the $TGTG$ formula for the Casimir interaction energy between two objects, for scalar fields as well as
electromagnetic fields. We consider both the case where the two objects are outside each other, and the case where one object is inside the other. We also discuss in Appendix \ref{A1} how this approach can be generalized to more than two objects. Along the way, we obtain prescription for computing the $\mathbb{T}$ and $\mathbb{G}$ matrices.

In the second part of this paper, we  illustrate the mode summation approach by considering the Casimir interactions between two spheres, two cylinders, a sphere and a plane, and a cylinder and a plane.
First we compute the $\mathbb{T}$-matrices for a plane, a sphere and a cylinder. Three cases are considered:   the object is imposed with Dirichlet, Neumann or Robin boundary conditions; the object is semitransparent; and the object is   magnetodielectric. As a matter of fact, scattering theory has been extensively used in classical and quantum field theories. Therefore, some of the $\mathbb{T}$-matrices  (up to some constants) have been well-known and it is quite impossible for us to provide complete references here.
We  apologize for not citing any earlier works.

The more technical part is the translation matrices. For two spheres or two cylinders, the translation matrices have been rather well-known for both scalar and electromagnetic interactions. For the sphere-plane and cylinder-plane configurations, the translation matrices for electromagnetic (but not scalar)   interactions  have been written down in \cite{4} implicitly without explicit derivation, as the multiplication of a change of basis matrix and a translation matrix for plane waves. To give a unified treatment, we generalize the operator approach in \cite{6} to find the translation matrices for the scalar and electromagnetic Casimir interactions   of cylinder-cylinder, sphere-plane and cylinder-plane configurations. From these, we can write down explicitly the $TGTG$ formulas for the scalar and electromagnetic interactions of each of these configuration.
We   obtain some new results that have not been considered before, such as the Casimir interaction energy between a semitransparent object and an object imposed with Dirichlet, Neumann or Robin boundary conditions; between a semitransparent sphere and a semitransparent plate; and between a semitransparent cylinder and a semitransparent plate.

The layout of this paper is as follows. In Section \ref{gen}, we present the general theory of mode summation approach for the derivation of the Casimir interaction energy between any two objects, when both are outside each other, and when one is inside the other. The electromagnetic case is discussed in detail. In Sections \ref{sca_plane}, \ref{sca_sphere} and \ref{sca_cylinder}, we compute the $\mathbb{T}$-matrices for a plane, a sphere and a cylinder, under various boundary conditions. In Section \ref{sphere_sphere}, we compute the translation matrices for the sphere-sphere configuration and obtain explicit formulas for the scalar and electromagnetic sphere-sphere interactions. In Sections \ref{sphere_plane}, \ref{cylinder_cylinder}, \ref{cylinder_plane}, we do the same for the sphere-plane, cylinder-cylinder, and cylinder-plane   configurations.

\section{General Theory}\label{gen}
In this section, we interpret   the $TGTG$ formula for the Casimir interaction between two objects from the point of view of mode summation approach. We consider both the scalar fields and the electromagnetic fields.

For scalar fields, the equation of motion is
\begin{equation}\label{eq3_21_1}
\left(\frac{1}{c^2}\frac{\pa^2}{\pa t^2}-\nabla^2+\frac{m^2c^2}{\hbar^2}+V(\mathbf{x})\right)\varphi =0.
\end{equation}
We consider two types of boundary conditions.

In the first case, the potential function $V(\mathbf{x})$ is zero and the field $\varphi$ satisfies
 the boundary condition
\begin{equation}
u\varphi +v\frac{\pa \varphi}{\pa n}\Biggr|_{S}=0
\end{equation}
on the boundary surface $S$ of the object. Here $n$ is a unit vector normal to the boundary $S$ and pointing away from the object. When $v=0$, we can take $u=1$, and this is the Dirichlet boundary condition. When $u=0$, we can take  $v=1$, and this is the Neumann boundary condition. In general, if $u$ and $v$ are both nonzero, we can take $v=1$, and this is the general Robin boundary condition with parameter $u$.

 In the second scenario,    $V(\mathbf{x})=\lambda\delta(\mathbf{x})$ is a Dirac delta potential function with support on the boundary of the object. We say that the boundary is semitransparent. Integrating the equation \eqref{eq3_21_1} across the boundary of the object, we find that the scalar field satisfies the boundary conditions
\begin{equation}\label{eq3_22_4}
\begin{split}
&\varphi\Bigr|_{S_+}-\varphi\Bigr|_{S_-}=0\\
&\frac{\pa\varphi}{\pa n}\biggr|_{S_+}-\frac{\pa\varphi}{\pa n}\Biggr|_{S_-}=\lambda \varphi.\end{split}
\end{equation} Here $S_+$ and $S_-$ denote respectively the outside and inside of the boundary surface $S$.

Let
\begin{equation*}
\varphi(\mathbf{x},t)=\int_{-\infty}^{\infty}d\omega \varphi(\mathbf{x}, \omega)e^{-i\omega t}.
\end{equation*}We find that $\varphi(\mathbf{x},k)$ satisfies the equation
\begin{equation}\label{eq4_23_1}
\nabla^2\varphi(\mathbf{x},k)=-k^2\varphi(\mathbf{x},k),
\end{equation}where the wave number $k$ and the frequency $\omega$ satisfy the dispersion relation
\begin{equation}\label{eq3_22_2}k=\sqrt{\frac{\omega^2}{c^2}-\frac{m^2c^2}{\hbar^2}}.\end{equation}
Fixing a coordinate system, we can express the solutions of the equation \eqref{eq4_23_1} as a linear combination of the regular wave $\varphi_{\alpha}^{\text{reg}}$ which is regular at the origin of the
coordinate system, and the outgoing wave $\varphi_{\alpha}^{\text{out}}$ which decreases to zero rapidly as $\mathbf{x}\rightarrow \infty$ and $k$ is replaced by $ik$. Here $\alpha$ labels the solutions.

For electromagnetic fields, the Maxwell's equations read as
\begin{equation*}\begin{split}
&\nabla\cdot\mathbf{D}=\rho_f,\hspace{1cm}\nabla\times\mathbf{E}+\frac{ \pa\mathbf{B} }{\pa t}=\mathbf{0},\\
&\nabla\cdot\mathbf{B}=0,\hspace{1cm}\nabla\times\mathbf{H}-\frac{\pa\mathbf{D}}{\pa t}=\mathbf{J}_f.
\end{split}\end{equation*} For a  magnetodielectric object, the charge density $\rho_f$ and the free current density $\mathbf{J}_f$  on the object are zero, and
we assume the linear relations
\begin{equation}\label{eq3_20_1}\begin{split}
\mathbf{D}(\mathbf{x},\omega)=&\varepsilon(\omega)\mathbf{E}(\mathbf{x},\omega),\\
\mathbf{H}(\mathbf{x},\omega)=&\frac{1}{\mu(\omega)}\mathbf{B}(\mathbf{x},\omega),
\end{split}\end{equation}  where $\varepsilon$ and $\mu$ are respectively the electric permittivity and magnetic permeability of the object.  Here
for $\mathbf{Z}=\mathbf{D}, \mathbf{E}, \mathbf{H}, \mathbf{B}$,  $$\mathbf{Z}(\mathbf{x},t)=\int_{-\infty}^{\infty}d\omega \,\mathbf{Z}(\mathbf{x},\omega)e^{-i\omega t}.$$
The boundary conditions are the continuities of $\displaystyle \varepsilon\mathbf{E}_{n}, \mathbf{E}_{\parallel}, \mathbf{B}_n, \frac{1}{\mu}\mathbf{B}_{\parallel}$
across the  boundary of the object. These conditions are not independent. The continuity of $\mathbf{E}_{\parallel}$ implies the continuity of $\mathbf{B}_n$, and the continuity of $\displaystyle\frac{1}{\mu}\mathbf{B}_{\parallel}$ implies the continuity of $\vep \mathbf{E}_n$.

As usual one can introduce a  vector potential   $\mathbf{A}$ that satisfies the gauge condition $\nabla\cdot\mathbf{A}=0$. In terms of $\mathbf{A}$,
\begin{equation*}
\mathbf{E}=-\frac{\pa\mathbf{A}}{\pa t},\hspace{1cm}\mathbf{B}=\nabla\times\mathbf{A}.
\end{equation*}
Then the Maxwell's equations are equivalent to the wave equation
\begin{equation}\label{eq3_20_9}
\nabla\times\nabla \times \mathbf{A}+\epsilon\mu\frac{\pa^2\mathbf{A}}{\pa t^2}=\mathbf{0}.
\end{equation}
Let
$$\mathbf{A}(\mathbf{x},t)=\int_{-\infty}^{\infty}d\omega \,\mathbf{A}(\mathbf{x},\omega)e^{-i\omega t}.$$
The wave equation can be written as an eigenvalue problem,
\begin{equation}\label{eq4_23_2}
\nabla\times\nabla \times \mathbf{A}(\mathbf{x},k)=k^2\mathbf{A}(\mathbf{x},k),
\end{equation}where the wave number $k$ and the frequency $\omega$ satisfy the dispersion relation
$$k=\sqrt{\varepsilon\mu}\omega.$$
The solutions of the equation \eqref{eq4_23_2} can be written in different bases. In the cases of rectangular, cylindrical and spherical bases, we can choose a distinct
direction and divide the solutions into transverse electric (TE) waves $\mathbf{A}^{\text{TE}}_{\alpha}$ and transverse magnetic (TM)
waves $\mathbf{A}^{\text{TM}}_{\alpha}$ parametrized by some parameter $\alpha$ and satisfy
\begin{equation}\label{eq3_20_2}
\frac{1}{k}\nabla\times \mathbf{A}^{\text{TE}}_{\alpha}=\mathbf{A}^{\text{TM}}_{\alpha},\hspace{1cm}\frac{1}{k}\nabla\times \mathbf{A}^{\text{TM}}_{\alpha}=
\mathbf{A}^{\text{TE}}_{\alpha}.
\end{equation}Moreover, the waves can be divided into regular waves $\mathbf{A}^{\text{TE,reg}}_{\alpha}$, $\mathbf{A}^{\text{TM,reg}}_{\alpha}$ that are
regular at the origin of the coordinate system  and outgoing waves $\mathbf{A}^{\text{TE,out}}_{\alpha}$, $\mathbf{A}^{\text{TM,out}}_{\alpha}$ that decrease
to zero rapidly when $\mathbf{x}\rightarrow  \infty$ and $k$ is replaced by $ik$.

Consider two objects O$_1$ and O$_2$. Choose appropriate coordinate systems with coordinate origins at $O$ and $O'$ respectively for each of the objects. Let $\mathbf{x}$ and $\mathbf{x}'$ be respectively the position vectors of a point
with respect to $O$ and $O'$. If $\mathbf{L}$ is the position vector of $O'$ with respect to $O$, then $\mathbf{x}=\mathbf{x}'+\mathbf{L}$.
In the following, we derive the $TGTG$ formula of the Casimir interaction energy between these two objects using mode summation approach.
We consider separately the case where the two objects are outside each other and the case
where one object is inside the other. We will focus on discussing the case of electromagnetic fields. The case of scalar fields is easier and can
be obtained in the same way.
\subsection{Two objects are outside each other}

Inside the object O$_1$, we express $\mathbf{A} $ in the coordinate system centered at $O$:
\begin{equation}\label{eq3_22_8}
\mathbf{A} =\int_{-\infty}^{\infty} d\omega \sum_{\alpha}\left(A_1^{\alpha} \mathbf{A}^{\text{TE,reg}}_{\alpha}(\mathbf{x},\omega)+C_1^{\alpha}
\mathbf{A}^{\text{TM,reg}}_{\alpha}(\mathbf{x},\omega)\right)e^{-i\omega t}.
\end{equation}
Inside the object O$_2$, we express $\mathbf{A} $ in the coordinate system centered at $O'$:
\begin{equation*}
\mathbf{A} =\int_{-\infty}^{\infty} d\omega \sum_{\beta}\left(A_2^{\beta} \mathbf{A}^{\text{TE,reg}}_{\beta}(\mathbf{x}',\omega)+C_2^{\beta}
\mathbf{A}^{\text{TM,reg}}_{\beta}(\mathbf{x}',\omega)\right)e^{-i\omega t}.
\end{equation*}In the region outside the two objects,
$\mathbf{A}$ can be expressed in both coordinate system. Close to the object O$_1$,
\begin{equation}\label{eq3_22_9}
\mathbf{A} =\int_{-\infty}^{\infty} d\omega \sum_{\alpha}\left(a_1^{\alpha} \mathbf{A}^{\text{TE,reg}}_{\alpha}(\mathbf{x},\omega)+
b_1^{\alpha} \mathbf{A}^{\text{TE,out}}_{\alpha}(\mathbf{x},\omega)+c_1^{\alpha} \mathbf{A}^{\text{TM,reg}}_{\alpha}(\mathbf{x},\omega)+d_1^{\alpha}
\mathbf{A}^{\text{TM,out}}_{\alpha}(\mathbf{x},\omega)\right)e^{-i\omega t}.
\end{equation}Close to the object O$_2$,
\begin{equation*}
\mathbf{A} =\int_{-\infty}^{\infty} d\omega \sum_{\beta}\left(a_2^{\beta} \mathbf{A}^{\text{TE,reg}}_{\beta}(\mathbf{x}',\omega)+
b_2^{\beta} \mathbf{A}^{\text{TE,out}}_{\beta}(\mathbf{x}',\omega)+c_2^{\beta} \mathbf{A}^{\text{TM,reg}}_{\beta}(\mathbf{x}',\omega)+d_2^{\beta}
\mathbf{A}^{\text{TM,out}}_{\beta}(\mathbf{x}',\omega)\right)e^{-i\omega t}.
\end{equation*}In fact, the regular waves close to O$_1$ are propagated from the outgoing waves outside O$_2$, and the regular waves close to O$_2$
are propagated from the outgoing waves outside O$_1$. More precisely, they are related by translations:
\begin{equation*}\begin{split}
\begin{pmatrix}\mathbf{A}^{\text{TE,out}}_{\beta}(\mathbf{x}',\omega)\\\mathbf{A}^{\text{TM,out}}_{\beta}(\mathbf{x}',\omega)\end{pmatrix}
=&\sum_{\alpha} \begin{pmatrix}U_{\alpha, \beta}^{\text{TE,TE}}(-\mathbf{L}) & U_{ \alpha, \beta}^{\text{TM,TE}} (-\mathbf{L})\\
 U_{ \alpha, \beta}^{\text{TE,TM}}(-\mathbf{L}) & U_{ \alpha, \beta}^{\text{TM,TM}}(-\mathbf{L}) \end{pmatrix}
 \begin{pmatrix}\mathbf{A}^{\text{TE,reg}}_{\alpha}(\mathbf{x},\omega)
\\\mathbf{A}^{\text{TM,reg}}_{\alpha}(\mathbf{x},\omega)\end{pmatrix},\\
\begin{pmatrix}\mathbf{A}^{\text{TE,out}}_{\alpha}(\mathbf{x},\omega)\\\mathbf{A}^{\text{TM,out}}_{\alpha}(\mathbf{x},\omega)\end{pmatrix}
=&\sum_{\beta} \begin{pmatrix}U_{\beta,\alpha}^{\text{TE,TE}}(\mathbf{L}) & U_{\beta,\alpha}^{\text{TM,TE}} (\mathbf{L})\\
 U_{\beta,\alpha}^{\text{TE,TM}}(\mathbf{L}) & U_{\beta,\alpha}^{\text{TM,TM}}(\mathbf{L}) \end{pmatrix}
 \begin{pmatrix}\mathbf{A}^{\text{TE,reg}}_{\beta}(\mathbf{x}',\omega)
\\\mathbf{A}^{\text{TM,reg}}_{\beta}(\mathbf{x}',\omega)\end{pmatrix}.
\end{split}\end{equation*}
From these, we find that
\begin{equation}\label{eq3_20_3}\begin{split}
\begin{pmatrix} a_1^{\alpha}\\c_1^{\alpha}\end{pmatrix}=&\sum_{\beta}\mathbb{U}_{\alpha,\beta}(-\mathbf{L})\begin{pmatrix} b_2^{\beta}\\d_2^{\beta}\end{pmatrix},\\
\begin{pmatrix} a_2^{\beta}\\c_2^{\beta}\end{pmatrix}=&\sum_{\alpha}\mathbb{U}_{\beta,\alpha}(\mathbf{L})\begin{pmatrix} b_1^{\alpha}\\d_1^{\alpha}\end{pmatrix}.
\end{split}\end{equation}Here $\mathbb{U}_{\alpha,\beta}$ is the $2\times 2$ translation matrix with components $U_{\alpha,\beta}^{\text{TE,TE}},
U_{\alpha,\beta}^{\text{TE,TM}}, U_{\alpha,\beta}^{\text{TM,TE}}, U_{\alpha,\beta}^{\text{TM,TM}}$.

Now, for $i=1,2$, the boundary conditions on the boundary of the object O$_i$ will give rise to four homogeneous relations between $a_i^{\alpha},
b_i^{\alpha}, c_i^{\alpha}, d_i^{\alpha}$ and
$A_i^{\alpha}, C_i^{\alpha}$. Eliminating $A_i^{\alpha}$ and $ C_i^{\alpha}$, we are left with two relations among $a_i^{\alpha}, b_i^{\alpha},
c_i^{\alpha}, d_i^{\alpha}$. From these we can solve for $b_i^{\alpha}, d_i^{\alpha}$ in terms of $a_i^{\alpha}, c_i^{\alpha}$ and write it in the matrix form
\begin{equation}\label{eq3_20_4}
\begin{pmatrix} b_i^{\alpha}\\d_i^{\alpha}\end{pmatrix}=-\mathbb{T}_i^{\alpha}\begin{pmatrix} a_i^{\alpha}\\c_i^{\alpha}\end{pmatrix}.
\end{equation}Up to a constant, $\mathbb{T}_i$ is the transition matrix or the Lippmann-Schwinger T-operator associated with object O$_i$. It is closely related to the scattering matrix of the object O$_i$ \cite{25,3,4, 52,13}.

From \eqref{eq3_20_3} and \eqref{eq3_20_4}, we have
\begin{equation*}\begin{split}
\begin{pmatrix} b_1^{\alpha}\\d_1^{\alpha}\end{pmatrix}=&-\mathbb{T}_1^{\alpha}\begin{pmatrix} a_1^{\alpha}\\c_1^{\alpha}\end{pmatrix}\\
=&-\mathbb{T}_1^{\alpha}\sum_{\beta}\mathbb{U}_{\alpha,\beta}(-\mathbf{L})\begin{pmatrix} b_2^{\beta}\\d_2^{\beta}\end{pmatrix}\\
=&\mathbb{T}_1^{\alpha}\sum_{\beta}\mathbb{U}_{\alpha,\beta}(-\mathbf{L})\mathbb{T}_2^{\beta}\begin{pmatrix} a_2^{\beta}\\c_2^{\beta}\end{pmatrix}\\
=&\mathbb{T}_1^{\alpha}\sum_{\beta}\mathbb{U}_{\alpha,\beta}(-\mathbf{L})\mathbb{T}_2^{\beta}\sum_{\alpha'}\mathbb{U}_{\beta,\alpha'}(\mathbf{L})
\begin{pmatrix} b_1^{\alpha'}\\d_1^{\alpha'}\end{pmatrix},
\end{split}\end{equation*}or in other words,
\begin{equation*}
\sum_{\alpha'}\left(\mathbb{I}_{\alpha,\alpha'}-\mathbb{T}_1^{\alpha}\sum_{\beta}\mathbb{U}_{\alpha,\beta}(-\mathbf{L})\mathbb{T}_2^{\beta}
\mathbb{U}_{\beta,\alpha'}(\mathbf{L})\right)\begin{pmatrix} b_1^{\alpha'}\\d_1^{\alpha'}\end{pmatrix}=\begin{pmatrix} 0\\0\end{pmatrix}.
\end{equation*}
Here $\mathbb{I}_{\alpha,\alpha'}$ is the $2\times 2$ identity matrix.
For nontrivial eigenmodes, the infinite vector with $\alpha$-component given by $\begin{pmatrix} b_1^{\alpha} &d_1^{\alpha}\end{pmatrix}^T$
cannot be identically zero, which holds if and only if $\det\left(\mathbb{I}-\mathbb{M} \right)=0$, where $\mathbb{I}$ is the identity matrix and
$\mathbb{M}=\mathbb{T}_1\mathbb{U}_{12}\mathbb{T}_2\mathbb{U}_{21}$. $\mathbb{T}_i$ is an infinite diagonal matrix whose $(\alpha,\alpha)$-component is
the $2\times 2$ matrix  $\mathbb{T}_i^{\alpha}$, $\mathbb{U}_{12}$ is an infinite matrix with $(\alpha,\beta)$-component given by
$ \mathbb{U}_{\alpha,\beta}(-\mathbf{L})$, and $\mathbb{U}_{21}$ is an infinite matrix with $(\beta,\alpha)$-component given by $ \mathbb{U}_{\beta,\alpha}(\mathbf{L})$. Therefore, we find that the eigenfrequencies of the system are the nonnegative solutions of the equation
\begin{equation*}
\det\left(\mathbb{I}-\mathbb{M}(\omega)\right)=0.
\end{equation*}
Hence, the Casimir interaction energy is given by
\begin{equation*}
E_{\text{Cas}}=\sum_{\substack{\omega>0\\\det\left(\mathbb{I}-\mathbb{M}(\omega)\right)=0}}\frac{\hbar\omega}{2}.\end{equation*}
Using the residue theorem, this can be computed by
\begin{equation}
\begin{split}
E_{\text{Cas}}=&\sum_{\omega>0}\frac{\hbar\omega}{2}\text{Res}_{\omega}\frac{d}{d\omega}\ln \det\left(\mathbb{I}-\mathbb{M}(\omega)\right)\\
=&-\frac{\hbar}{2}\frac{1}{2\pi i}\int_{-i\infty}^{i\infty}d\omega\,\omega \frac{d}{d\omega}\ln \det\left(\mathbb{I}-\mathbb{M}(\omega)\right)\\
=&-\frac{\hbar}{2\pi}\int_0^{\infty} d\xi \,\xi \frac{d}{d\xi}\text{Tr}\,\ln  \left(\mathbb{I}-\mathbb{M}(i\xi)\right).
\end{split}
\end{equation}
Integration by parts give
\begin{equation}\label{eq3_20_7}\begin{split}
E_{\text{Cas}}
=&\frac{\hbar}{2\pi}\int_0^{\infty} d\xi \text{Tr}\,\ln  \left(\mathbb{I}-\mathbb{M}(i\xi)\right),
\end{split}
\end{equation}
where
$$\mathbb{M}(i\xi)=\mathbb{T}_1(i\xi)\mathbb{U}_{12}(i\xi)\mathbb{T}_2(i\xi)\mathbb{U}_{21}(i\xi).$$
Eq. \eqref{eq3_20_7} is called the $TGTG$ formula for the Casimir interaction energy.

In the case of a scalar field,
we let
\begin{equation}\label{eq3_22_5}
\varphi=\int_{-\infty}^{\infty}d\omega \sum_{\alpha}A_1^{\alpha}\varphi_{\alpha}^{\text{reg}}(\mathbf{x},\omega)e^{-i\omega t}
\end{equation}inside object O$_1$,
\begin{equation*}
\varphi=\int_{-\infty}^{\infty}d\omega \sum_{\beta}A_2^{\beta}\varphi_{\beta}^{\text{reg}}(\mathbf{x}',\omega)e^{-i\omega t}
\end{equation*}inside object O$_2$. Outside O$_1$ and O$_2$,
\begin{equation}\label{eq3_22_1}\begin{split}
\varphi=&\int_{-\infty}^{\infty}d\omega \sum_{\alpha}\left(a_1^{\alpha}\varphi_{\alpha}^{\text{reg}}(\mathbf{x},\omega)
+b_1^{\alpha}\varphi_{\alpha}^{\text{out}}(\mathbf{x},\omega)\right)e^{-i\omega t}\\
=&\int_{-\infty}^{\infty}d\omega \sum_{\beta}\left(a_2^{\beta}\varphi_{\beta}^{\text{reg}}(\mathbf{x}',\omega)
+b_2^{\beta}\varphi_{\beta}^{\text{out}}(\mathbf{x}',\omega)\right)e^{-i\omega t}.
\end{split}\end{equation}Then the Casimir energy is given by \eqref{eq3_20_7}, where the matrix $\mathbb{T}_i$ is obtained from the boundary conditions
on object O$_i$. It is diagonal with $(\alpha,\alpha)$-component $T_i^{\alpha}$ being a scalar relating $b_i^{\alpha}$ with $a_i^{\alpha}$:
\begin{equation*}
b_i^{\alpha}=-T_i^{\alpha}a_i^{\alpha}.
\end{equation*}
For the translation matrices $\mathbb{U}_{12}$ and $\mathbb{U}_{21}$, their components are defined by the relations:
\begin{equation*}\begin{split}
\varphi_{\beta}^{\text{out}}(\mathbf{x}',\omega)=&\sum_{\alpha}U_{\alpha,\beta}(-\mathbf{L})\varphi_{\alpha}^{\text{reg}}(\mathbf{x},\omega),\\
\varphi_{\alpha}^{\text{out}}(\mathbf{x},\omega)=&\sum_{\beta}U_{\beta,\alpha}(\mathbf{L})\varphi_{\beta}^{\text{reg}}(\mathbf{x}',\omega).
\end{split}\end{equation*}

\subsection{One object is inside the other}
Assume that object O$_1$ lies inside object O$_2$.
Inside the object O$_1$, we express $\mathbf{A} $ in the coordinate system centered at $O$:
\begin{equation*}
\mathbf{A} =\int_{-\infty}^{\infty} d\omega \sum_{\alpha}\left(A_1^{\alpha} \mathbf{A}^{\text{TE,reg}}_{\alpha}(\mathbf{x},\omega)+C_1^{\alpha}
\mathbf{A}^{\text{TM,reg}}_{\alpha}(\mathbf{x},\omega)\right)e^{-i\omega t}.
\end{equation*}
Outside the object O$_2$,
we express $\mathbf{A} $ in the coordinate system centered at $O'$:
\begin{equation*}
\mathbf{A} =\int_{-\infty}^{\infty} d\omega \sum_{\beta}\left(B_2^{\beta} \mathbf{A}^{\text{TE,out}}_{\beta}(\mathbf{x}',\omega)+D_2^{\beta}
\mathbf{A}^{\text{TM,out}}_{\beta}(\mathbf{x}',\omega)\right)e^{-i\omega t}.
\end{equation*}
In the region between O$_1$ and O$_2$,
\begin{equation*}
\mathbf{A} =\int_{-\infty}^{\infty} d\omega \sum_{\alpha}\left(a_1^{\alpha} \mathbf{A}^{\text{TE,reg}}_{\alpha}(\mathbf{x},\omega)+
b_1^{\alpha} \mathbf{A}^{\text{TE,out}}_{\alpha}(\mathbf{x},\omega)+c_1^{\alpha} \mathbf{A}^{\text{TM,reg}}_{\alpha}(\mathbf{x},\omega)+d_1^{\alpha}
\mathbf{A}^{\text{TM,out}}_{\alpha}(\mathbf{x},\omega)\right)e^{-i\omega t},
\end{equation*}which can be re-expanded with respect to $O'$ as
\begin{equation*}
\mathbf{A} =\int_{-\infty}^{\infty} d\omega \sum_{\beta}\left(a_2^{\beta} \mathbf{A}^{\text{TE,reg}}_{\beta}(\mathbf{x}',\omega)+
b_2^{\beta} \mathbf{A}^{\text{TE,out}}_{\beta}(\mathbf{x}',\omega)+c_2^{\beta} \mathbf{A}^{\text{TM,reg}}_{\beta}(\mathbf{x}',\omega)+d_2^{\beta}
\mathbf{A}^{\text{TM,out}}_{\beta}(\mathbf{x}',\omega)\right)e^{-i\omega t},
\end{equation*}using the translation formulas
\begin{equation*}\begin{split}
\begin{pmatrix}\mathbf{A}^{\text{TE,reg}}_{\beta}(\mathbf{x}',\omega)\\\mathbf{A}^{\text{TM,reg}}_{\beta}(\mathbf{x}',\omega)\end{pmatrix}
=&\sum_{\alpha} \begin{pmatrix}V_{\alpha, \beta}^{\text{TE,TE}}(-\mathbf{L}) & V_{ \alpha, \beta}^{\text{TM,TE}} (-\mathbf{L})\\
 V_{ \alpha, \beta}^{\text{TE,TM}}(-\mathbf{L}) & V_{ \alpha, \beta}^{\text{TM,TM}}(-\mathbf{L}) \end{pmatrix}
 \begin{pmatrix}\mathbf{A}^{\text{TE,reg}}_{\alpha}(\mathbf{x},\omega)
\\\mathbf{A}^{\text{TM,reg}}_{\alpha}(\mathbf{x},\omega)\end{pmatrix}\\
\begin{pmatrix}\mathbf{A}^{\text{TE,out}}_{\alpha}(\mathbf{x},\omega)\\\mathbf{A}^{\text{TM,out}}_{\alpha}(\mathbf{x},\omega)\end{pmatrix}
=&\sum_{\beta} \begin{pmatrix}W_{\beta,\alpha}^{\text{TE,TE}}(\mathbf{L}) & W_{\beta,\alpha}^{\text{TM,TE}} (\mathbf{L})\\
W_{\beta,\alpha}^{\text{TE,TM}}(\mathbf{L}) & W_{\beta,\alpha}^{\text{TM,TM}}(\mathbf{L}) \end{pmatrix}
 \begin{pmatrix}\mathbf{A}^{\text{TE,out}}_{\beta}(\mathbf{x}',\omega)
\\\mathbf{A}^{\text{TM,out}}_{\beta}(\mathbf{x}',\omega)\end{pmatrix}.
\end{split}\end{equation*}
In other words,
\begin{equation*}
\begin{pmatrix}
a_1^{\alpha}\\c_1^{\alpha}
\end{pmatrix}=\sum_{\beta}\mathbb{V}_{\alpha,\beta}(-\mathbf{L})\begin{pmatrix}
a_2^{\beta}\\c_2^{\beta}
\end{pmatrix},\hspace{1cm}
\begin{pmatrix}
b_2^{\beta}\\d_2^{\beta}
\end{pmatrix}=\sum_{\alpha}\mathbb{W}_{\alpha,\beta}( \mathbf{L})\begin{pmatrix}
b_1^{\alpha}\\d_1^{\alpha}
\end{pmatrix}.
\end{equation*}
As before, the boundary conditions on the boundary of O$_1$ gives
\begin{equation}\label{eq3_20_5}
\begin{pmatrix} b_1^{\alpha}\\d_1^{\alpha}\end{pmatrix}=-\mathbb{T}_1^{\alpha}\begin{pmatrix} a_1^{\alpha}\\c_1^{\alpha}\end{pmatrix}.
\end{equation}
The  matrix $\mathbb{T}$ is the same as in the case when two objects are outside each other.
On the other hand, by eliminating $B_2^{\beta}$ and $D_2^{\beta}$, the boundary conditions on the boundary of O$_2$ give rise to a relation
\begin{equation}\label{eq3_20_6}
\begin{pmatrix} a_2^{\beta}\\c_2^{\beta}\end{pmatrix}=-\widetilde{\mathbb{T}}_2^{\beta}\begin{pmatrix} b_2^{\beta}\\d_2^{\beta}\end{pmatrix}.
\end{equation}
Therefore,
\begin{equation*}\begin{split}
\begin{pmatrix} b_1^{\alpha}\\d_1^{\alpha}\end{pmatrix}=&-\mathbb{T}_1^{\alpha}\begin{pmatrix} a_1^{\alpha}\\c_1^{\alpha}\end{pmatrix}\\
=&-\mathbb{T}_1^{\alpha}\sum_{\beta}\mathbb{V}_{\alpha,\beta}(-\mathbf{L})\begin{pmatrix} a_2^{\beta}\\c_2^{\beta}\end{pmatrix}\\
=&\mathbb{T}_1^{\alpha}\sum_{\beta}\mathbb{V}_{\alpha,\beta}(-\mathbf{L})\widetilde{\mathbb{T}}_2^{\beta}\begin{pmatrix} b_2^{\beta}\\d_2^{\beta}\end{pmatrix}\\
=&\mathbb{T}_1^{\alpha}\sum_{\beta}\mathbb{V}_{\alpha,\beta}(-\mathbf{L})\widetilde{\mathbb{T}}_2^{\beta}\sum_{\alpha'}\mathbb{W}_{\beta,\alpha'}(\mathbf{L})
\begin{pmatrix} b_1^{\alpha'}\\d_1^{\alpha'}\end{pmatrix}.
\end{split}\end{equation*}
As in the case where   two objects are outside each other, one then obtains the formula \eqref{eq3_20_7} for the Casimir energy, where
now the matrix $\mathbb{M}$ is given by
\begin{equation}\label{eq3_21_2}
\mathbb{M}=\mathbb{T}_1\mathbb{V}_{12}(-\mathbf{L})\widetilde{\mathbb{T}}_2\mathbb{W}_{21}(\mathbf{L}).
\end{equation}

In the case of a scalar field,
we let
\begin{equation*}
\varphi=\int_{-\infty}^{\infty}d\omega \sum_{\alpha}A_1^{\alpha}\varphi_{\alpha}^{\text{reg}}(\mathbf{x},\omega)e^{-i\omega t}
\end{equation*}inside object O$_1$,
\begin{equation}\label{eq3_22_7}
\varphi=\int_{-\infty}^{\infty}d\omega \sum_{\beta}B_2^{\beta}\varphi_{\beta}^{\text{out}}(\mathbf{x}',\omega)e^{-i\omega t}
\end{equation}outside object O$_2$. In the region between  O$_1$ and O$_2$,
\begin{equation}\begin{split}\label{eq3_22_3}
\varphi=&\int_{-\infty}^{\infty}d\omega \sum_{\alpha}\left(a_1^{\alpha}\varphi_{\alpha}^{\text{reg}}(\mathbf{x},\omega)
+b_1^{\alpha}\varphi_{\alpha}^{\text{out}}(\mathbf{x},\omega)\right)e^{-i\omega t}\\
=&\int_{-\infty}^{\infty}d\omega \sum_{\beta}\left(a_2^{\beta}\varphi_{\beta}^{\text{reg}}(\mathbf{x}',\omega)
+b_2^{\beta}\varphi_{\beta}^{\text{out}}(\mathbf{x}',\omega)\right)e^{-i\omega t}.
\end{split}\end{equation}The Casimir energy is given by \eqref{eq3_20_7},
with $\mathbb{M}$ given by \eqref{eq3_21_2}.
The matrices $\mathbb{T}_1$ and $\widetilde{\mathbb{T}}_2$ are obtained from the boundary conditions on objects O$_1$ and O$_2$ respectively.
They are diagonal with
\begin{equation*}
b_1^{\alpha}=-T_1^{\alpha}a_1^{\alpha},\hspace{1cm} a_2^{\beta}=-\widetilde{T}_2^{\beta}b_2^{\beta}.
\end{equation*}
For the translation matrices $\mathbb{V}_{12}$ and $\mathbb{W}_{21}$, they are defined by the relations:
\begin{equation*}
\begin{split}
\varphi_{\beta}^{\text{reg}}(\mathbf{x}',\omega)=&\sum_{\alpha}V_{\alpha,\beta}(-\mathbf{L})\varphi_{\alpha}^{\text{reg}}(\mathbf{x},\omega),\\
\varphi_{\alpha}^{\text{out}}(\mathbf{x},\omega)=&\sum_{\beta}W_{\beta,\alpha}(\mathbf{L})\varphi_{\beta}^{\text{out}}(\mathbf{x}',\omega).\end{split}
\end{equation*}

Using the mode summation approach, we have shown that the Casimir interaction energy between two objects can be expressed as the $TGTG$ formula \eqref{eq3_20_7}. Our approach here is somehow formal in contrast to the mode summation approach employed  in \cite{18, 19} for the Casimir interaction   between two eccentric cylinders. The Casimir self energies of each of the objects do not appear in our approach, and we obtain immediately the Casimir interaction energy. Nevertheless, we would like to emphasize that our approach can also be made rigorous as in \cite{18,19}.

In the above and in the following, the Casimir energy is understood as the Casimir energy at zero temperature. We will not elaborate on the finite temperature Casimir free energy as it can be easily derived from the zero temperature Casimir energy using Matsubara formalism. More precisely, the Casimir free energy can be expressed in the form of $TGTG$ formula by changing the integration over $\xi$ in the formula for the zero temperature Casimir energy \eqref{eq3_20_7} to summation over Matsubara frequencies $\xi_p=2\pi p k_BT/\hbar$. More precisely, the Casimir free energy is given by
\begin{equation}\label{eq5_7_1}
\begin{split}
\mathfrak{F}_{\text{Cas}}=& k_BT\sum_{p=0}^{\infty}\!' \text{Tr}\,\ln  \left(\mathbb{I}-\mathbb{M}(i\xi_p)\right),
\end{split}
\end{equation}where the prime $'$ over the summation means that the term with $p=0$ comes with a weight $1/2$.

The main ingredients of the $TGTG$ formulas are the $\mathbb{T}, \widetilde{\mathbb{T}}$ matrices and the translation matrices $\mathbb{U}, \mathbb{V}, \mathbb{W}$. Notice that the $\mathbb{T}$ or $ \widetilde{\mathbb{T}}$ matrix of an object
does not depend on the other objects. It is derived solely from the boundary conditions imposed on that single object. If we consider the Casimir interaction between
the object and an object outside it, we need to find the  $\mathbb{T}$-matrix of the object. If we  consider the Casimir interaction between the object and
an object inside it, we need to find the  $\widetilde{\mathbb{T}}$-matrix of the object. In the following sections, we will find the $\mathbb{T}$ and $ \widetilde{\mathbb{T}}$ matrices of
planes, spheres and cylinders under various boundary conditions. We will then discuss the translation matrices $\mathbb{U}, \mathbb{V}, \mathbb{W}$ between different pairs of objects.

In this section, we focus on Casimir interaction between two objects. As in \cite{5,4}, one can in fact consider Casimir interaction between multiple objects. In Appendix \ref{A1}, we show how the approach in this section can be generalized to more than two objects. In that case, the Casimir interaction can still be expressed in the form \eqref{eq3_20_7} but the matrix $\mathbb{M}$ has a different representation. However, the building blocks of the matrix $\mathbb{M}$ are still the $\mathbb{T}, \widetilde{\mathbb{T}}$ matrices and the translation matrices.

\section{The $ \widetilde{\mathbb{T}}$ matrix of a plane}\label{sca_plane}
Choosing $\mathbf{e}_z$ as the distinct direction, plane wave basis are parametrized by $\mathbf{k}_{\perp}=(k_x,k_y)\in \mathbb{R}^2$, with
\begin{equation*}\begin{split}
\varphi_{\mathbf{k}_{\perp}}^{\substack{\text{reg}\\\text{out}}}(\mathbf{x},k)=& e^{ik_xx+ik_yy\mp i\sqrt{k^2-k_{\perp}^2}z},
\\
\mathbf{A}_{\mathbf{k}_{\perp}}^{\text{TE}, \substack{\text{reg}\\\text{out}}}(\mathbf{x},k)=&\frac{1}{k_{\perp}}\nabla\times
\varphi_{\mathbf{k}_{\perp}}^{\substack{\text{reg}\\\text{out}}}(\mathbf{x},k)\mathbf{e}_z\\
=& \frac{1}{k_{\perp}}e^{ik_xx+ik_yy\mp i\sqrt{k^2-k_{\perp}^2}z}
\left(ik_y\mathbf{e}_x-ik_x\mathbf{e}_y\right),\\
\mathbf{A}_{\mathbf{k}_{\perp}}^{\text{TM}, \substack{\text{reg}\\\text{out}}}(\mathbf{x},k)=&\frac{1}{kk_{\perp}}\nabla\times\nabla\times
\varphi_{\mathbf{k}_{\perp}}^{\substack{\text{reg}\\\text{out}}}(\mathbf{x},k)\mathbf{e}_z\\
=& \frac{1}{kk_{\perp}}e^{ik_xx+ik_yy\mp i\sqrt{k^2-k_{\perp}^2}z}
\left(\pm k_x\sqrt{k^2-k_{\perp}^2}\mathbf{e}_x \pm k_y\sqrt{k^2-k_{\perp}^2}\mathbf{e}_y+k_{\perp}^2\mathbf{e}_z\right).
\end{split}\end{equation*}
Here $k_{\perp}=\sqrt{k_x^2+k_y^2}$.

Assume that a plane of dimension $H\times H$ is located at $z=0$. In the following, when we consider the Casimir effect between a sphere and a plane or a cylinder
and a plane, we always regard the sphere and the cylinder as inside the plane. Hence, we only compute the    $\widetilde{\mathbb{T}}$-matrix
for a plane. Let a scalar field be represented as
\begin{equation}\label{eq3_23_5}
\varphi =H^2\int_{-\infty}^{\infty}d\omega \int_{-\infty}^{\infty}\frac{dk_x}{2\pi}\int_{-\infty}^{\infty}
\frac{dk_y}{2\pi} \left(a^{\mathbf{k}_{\perp}}\varphi_{\mathbf{k}_{\perp}}^{\text{reg}}(\mathbf{x},k)+b^{\mathbf{k}_{\perp}}
\varphi_{\mathbf{k}_{\perp}}^{\text{out}}(\mathbf{x},k)\right)
e^{-i\omega t}
\end{equation}for $z<0$ (inside), and by
\begin{equation}\label{eq3_23_6}
\varphi =H^2\int_{-\infty}^{\infty}d\omega \int_{-\infty}^{\infty}\frac{dk_x}{2\pi}\int_{-\infty}^{\infty}
\frac{dk_y}{2\pi} B^{\mathbf{k}_{\perp}}\varphi_{\mathbf{k}_{\perp}}^{\text{out}}(\mathbf{x},k)e^{-i\omega t},
\end{equation} for $z>0$ (outside).
First consider the boundary condition
$u\varphi+v\pa_z\varphi=0$ on the plane. In this case, the field   outside does not contribute to $\widetilde{\mathbb{T}}$.
The boundary condition gives
\begin{equation*}
a^{\mathbf{k}_{\perp}}\left(u-i\sqrt{k^2-k_{\perp}^2} v\right)+b^{\mathbf{k}_{\perp}}\left(u+i\sqrt{k^2-k_{\perp}^2} v\right)=0.
\end{equation*}
where $k$ is given by \eqref{eq3_22_2}. Passing to the imaginary frequency $\omega=i\xi$, we find that
\begin{equation*}
\widetilde{T}^{\mathbf{k}_{\perp}}(i\xi)=-\frac{a^{\mathbf{k}_{\perp}}}{b^{\mathbf{k}_{\perp}}}=\frac{u-v\sqrt{\kappa^2+k_{\perp}^2}}{u+v\sqrt{\kappa^2+k_{\perp}^2}},
\end{equation*}
where
\begin{equation}\label{eq3_22_6}\kappa=\sqrt{\frac{\xi^2}{c^2}+\frac{m^2c^2}{\hbar^2}}.\end{equation}

For a semitransparent plane, the boundary conditions \eqref{eq3_22_4} give
\begin{equation*}\begin{split}
&a^{\mathbf{k_{\perp}}}+b^{\mathbf{k}_{\perp}}=B^{\mathbf{k}_{\perp}},\\
&   i\sqrt{k^2-k_{\perp}^2}B^{\mathbf{k}_{\perp}}+i\sqrt{k^2-k_{\perp}^2} \left(a^{\mathbf{k}_{\perp}}-b^{\mathbf{k}_{\perp}} \right)
=\lambda B^{\mathbf{k}_{\perp}}.
\end{split}\end{equation*}Eliminating $B^{\mathbf{k}_{\perp}}$ and passing to imaginary frequency, we find that
\begin{equation*}
\widetilde{T}^{\mathbf{k}_{\perp}}(i\xi)=-\frac{a^{\mathbf{k}_{\perp}}}{b^{\mathbf{k}_{\perp}}}=\frac{\lambda}{\lambda+2\sqrt{\kappa^2+k_{\perp}^2}},
\end{equation*}with $\kappa$ given by \eqref{eq3_22_6}.

For an electromagnetic field, let
\begin{equation}\label{eq3_23_7}
\mathbf{A}=H^2\int_{-\infty}^{\infty}d\omega \int_{-\infty}^{\infty}\frac{dk_x}{2\pi}\int_{-\infty}^{\infty}
\frac{dk_y}{2\pi} \left(a^{\mathbf{k}_{\perp}}\mathbf{A}_{\mathbf{k}_{\perp}}^{\text{TE, reg}}(\mathbf{x},k)
+b^{\mathbf{k}_{\perp}}\mathbf{A}_{\mathbf{k}_{\perp}}^{\text{TE, out}}(\mathbf{x},k)
+c^{\mathbf{k}_{\perp}}\mathbf{A}_{\mathbf{k}_{\perp}}^{\text{TM, reg}}(\mathbf{x},k)
+d^{\mathbf{k}_{\perp}}\mathbf{A}_{\mathbf{k}_{\perp}}^{\text{TM, out}}(\mathbf{x},k)\right)e^{-i\omega t}
\end{equation}
for $z<0$ (inside), and
\begin{equation}\label{eq3_23_8}
\mathbf{A}=H^2\int_{-\infty}^{\infty}d\omega \int_{-\infty}^{\infty}\frac{dk_x}{2\pi}\int_{-\infty}^{\infty}
\frac{dk_y}{2\pi} \left(B^{\mathbf{k}_{\perp}}\mathbf{A}_{\mathbf{k}_{\perp}}^{\text{TE, out}}(\mathbf{x},k)
+D^{\mathbf{k}_{\perp}}\mathbf{A}_{\mathbf{k}_{\perp}}^{\text{TM, out}}(\mathbf{x},k)\right)e^{-i\omega t}
\end{equation}for $z>0$ (outside).
Assume that the inside of the plane is filled with material with permittivity $\vep_i$ and permeability $\mu_i$, and the outside
is filled with material with permittivity $\vep_e$ and permeability $\mu_e$. Then the continuities of $\varepsilon\mathbf{E}_n, \mathbf{E}_{\parallel},
\mathbf{B}_n$ and $\mu^{-1}\mathbf{B}_{\parallel}$ give
\begin{equation*}\begin{split}
&a^{\mathbf{k}_{\perp}}+ b^{\mathbf{k}_{\perp}}=B^{\mathbf{k}_{\perp}},\\
&\frac{\sqrt{k_i^2-k_{\perp}^2}}{\mu_i}\left(a^{\mathbf{k}_{\perp}}- b^{\mathbf{k}_{\perp}} \right)=-\frac{1}{\mu_e}B^{\mathbf{k}_{\perp}}
\sqrt{k_e^2-k_{\perp}^2},\\
&\frac{\sqrt{k_i^2-k_{\perp}^2}}{k_i}\left(c^{\mathbf{k}_{\perp}}- d^{\mathbf{k}_{\perp}}\right)=-D^{\mathbf{k}_{\perp}}\frac{\sqrt{k_e^2-k_{\perp}^2}}{k_e},\\
&\frac{\vep_i}{k_i}\left(c^{\mathbf{k}_{\perp}}+ d^{\mathbf{k}_{\perp}}\right)=\frac{\vep_e}{k_e}D^{\mathbf{k}_{\perp}},
\end{split}\end{equation*}where \begin{equation}\label{eq3_23_9}
\begin{split}
k_e=&\sqrt{\varepsilon_e\mu_e}\omega=n_e\frac{\omega}{c},\\ k_i=&\sqrt{\varepsilon_i\mu_i}\omega=n_i\frac{\omega}{c}.\end{split}
\end{equation}Eliminating $B^{\mathbf{k}_{\perp}}$ and $D^{\mathbf{k}_{\perp}}$ gives
\begin{equation*}
\begin{split}
\frac{a^{\mathbf{k}_{\perp}}}{b^{\mathbf{k}_{\perp}}}=&-\frac{\mu_i\sqrt{k_e^2-k_{\perp}^2}-\mu_e\sqrt{k_i^2-k_{\perp}^2}}
{\mu_i\sqrt{k_e^2-k_{\perp}^2}+\mu_e\sqrt{k_i^2-k_{\perp}^2}},\\
\frac{c^{\mathbf{k}_{\perp}}}{d^{\mathbf{k}_{\perp}}}=&-\frac{\vep_i\sqrt{k_e^2-k_{\perp}^2}-\vep_e\sqrt{k_i^2-k_{\perp}^2}}
{\vep_i\sqrt{k_e^2-k_{\perp}^2}+\vep_e\sqrt{k_i^2-k_{\perp}^2}}.
\end{split}
\end{equation*}
Hence
\begin{equation}\label{eq4_4_1}
\widetilde{\mathbb{T}}^{\mathbf{k}_{\perp}}=\begin{pmatrix} \widetilde{T}_{\mathbf{k}_{\perp}}^{\text{TE}}& 0\\0 &
\widetilde{T}_{\mathbf{k}_{\perp}}^{\text{TM}}\end{pmatrix}
\end{equation}is a diagonal matrix with elements
\begin{equation}\label{eq4_3_14}
\begin{split}
\widetilde{T}_{\mathbf{k}_{\perp}}^{\text{TE}}(i\xi)=&\frac{\mu_i\sqrt{\kappa_e^2+k_{\perp}^2}-\mu_e\sqrt{\kappa_i^2+k_{\perp}^2}}
{\mu_i\sqrt{\kappa_e^2+k_{\perp}^2}+\mu_e\sqrt{\kappa_i^2+k_{\perp}^2}},\\
\widetilde{T}_{\mathbf{k}_{\perp}}^{\text{TM}}(i\xi)=&\frac{\vep_i\sqrt{\kappa_e^2+k_{\perp}^2}-\vep_e\sqrt{\kappa_i^2+k_{\perp}^2}}
{\vep_i\sqrt{\kappa_e^2+k_{\perp}^2}+\vep_e\sqrt{\kappa_i^2+k_{\perp}^2}},
\end{split}
\end{equation}where \begin{equation}\label{eq3_25_1}
\kappa_e= n_e(i\xi)\frac{\xi}{c},\quad \kappa_i= n_i(i\xi) \frac{\xi}{c}.
\end{equation}

\section{The $\mathbb{T}, \widetilde{\mathbb{T}}$ matrices of a sphere}\label{sca_sphere}
  In spherical coordinates  $x=r\sin\theta\cos\phi, y=r\sin\theta\sin\phi, z=r\cos\theta$, the regular waves and outgoing waves solutions
  of the  equation \eqref{eq4_23_1} are
\begin{equation}\label{eq3_20_10}
\varphi_{l m}^{\text{reg}}(\mathbf{x},k) =\mathcal{C}_{l}^{\text{reg}} j_l(kr)Y_{lm}(\theta,\phi),\quad \varphi_{l m}^{\text{out}}(\mathbf{x},k)=
\mathcal{C}_{l}^{\text{out}} h_l^{(1)}(kr)Y_{lm}(\theta,\phi),
\end{equation}where $l=0,1,2,\ldots$, $-l\leq m\leq l$,
$$\mathcal{C}_{l}^{\text{reg}}=i^{-l},\quad \mathcal{C}_{l}^{\text{out}}=\frac{\pi}{2}i^{l+2},$$$$j_l(z)=\sqrt{\frac{\pi}{2z}}J_{l+\frac{1}{2}}(z),
\hspace{1cm} h_l^{(1)}(z)=\sqrt{\frac{\pi}{2z}}H_{l+\frac{1}{2}}^{(1)}(z)$$ are the spherical Bessel functions, and $Y_{lm}(\theta,\phi)$ are spherical harmonics given by
\begin{equation*}
Y_{lm}(\theta,\phi)=\sqrt{\frac{2l+1}{4\pi}\frac{(l-m)!}{(l+m)!}}P_l^m(\cos\theta)e^{im\phi}.
\end{equation*}Here $P_l^m(z)$ are the associated Legendre functions. The constants $\mathcal{C}_{l}^{\text{reg}}$ and $\mathcal{C}_{l}^{\text{out}}$ are chosen so that
\begin{equation*}
\mathcal{C}_{l}^{\text{reg}} j_l(i\zeta) =\sqrt{\frac{\pi}{2\zeta}}I_{l+\frac{1}{2}}(\zeta),\quad \mathcal{C}_{l}^{\text{out}} h_l^{(1)}(i\zeta) =\sqrt{\frac{\pi}{2\zeta}}K_{l+\frac{1}{2}}(\zeta).
\end{equation*}

For the vector wave equation \eqref{eq3_20_9}, choosing  $\mathbf{e}_r$ as the distinct direction, the TE and TM waves are
\begin{equation*}\begin{split}
\mathbf{A}_{lm}^{\text{TE},*}(\mathbf{x},k)=&\frac{1}{\sqrt{l(l+1)}} \nabla\times \varphi_{l m}^{*}(\mathbf{x},k) r\mathbf{e}_r \\
=&\frac{\mathcal{C}_{l}^*}{\sqrt{l(l+1)}}f_l^*(kr)\left(\frac{im}{\sin\theta}Y_{lm}(\theta,\phi)\mathbf{e}_{\theta}-\frac{\pa Y_{lm}(\theta,\phi)}{\pa\theta}
\mathbf{e}_{\phi}\right),\\
\mathbf{A}_{lm}^{\text{TM},*}(\mathbf{x},k)=&\frac{1}{k\sqrt{l(l+1)}}\nabla\times \nabla\times \varphi_{l m}^{*}(\mathbf{x},k) r\mathbf{e}_r \\
=&\mathcal{C}_{l}^*\left(\frac{\sqrt{l(l+1)}}{kr}f_l^*(kr)Y_{lm}(\theta,\phi)\mathbf{e}_r+\frac{1}{\sqrt{l(l+1)}}\frac{1}{kr}\frac{d}{dr}\left(rf_l^*(kr)\right)
\left[\frac{\pa Y_{lm}(\theta,\phi)}{\pa\theta}\mathbf{e}_{\theta}+\frac{im}{\sin\theta}Y_{lm}(\theta,\phi)\mathbf{e}_{\phi}\right]\right),
\end{split}\end{equation*}where $l=1,2,\ldots,$ $-l\leq m\leq l$. Here $*=$ reg or out, with $f_l^{\text{reg}}(z)=j_l(z)$ and $f_l^{\text{out}}(z)=h_l^{(1)}(z)$.

Now we consider the $\mathbb{T}, \widetilde{\mathbb{T}}$ matrices for various scenarios.  Consider a sphere of radius $R$ centered at the origin.
To compute the  $\mathbb{T}$-matrix, let a scalar field be represented as
\begin{equation*}
\varphi=\int_{-\infty}^{\infty} d\omega\sum_{l=0}^{\infty}\sum_{m=-l}^l A^{lm}\varphi_{lm}^{\text{reg}}(\mathbf{x}, k)e^{-i\omega t}
\end{equation*}inside the sphere, and as
\begin{equation}\label{eq3_22_11}
\varphi=\int_{-\infty}^{\infty} d\omega\sum_{l=0}^{\infty}\sum_{m=-l}^l \left(a^{lm}\varphi_{lm}^{\text{reg}}(\mathbf{x}, k)
+b^{lm}\varphi_{lm}^{\text{out}}(\mathbf{x}, k)\right)e^{-i\omega t}
\end{equation}outside the sphere.
To compute the $\widetilde{\mathbb{T}}$-matrix, a scalar field is represented as \eqref{eq3_22_11}  inside the sphere, and as
\begin{equation*}
\varphi=\int_{-\infty}^{\infty} d\omega\sum_{l=0}^{\infty}\sum_{m=-l}^l B^{lm}\varphi_{lm}^{\text{out}}(\mathbf{x}, k)e^{-i\omega t},
\end{equation*}outside the sphere.

If  Dirichlet, Neumann or Robin boundary condition   is imposed on the sphere, and we are considering the Casimir interaction between
this sphere and other objects outside the sphere, the field inside the sphere does not contribute to the Casimir interaction energy.   With the boundary condition
$ u\varphi+vr\pa_r\varphi=0$ on the sphere,
we find that
\begin{equation*}
a^{lm} \mathcal{C}_{l }^{\text{reg}}\left(uj_l(kR)+vkRj_l'(kR)\right)+b^{lm}\mathcal{C}_{l }^{\text{out}} \left(uh_l^{(1)}(kR)+vkRh_l^{(1)'}(kR)\right)=0.
\end{equation*}  Passing to imaginary frequency, we have
\begin{equation*}
\begin{split}
T^{lm}(i\xi)=&-\frac{b^{lm}}{a^{lm}}= \frac{\left(u-\frac{v}{2}\right)I_{l+\frac{1}{2}}(\kappa R)
+v\kappa R I_{l+\frac{1}{2}}'(\kappa R)}{\left(u-\frac{v}{2}\right)K_{l+\frac{1}{2}}(\kappa R)
+v\kappa R K_{l+\frac{1}{2}}'(\kappa R)},
\end{split}\end{equation*}
where $\kappa$ is given by \eqref{eq3_22_6}.

If we are considering the Casimir interaction of the sphere with other objects inside the sphere, the field outside the sphere does not contribute.  Using the
boundary condition $ u\varphi+vr\pa_r\varphi=0$, we find that
\begin{equation*}
\widetilde{T}^{lm}(i\xi)=\left(T^{lm}\right)^{-1}(i\xi)= \frac{\left(u-\frac{v}{2}\right)K_{l+\frac{1}{2}}(\kappa R)
+v\kappa R K_{l+\frac{1}{2}}'(\kappa R)}{\left(u-\frac{v}{2}\right)I_{l+\frac{1}{2}}(\kappa R)
+v\kappa R I_{l+\frac{1}{2}}'(\kappa R)}.
\end{equation*}

If the sphere is  semitransparent, i.e., it satisfies the boundary conditions \eqref{eq3_22_4}, then for   $T^{lm}$, the boundary conditions give
\begin{equation*}
\begin{split}
&a^{lm} \mathcal{C}_{l }^{\text{reg}}j_l(kR)+b^{lm}\mathcal{C}_{l }^{\text{out}} h_l^{(1)}(kR)=A^{lm} \mathcal{C}_{l }^{\text{reg}}j_l(kR),\\
&a^{lm} \mathcal{C}_{l }^{\text{reg}}kj_l'(kR)+b^{lm}\mathcal{C}_{l }^{\text{out}} kh_l^{(1)\prime}(kR)-A^{lm}\mathcal{C}_{l }^{\text{reg}} kj_l'(kR)=\lambda  A^{lm}
\mathcal{C}_{l }^{\text{reg}}j_l(kR).
\end{split}\end{equation*}Eliminating $A^{lm}$, we find that
\begin{equation*}
\begin{split}
T^{lm}=&-\frac{b^{lm}}{a^{lm}}=\frac{\mathcal{C}_{l }^{\text{reg}} }{\mathcal{C}_{l }^{\text{out}} }\frac{\lambda j_l(kR)^2}{kj_l'(kR)h_l^{(1)}(kR)-kh_l^{(1)\prime}(kR)
j_l(kR)+\lambda j_l(kR)h_l^{(1)}(kR)} .
\end{split}\end{equation*}Using the identity
\begin{equation*}
j_l'(z)h_l^{(1)}(z)-h_l^{(1)\prime}(z)j_l(z)=-\frac{i}{z^2},
\end{equation*}and passing to imaginary frequency, we find that
\begin{equation*}
T^{lm}(i\xi)= \frac{\lambda R I_{l+\frac{1}{2}}(\kappa R)^2}{
1+\lambda R I_{l+\frac{1}{2}}(\kappa R)K_{l+\frac{1}{2}}(\kappa R)},
\end{equation*}where $\kappa$ is given by \eqref{eq3_22_6}.

For   $\widetilde{T}^{lm}$,   the boundary conditions give
\begin{equation*}
\begin{split}
&a^{lm} \mathcal{C}_{l }^{\text{reg}}j_l(kR)+b^{lm}\mathcal{C}_{l }^{\text{out}} h_l^{(1)}(kR)=B^{lm}\mathcal{C}_{l }^{\text{out}}  h_l^{(1)}(kR),\\
&B^{lm}\mathcal{C}_{l }^{\text{out}} kh_l^{(1)\prime}(kR)-a^{lm}\mathcal{C}_{l }^{\text{reg}} kj_l'(kR)-b^{lm}\mathcal{C}_{l }^{\text{out}} kh_l^{(1)\prime}(kR)=
\lambda B^{lm}\mathcal{C}_{l }^{\text{out}}h_l^{(1)}(kR).
\end{split}\end{equation*}
As before, we find that
\begin{equation*}
\widetilde{T}^{lm}(i\xi)= \frac{\lambda R K_{l+\frac{1}{2}}(\kappa R)^2}{
1+\lambda R I_{l+\frac{1}{2}}(\kappa R)K_{l+\frac{1}{2}}(\kappa R)}.
\end{equation*}

  Now consider the electromagnetic case.
For the   $\mathbb{T}$-matrix, an electromagnetic field is represented as
\begin{equation*}
\mathbf{A}=\int_{-\infty}^{\infty} d\omega\sum_{l=1}^{\infty}\sum_{m=-l}^l \left(A^{lm}\mathbf{A}_{lm}^{\text{TE, reg}}(\mathbf{x}, k)
+C^{lm}\mathbf{A}_{lm}^{\text{TM, reg}}(\mathbf{x}, k)\right)e^{-i\omega t}
\end{equation*}inside the sphere, and as
\begin{equation}\label{eq3_22_12}
\mathbf{A}=\int_{-\infty}^{\infty} d\omega\sum_{l=1}^{\infty}\sum_{m=-l}^l \left(a^{lm}\mathbf{A}_{lm}^{\text{TE, reg}}(\mathbf{x}, k)
+b^{lm}\mathbf{A}_{lm}^{\text{TE, out}}(\mathbf{x}, k)
+c^{lm}\mathbf{A}_{lm}^{\text{TM, reg}}(\mathbf{x}, k)+d^{lm}\mathbf{A}_{lm}^{\text{TM, out}}(\mathbf{x}, k)\right)e^{-i\omega t}
\end{equation}outside the sphere. For the  $\widetilde{\mathbb{T}}$-matrix, an electromagnetic field is represented as \eqref{eq3_22_12} inside the sphere
and as
\begin{equation*}
\mathbf{A}=\int_{-\infty}^{\infty} d\omega\sum_{l=1}^{\infty}\sum_{m=-l}^l \left(B^{lm}\mathbf{A}_{lm}^{\text{TE, out}}(\mathbf{x}, k)
+D^{lm}\mathbf{A}_{lm}^{\text{TM, out}}(\mathbf{x}, k)\right)e^{-i\omega t}
\end{equation*} outside the sphere.

For a magnetodielectric sphere whose inside is filled with material with permittivity $\varepsilon_i$ and permeability $\mu_i$, and whose outside is a medium
  with permittivity $\varepsilon_e$ and permeability $\mu_e$, the continuities of $\varepsilon\mathbf{E}_n,
  \mathbf{E}_{\parallel}, \mathbf{B}_n$ and $\mu^{-1}\mathbf{B}_{\parallel}$ give
\begin{equation*}
\begin{split}
&a^{lm}\mathcal{C}_{l }^{\text{reg}}j_l(k_eR)+b^{lm}\mathcal{C}_{l }^{\text{out}} h_l^{(1)}(k_eR)=A^{lm}\mathcal{C}_{l }^{\text{reg}}j_l(k_iR),\\
&a^{lm}\mathcal{C}_{l }^{\text{reg}}\frac{1}{\mu_e}\Bigl(j_l(k_eR)+k_eRj_l'(k_eR)\Bigr)+b^{lm}\mathcal{C}_{l }^{\text{out}}\frac{1}{\mu_e}
\left(h_l^{(1)}(k_eR)+k_eRh_l^{(1)\prime}(k_eR)\right)=A^{lm}\mathcal{C}_{l }^{\text{reg}}\frac{1}{\mu_i}\Bigl(j_l(k_iR)+k_iRj_l'(k_iR)\Bigr),\\
&c^{lm}\mathcal{C}_{l }^{\text{reg}}\frac{\varepsilon_e}{k_e}j_l(k_eR)+d^{lm}\mathcal{C}_{l }^{\text{out}} \frac{\varepsilon_e}{k_e}h_l^{(1)}(k_eR)=
C^{lm}\mathcal{C}_{l }^{\text{reg}}\frac{\varepsilon_i}{k_i}j_l(k_iR),\\
&c^{lm}\mathcal{C}_{l }^{\text{reg}}\frac{1}{k_e}\Bigl(j_l(k_eR)+k_eRj_l'(k_eR)\Bigr)+d^{lm}\mathcal{C}_{l }^{\text{out}}\frac{1}{k_e}
\left(h_l^{(1)}(k_eR)+k_eRh_l^{(1)\prime}(k_eR)\right)=C^{lm}\mathcal{C}_{l }^{\text{reg}}\frac{1}{k_i}\Bigl(j_l(k_iR)+k_iRj_l'(k_iR)\Bigr).
\end{split}\end{equation*}
Here $k_i$ and $k_e$ are given by \eqref{eq3_23_9}.
Eliminating $A^{lm}$ and $C^{lm}$, we find that
\begin{equation*}
\begin{pmatrix} b^{lm}\\d^{lm}\end{pmatrix}=-\mathbb{T}^{lm}\begin{pmatrix} a^{lm}\\c^{lm}\end{pmatrix},
\end{equation*}where
\begin{equation*}
\mathbb{T}^{lm}=\begin{pmatrix} T_{lm}^{\text{TE}}& 0\\0 & T_{lm}^{\text{TM}}\end{pmatrix}
\end{equation*}is a diagonal matrix with elements
\begin{equation*}
\begin{split}
T_{lm}^{\text{TE}}(i\xi)=&
\frac{\mu_e I_{l+\frac{1}{2}}(\kappa_e R)\left(\frac{1}{2}I_{l+\frac{1}{2}}(\kappa_iR)+\kappa_iRI_{l+\frac{1}{2}}'(\kappa_iR)\right)-
\mu_i I_{l+\frac{1}{2}}(\kappa_i R)\left(\frac{1}{2}I_{l+\frac{1}{2}}(\kappa_eR)+\kappa_eRI_{l+\frac{1}{2}}'(\kappa_eR)\right)}{\mu_e K_{l+\frac{1}{2}}(\kappa_e R)
\left(\frac{1}{2}I_{l+\frac{1}{2}}(\kappa_iR)+\kappa_iRI_{l+\frac{1}{2}}'(\kappa_iR)\right)-
\mu_i I_{l+\frac{1}{2}}(\kappa_i R)\left(\frac{1}{2}K_{l+\frac{1}{2}}(\kappa_eR)+\kappa_eRK_{l+\frac{1}{2}}'(\kappa_eR)\right)},\\
T_{lm}^{\text{TM}}(i\xi)=&
\frac{\varepsilon_e I_{l+\frac{1}{2}}(\kappa_e R)\left(\frac{1}{2}I_{l+\frac{1}{2}}(\kappa_iR)+\kappa_iRI_{l+\frac{1}{2}}'(\kappa_iR)\right)-
\varepsilon_i I_{l+\frac{1}{2}}(\kappa_i R)\left(\frac{1}{2}I_{l+\frac{1}{2}}(\kappa_eR)+\kappa_eRI_{l+\frac{1}{2}}'(\kappa_eR)\right)}{\varepsilon_e K_{l+\frac{1}{2}}
(\kappa_e R)\left(\frac{1}{2}I_{l+\frac{1}{2}}(\kappa_iR)+\kappa_iRI_{l+\frac{1}{2}}'(\kappa_iR)\right)-
\varepsilon_i I_{l+\frac{1}{2}}(\kappa_i R)\left(\frac{1}{2}K_{l+\frac{1}{2}}(\kappa_eR)+\kappa_eRK_{l+\frac{1}{2}}'(\kappa_eR)\right)}.
\end{split}\end{equation*}
Here $\kappa_i$ and $\kappa_e$ are given by \eqref{eq3_25_1}.

For the   $\widetilde{\mathbb{T}}$-matrix, we find in a similar way that $\widetilde{\mathbb{T}}^{lm}$ is a diagonal matrix with elements
\begin{equation*}
\begin{split}
\widetilde{T}_{lm}^{\text{TE}}(i\xi)=&
\frac{\mu_i K_{l+\frac{1}{2}}(\kappa_i R)\left(\frac{1}{2}K_{l+\frac{1}{2}}(\kappa_eR)+\kappa_eRK_{l+\frac{1}{2}}'(\kappa_eR)\right)-
\mu_e K_{l+\frac{1}{2}}(\kappa_e R)\left(\frac{1}{2}K_{l+\frac{1}{2}}(\kappa_iR)+\kappa_iRK_{l+\frac{1}{2}}'(\kappa_iR)\right)}{\mu_i I_{l+\frac{1}{2}}(\kappa_i R)
\left(\frac{1}{2}K_{l+\frac{1}{2}}(\kappa_eR)+\kappa_eRK_{l+\frac{1}{2}}'(\kappa_eR)\right)-
\mu_e K_{l+\frac{1}{2}}(\kappa_e R)\left(\frac{1}{2}I_{l+\frac{1}{2}}(\kappa_iR)+\kappa_iRI_{l+\frac{1}{2}}'(\kappa_iR)\right)},\\
\widetilde{T}_{lm}^{\text{TM}}(i\xi)=&
\frac{\varepsilon_i K_{l+\frac{1}{2}}(\kappa_i R)\left(\frac{1}{2}K_{l+\frac{1}{2}}(\kappa_eR)+\kappa_eRK_{l+\frac{1}{2}}'(\kappa_eR)\right)-
\varepsilon_e K_{l+\frac{1}{2}}(\kappa_e R)\left(\frac{1}{2}K_{l+\frac{1}{2}}(\kappa_iR)+\kappa_iRK_{l+\frac{1}{2}}'(\kappa_iR)\right)}{\varepsilon_i
I_{l+\frac{1}{2}}(\kappa_i R)\left(\frac{1}{2}K_{l+\frac{1}{2}}(\kappa_eR)+\kappa_eRK_{l+\frac{1}{2}}'(\kappa_eR)\right)-
\varepsilon_e K_{l+\frac{1}{2}}(\kappa_e R)\left(\frac{1}{2}I_{l+\frac{1}{2}}(\kappa_iR)+\kappa_iRI_{l+\frac{1}{2}}'(\kappa_iR)\right)}.
\end{split}\end{equation*}

\section{The $\mathbb{T}, \widetilde{\mathbb{T}}$ matrices of a cylinder}\label{sca_cylinder}
In cylindrical coordinates $x=\rho\cos\phi, y=\rho\sin\phi, z=z$, the regular waves and outgoing waves   are parametrized by $n\in\mathbb{Z}$ and $k_z\in\mathbb{R}$.
They are given by
\begin{equation*}
\begin{split}
\varphi_{nk_z}^{\text{reg}}(\mathbf{x},k)=&\mathcal{C}_{n }^{\text{reg}}J_n(k_{\perp}\rho)e^{in\phi+ik_zz},\\
\varphi_{nk_z}^{\text{out}}(\mathbf{x},k)=&\mathcal{C}_{n }^{\text{out}}H_n^{(1)}(k_{\perp}\rho)e^{in\phi+ik_zz},
\end{split}
\end{equation*}
and
\begin{equation*}
\begin{split}
\mathbf{A}_{nk_z}^{\text{TE}, *}(\mathbf{x},k)=&\frac{1}{k_{\perp}}\nabla\times \varphi_{nk_z}^*\mathbf{e}_z\\
=& \mathcal{C}_n^*\left(\frac{in}{k_{\perp}\rho }f_n^*(k_{\perp}\rho)\mathbf{e}_{\rho}-f_n^{*\prime}(k_{\perp}\rho)\mathbf{e}_{\phi}\right)e^{in\phi+ik_zz},\\
\mathbf{A}_{nk_z}^{\text{TM}, *}(\mathbf{x},k)=&\frac{1}{kk_{\perp}}\nabla\times\nabla\times \varphi_{nk_z}^*\mathbf{e}_z\\
=&\mathcal{C}_n^*\left(\frac{ik_z}{k}f_n^{*\prime}(k_{\perp}\rho)\mathbf{e}_{\rho}-\frac{nk_z}{kk_{\perp}\rho}f_n^*(k_{\perp}\rho)\mathbf{e}_{\phi}+\frac{k_{\perp}}{k}
f_n^*(k_{\perp}\rho)\mathbf{e}_z\right)e^{in\phi+ik_zz}.
\end{split}
\end{equation*}
Here
\begin{equation*}
\begin{split}
k_{\perp}=&\sqrt{k^2-k_z^2},\\
\mathcal{C}_n^{\text{reg}}=&i^{-n},\quad\mathcal{C}_n^{\text{out}}= \frac{\pi  }{2}i^{n+1},\\
f_n^{\text{reg}}(z)=&J_n(z),\quad f_n^{\text{out}}=H_n^{(1)}(z).
\end{split}
\end{equation*}
The constants $\mathcal{C}_n^{\text{reg}}$ and $\mathcal{C}_n^{\text{out}}$ are chosen so that
\begin{equation*}
\mathcal{C}_n^{\text{reg}}J_n(i\zeta)=I_n(\zeta),\hspace{1cm}\mathcal{C}_n^{\text{out}}H_n^{(1)}(i\zeta)=K_n(\zeta).
\end{equation*}

Consider a     cylinder $\rho=R$ of radius $R$ and length $H$.  For a scalar field, to find the  $\mathbb{T}$-matrix, let
\begin{equation*}
\varphi=H\int_{-\infty}^{\infty}d\omega \int_{-\infty}^{\infty}\frac{dk_z}{2\pi}\sum_{n=-\infty}^{\infty}A^{nk_z}\varphi_{nk_z}^{\text{reg}}(\mathbf{x},k)e^{-i\omega t}
\end{equation*} inside the cylinder, and
\begin{equation}\label{eq3_25_2}
\varphi=H\int_{-\infty}^{\infty}d\omega \int_{-\infty}^{\infty}\frac{dk_z}{2\pi}\sum_{n=-\infty}^{\infty}\left(
a^{nk_z}\varphi_{nk_z}^{\text{reg}}(\mathbf{x},k)+b^{nk_z}\varphi_{nk_z}^{\text{out}}(\mathbf{x},k)\right)e^{-i\omega t}
\end{equation}outside the cylinder. For the   $\widetilde{\mathbb{T}}$-matrix, let the scalar field be represented by \eqref{eq3_25_2} inside the cylinder, and by
\begin{equation*}
\varphi=H\int_{-\infty}^{\infty}d\omega \int_{-\infty}^{\infty}\frac{dk_z}{2\pi}\sum_{n=-\infty}^{\infty}B^{nk_z}\varphi_{nk_z}^{\text{out}}(\mathbf{x},k)e^{-i\omega t}
\end{equation*} outside the cylinder.

First let us consider a scalar field with boundary condition $u\varphi+\rho\pa_{\rho}\varphi=0$ on the cylinder.
This gives
\begin{equation*}
a^{nk_z}\mathcal{C}_n^{\text{reg}}\left(uJ_n(k_{\perp}R)+vk_{\perp}RJ_n'(k_{\perp}R)\right)+b^{nk_z}\mathcal{C}_n^{\text{out}}\left(uH_n(k_{\perp}R)+
vk_{\perp}RH_n^{(1)\prime}(k_{\perp}R)\right)=0.
\end{equation*}
Therefore,
\begin{equation*}
\begin{split}
T^{nk_z}(i\xi)=-\frac{b^{nk_z}}{a^{nk_z}}=\frac{uI_n(\gamma R)+v\gamma RI_n'(\gamma R)}{uK_n(\gamma R)+v\gamma RH_n^{(1)\prime}(\gamma R)},\\
\widetilde{T}^{nk_z}(i\xi)=-\frac{a^{nk_z}}{b^{nk_z}}=\frac{uK_n(\gamma R)+v\gamma RH_n^{(1)\prime}(\gamma R)}{uI_n(\gamma R)+v\gamma RI_n'(\gamma R)},
\end{split}
\end{equation*}where
\begin{equation*}
\gamma=\sqrt{\frac{\xi^2}{c^2}+k_z^2+\frac{m^2c^2}{\hbar^2}}.
\end{equation*}
Next consider the case of a semitransparent cylinder. For the  $\mathbb{T}$-matrix, the boundary conditions \eqref{eq3_22_4} imply that
\begin{equation*}
\begin{split}
&a^{nk_z}\mathcal{C}_n^{\text{reg}}J_n(k_{\perp}R)+b^{nk_z}\mathcal{C}_n^{\text{out}}H_n^{(1)}(k_{\perp}R)=A^{nk_z}\mathcal{C}_n^{\text{reg}}J_n(k_{\perp}R),\\
&a^{nk_z}\mathcal{C}_n^{\text{reg}}k_{\perp}J_n'(k_{\perp}R)+b^{nk_z}\mathcal{C}_n^{\text{out}}k_{\perp}H_n^{(1)\prime}(k_{\perp}R)-A^{nk_z}
\mathcal{C}_n^{\text{reg}}k_{\perp}J_n'(k_{\perp}R)=\lambda A^{nk_z}\mathcal{C}_n^{\text{reg}}J_n(k_{\perp}R).
\end{split}
\end{equation*}Eliminating $A^{nk_z}$ and passing to imaginary frequency, we find that
\begin{equation*}
\begin{split}
T^{nk_z}=-\frac{b^{nk_z}}{a^{nk_z}}=\frac{\lambda RI_n(\gamma R)^2}{1+\lambda R I_n(\gamma R)K_n(\gamma R)}.
\end{split}
\end{equation*}Similarly, we find that
\begin{equation*}
\begin{split}
\widetilde{T}^{nk_z}= \frac{\lambda RK_n(\gamma R)^2}{1+\lambda R I_n(\gamma R)K_n(\gamma R)}.
\end{split}
\end{equation*}

The $\mathbb{T}$ and $ \widetilde{\mathbb{T}}$ matrices for electromagnetic fields are more complicated. Assume that a magnetodielectric cylinder has permittivity $\vep_i$ and permeability $\mu_i$,
and it is surrounded by medium of permittivity $\vep_e$ and permeability $\mu_e$. For the $\mathbb{T}$-matrix, the continuities of $\vep \mathbf{E}_n,
\mathbf{E}_{\parallel}, \mathbf{B}_n, \mu^{-1}\mathbf{B}_{\parallel}$ across the surface give
\begin{equation}\label{eq3_27_1}
c^{nk_z}\mathcal{C}_n^{\text{reg}}\frac{k_{\perp, e}}{k_e}J_n(k_{\perp,e}R)+d^{nk_z}\mathcal{C}_n^{\text{out}} \frac{k_{\perp, e}}{k_e}H_n^{(1)}(k_{\perp,e}R)=C^{nk_z}
\mathcal{C}_n^{\text{reg}}\frac{k_{\perp, i}}{k_i}J_n(k_{\perp,i}R),
\end{equation}\begin{equation}\label{eq3_27_2}\begin{split}
a^{nk_z}\mathcal{C}_n^{\text{reg}}J_n'(k_{\perp,e}R)+b^{nk_z}\mathcal{C}_n^{\text{out}}H_n^{(1)\prime}(k_{\perp,e}R)+
c^{nk_z}\mathcal{C}_n^{\text{reg}}\frac{nk_z}{k_ek_{\perp,e}R}J_n(k_{\perp,e}R)+d^{nk_z}\mathcal{C}_n^{\text{out}}\frac{nk_z}{k_ek_{\perp,e}R}H_n^{(1)}(k_{\perp,e}R)\\
=A^{nk_z}\mathcal{C}_n^{\text{reg}}J_n'(k_{\perp,i}R)+C^{nk_z} \mathcal{C}_n^{\text{reg}}\frac{nk_z}{k_ik_{\perp,i}R}J_n(k_{\perp,i}R),\end{split}
\end{equation}\begin{equation}\label{eq3_27_3}
a^{nk_z}\mathcal{C}_n^{\text{reg}}\frac{k_{\perp, e}}{\mu_e}J_n(k_{\perp,e}R)+b^{nk_z}\mathcal{C}_n^{\text{out}}\frac{k_{\perp, e}}{\mu_e}H_n^{(1)}(k_{\perp,e}R)=A^{nk_z}
\mathcal{C}_n^{\text{reg}}\frac{k_{\perp, i}}{\mu_i}J_n(k_{\perp,i}R),
\end{equation}\begin{equation}\label{eq3_27_4}\begin{split}
a^{nk_z}\mathcal{C}_n^{\text{reg}}\frac{nk_z}{\mu_ek_{\perp,e}R}J_n(k_{\perp,e}R)+b^{nk_z}\mathcal{C}_n^{\text{out}}\frac{nk_z}{\mu_ek_{\perp,e}R}H_n^{(1)}(k_{\perp,e}R)
+c^{nk_z}\mathcal{C}_n^{\text{reg}}\frac{k_e}{\mu_e}J_n'(k_{\perp,e}R)+d^{nk_z}\mathcal{C}_n^{\text{out}}\frac{k_e}{\mu_e}H_n^{(1)\prime}(k_{\perp,e}R)\\
=A^{nk_z}\mathcal{C}_n^{\text{reg}}\frac{nk_z}{\mu_ik_{\perp,i}R}J_n(k_{\perp,i}R)+C^{nk_z} \mathcal{C}_n^{\text{reg}}\frac{k_i}{\mu_i}J_n'(k_{\perp,i}R),
\end{split}\end{equation}where
$k_i$ and $k_e$ are defined in \eqref{eq3_23_9}. The first two conditions come  from the continuities of $\mathbf{E}_{\parallel}$. They imply the continuity of
$\mathbf{B}_n$. The last two conditions come from the continuities of $\mu^{-1}\mathbf{B}_{\parallel}$. They imply the continuity of $\vep\mathbf{E}_n$.
By eliminating $A^{nk_z}$ and $C^{nk_z}$, we can solve $b^{nk_z}$ and $d^{nk_z}$ in terms of $a^{nk_z}$ and $c^{nk_z}$, which can be presented in the form:
\begin{equation*}
\begin{pmatrix}
b^{nk_z}\\ d^{nk_z}\end{pmatrix}=-\mathbb{T}^{nk_z}\begin{pmatrix}
a^{nk_z}\\ c^{nk_z}\end{pmatrix}=-\begin{pmatrix}
T_{nk_z}^{\text{TE,TE}}& T_{nk_z}^{\text{TE,TM}}\\ T_{nk_z}^{\text{TM,TE}}& T_{nk_z}^{\text{TM,TM}}\end{pmatrix}\begin{pmatrix}
b^{nk_z}\\ d^{nk_z}\end{pmatrix},
\end{equation*}where
\begin{equation*}
\begin{split}
T_{nk_z}^{\text{TE,TE}}(i\xi)=&\frac{1}{\Delta_{nk_z}}\left(\Delta_{nk_z}^{\text{TE,1}}\Delta_{nk_z}^{\text{TM,2}}+\frac{n^2k_z^2\xi^2}{\gamma_i^2\gamma_e^2 R^2c^4}
(n_e^2-n_i^2)^2I_n(\gamma_eR)K_n(\gamma_eR) I_n(\gamma_iR)^2\right),\\
T_{nk_z}^{\text{TM,TM}}(i\xi)=&\frac{1}{\Delta_{nk_z}}\left(\Delta_{nk_z}^{\text{TE,2}}\Delta_{nk_z}^{\text{TM,1}}+\frac{n^2k_z^2\xi^2}{\gamma_i^2\gamma_e^2 R^2c^4}
(n_e^2-n_i^2)^2I_n(\gamma_eR)K_n(\gamma_eR) I_n(\gamma_iR)^2\right),\\
T_{nk_z}^{\text{TE,TM}}(i\xi)=&T_{nk_z}^{\text{TM,TE}}(i\xi)=-\frac{1}{\Delta_{nk_z}}\frac{ink_z\kappa_e}{\gamma_e^2R^2c^2}(n_e^2-n_i^2)I_n(\gamma_iR)^2,\\
\Delta_{nk_z}^{\text{TE,1}}=&\gamma_i\mu_eI_n(\gamma_iR)I_n'(\gamma_eR)-\gamma_e\mu_iI_n(\gamma_eR)I_n'(\gamma_iR),\\
\Delta_{nk_z}^{\text{TE,2}}=&\gamma_i\mu_eI_n(\gamma_iR)K_n'(\gamma_eR)-\gamma_e\mu_iK_n(\gamma_eR)I_n'(\gamma_iR),\\
\Delta_{nk_z}^{\text{TM,1}}=&\gamma_i\vep_eI_n(\gamma_iR)I_n'(\gamma_eR)-\gamma_e\vep_iI_n(\gamma_eR)I_n'(\gamma_iR),\\
\Delta_{nk_z}^{\text{TM,2}}=&\gamma_i\vep_eI_n(\gamma_iR)K_n'(\gamma_eR)-\gamma_e\vep_iK_n(\gamma_eR)I_n'(\gamma_iR),\\
\Delta_{nk_z}=&\Delta_{nk_z}^{\text{TE,2}}\Delta_{nk_z}^{\text{TM,2}}+\frac{n^2k_z^2\xi^2}{\gamma_i^2\gamma_e^2 R^2c^4}(n_e^2-n_i^2)^2K_n(\gamma_eR)^2I_n(\gamma_iR)^2,\\
\gamma_e=&\sqrt{\kappa_e^2+k_z^2},\quad \gamma_i=\sqrt{\kappa_i^2+k_z^2},
\end{split}
\end{equation*}where $\kappa_i$ and $\kappa_e$ are given by \eqref{eq3_25_1}.

For the   $\widetilde{\mathbb{T}}$-matrix, we find in a similar way that
\begin{equation*}
\begin{split}
\widetilde{T}_{nk_z}^{\text{TE,TE}}(i\xi)=&\frac{1}{\widetilde{\Delta}_{nk_z}}\left(\widetilde{\Delta}_{nk_z}^{\text{TE,1}}\widetilde{\Delta}_{nk_z}^{\text{TM,2}}
+\frac{n^2k_z^2\xi^2}
{\gamma_i^2\gamma_e^2 R^2c^4}(n_e^2-n_i^2)^2I_n(\gamma_iR)K_n(\gamma_iR) K_n(\gamma_eR)^2\right),\\
\widetilde{T}_{nk_z}^{\text{TM,TM}}(i\xi)=&\frac{1}{\widetilde{\Delta}_{nk_z}}\left(\widetilde{\Delta}_{nk_z}^{\text{TE,2}}\widetilde{\Delta}_{nk_z}^{\text{TM,1}}
+\frac{n^2k_z^2\xi^2}{\gamma_i^2\gamma_e^2 R^2c^4}(n_e^2-n_i^2)^2I_n(\gamma_iR)K_n(\gamma_iR) K_n(\gamma_eR)^2\right),\\
\widetilde{T}_{nk_z}^{\text{TE,TM}}(i\xi)=&\widetilde{T}_{nk_z}^{\text{TM,TE}}(i\xi)=-\frac{1}{\widetilde{\Delta}_{nk_z}}\frac{ink_z\kappa_i}{\gamma_i^2R^2c^2}
(n_e^2-n_i^2)K_n(\gamma_eR)^2,\\
\widetilde{\Delta}_{nk_z}^{\text{TE,1}}=&\gamma_e\mu_iK_n(\gamma_eR)K_n'(\gamma_iR)-\gamma_i\mu_eK_n(\gamma_iR)K_n'(\gamma_eR),\\
\widetilde{\Delta}_{nk_z}^{\text{TE,2}}=&\gamma_e\mu_iK_n(\gamma_eR)I_n'(\gamma_iR)-\gamma_i\mu_eI_n(\gamma_iR)K_n'(\gamma_eR),\\
\widetilde{\Delta}_{nk_z}^{\text{TM,1}}=&\gamma_e\vep_iK_n(\gamma_eR)K_n'(\gamma_iR)-\gamma_i\vep_eK_n(\gamma_iR)K_n'(\gamma_eR),\\
\widetilde{\Delta}_{nk_z}^{\text{TM,2}}=&\gamma_e\vep_iK_n(\gamma_eR)I_n'(\gamma_iR)-\gamma_i\vep_eI_n(\gamma_iR)K_n'(\gamma_eR),\\
\widetilde{\Delta}_{nk_z}=&\widetilde{\Delta}_{nk_z}^{\text{TE,2}}\widetilde{\Delta}_{nk_z}^{\text{TM,2}}+\frac{n^2k_z^2\xi^2}
{\gamma_i^2\gamma_e^2 R^2c^4}(n_e^2-n_i^2)^2K_n(\gamma_eR)^2I_n(\gamma_iR)^2.
\end{split}
\end{equation*}
Notice that the   $\mathbb{T}$ and $\widetilde{\mathbb{T}}$ matrices are diagonal if and only if $n_i=n_e$, i.e., the media inside and outside the
cylinder have the same refractive index. In this case,
\begin{equation*}
\begin{split}
T_{nk_z}^{\text{TE,TE}}=&\frac{\gamma_i\mu_eI_n(\gamma_iR)I_n'(\gamma_eR)-\gamma_e\mu_iI_n(\gamma_eR)I_n'(\gamma_iR)}
{\gamma_i\mu_eI_n(\gamma_iR)K_n'(\gamma_eR)-\gamma_e\mu_iK_n(\gamma_eR)I_n'(\gamma_iR)},\\
T_{nk_z}^{\text{TM,TM}}=&\frac{\gamma_i\vep_eI_n(\gamma_iR)I_n'(\gamma_eR)-\gamma_e\vep_iI_n(\gamma_eR)I_n'(\gamma_iR)}{\gamma_i\vep_eI_n(\gamma_iR)
K_n'(\gamma_eR)-\gamma_e\vep_iK_n(\gamma_eR)I_n'(\gamma_iR)},\\
\widetilde{T}_{nk_z}^{\text{TE,TE}}=&\frac{\gamma_e\mu_iK_n(\gamma_eR)K_n'(\gamma_iR)-\gamma_i\mu_eK_n(\gamma_iR)K_n'(\gamma_eR)}{\gamma_e\mu_i
K_n(\gamma_eR)I_n'(\gamma_iR)-\gamma_i\mu_eI_n(\gamma_iR)K_n'(\gamma_eR)},\\
\widetilde{T}_{nk_z}^{\text{TM,TM}}=&\frac{\gamma_e\vep_iK_n(\gamma_eR)K_n'(\gamma_iR)-\gamma_i\vep_eK_n(\gamma_iR)K_n'(\gamma_eR)}{\gamma_e\vep_i
K_n(\gamma_eR)I_n'(\gamma_iR)-\gamma_i\vep_eI_n(\gamma_iR)K_n'(\gamma_eR)}.
\end{split}
\end{equation*}

\section{Casimir effect between two spheres}\label{sphere_sphere}
In this section, we derive the $TGTG$ formula for the Casimir energy between two spheres using the prescription in Section \ref{gen}. Since we have discussed the   $\mathbb{T}$ and $\widetilde{\mathbb{T}}$ matrices in Section \ref{sca_sphere}, now we only have to find the translation matrices $\mathbb{U}, \mathbb{V}$ and $\mathbb{W}$. These matrices have been derived in a numbers of works \cite{6,53,54,55,56,57}.
 Since the operator approach used in \cite{6} can be used for other scenarios, we will discuss in detail   the approach of \cite{6}.

Consider two spheres, one is centered at $O=(0,0,0)$ with radius $R_1$ and one is centered at $O'=(0,0,L)$ with radius $R_2$. Assume that $R_2>R_1$.

Define a differential operator $\mathcal{P}_{lm}$ by
\begin{equation*}
\begin{split}
\mathcal{P}_{lm}=&(-1)^m \sqrt{\frac{2l+1}{4\pi}\frac{(l-m)!}{(l+m)!}}\left(\frac{\pa_x+i\pa_y}{ik}\right)^mP_l^{(m)}\left(\frac{\pa_z}{ik}\right), \\
\mathcal{P}_{l,-m}=&  \sqrt{\frac{2l+1}{4\pi}\frac{(l-m)!}{(l+m)!}}\left(\frac{\pa_x-i\pa_y}{ik}\right)^mP_l^{(m)}\left(\frac{\pa_z}{ik}\right).
\end{split}
\end{equation*}Here $m\geq 0$,  $P_l(z)$ is the Legendre polynomial of degree $l$ and $P_l^{(m)}(z)$ is its $m$-times derivative. It follows from the definition of
spherical harmonics \cite{17} that
\begin{equation*}
\mathcal{P}_{lm}e^{i\mathbf{k}\cdot\mathbf{r}}=Y_{lm}(\theta_k,\phi_k)e^{i\mathbf{k}\cdot\mathbf{r}}.
\end{equation*}Here $\mathbf{k}=  k_x\mathbf{e}_x +k_y\mathbf{e}_y+ k_z\mathbf{e}_z$, $\mathbf{r}=x\mathbf{e}_x+y\mathbf{e}_y+z\mathbf{e}_z$,
$k_x=k\sin\theta_k\cos\phi_k, k_y=k\sin\theta_k\sin\phi_k, k_z=k\cos\theta_k$. On the other hand, it can be proved by induction that
\begin{equation}\label{eq5_2_1}
\begin{split}
\mathcal{P}_{lm}j_0(kr)=i^lj_l(kr)Y_{lm}(\theta,\phi),\hspace{1cm}\mathcal{P}_{lm}h_0^{(1)}(kr)=i^lh_l^{(1)}(kr)Y_{lm}(\theta,\phi).
\end{split}
\end{equation}
Using the integral representations
\begin{equation*}
\begin{split}
j_0(kr)=&\frac{\sin (kr)}{kr}=\frac{1}{4\pi}\int_0^{2\pi}d\phi_k\int_0^{\pi}d\theta_k \sin\theta_k e^{i\mathbf{k}\cdot\mathbf{r}},\\
h_0(kr)=&\frac{\exp(ikr)}{ikr}=\frac{1}{2\pi}\int_{-\infty}^{\infty} dk_x\int_{-\infty}^{\infty} dk_y \frac{e^{ik_xx+ik_yy \pm i\sqrt{k^2-k_x^2-k_y^2}z}}
{k\sqrt{k^2-k_x^2-k_y^2}},\quad z\gtrless 0,
\end{split}
\end{equation*}where $\sqrt{k^2-k_x^2-k_y^2}=i\sqrt{k_x^2+k_y^2-k^2}$ if $k_x^2+k_y^2>k^2$, we have
\begin{equation}\label{eq5_9_1}
\begin{split}
\varphi_{lm}^{\text{reg}}(\mathbf{x},k)
=&\frac{\mathcal{C}_l^{\text{reg}}}{4\pi i^l}\mathcal{P}_{lm}
\int_0^{2\pi}d\phi_k\int_0^{\pi}d\theta_k \sin\theta_k e^{i\mathbf{k}\cdot\mathbf{r}}\\
=&\frac{1}{4\pi i^l}\int_0^{2\pi}d\phi_k\int_0^{\pi}d\theta_k \sin\theta_k Y_{lm}(\theta_k,\phi_k)e^{i\mathbf{k}\cdot\mathbf{r}},\end{split}
\end{equation}and similarly
\cite{15}:
\begin{equation}\label{eq4_2_3}
\begin{split}
\varphi_{lm}^{\text{out}}(\mathbf{x},k)
=&\frac{\mathcal{C}_l^{\text{out}}}{2\pi i^l}
\int_{-\infty}^{\infty} dk_x\int_{-\infty}^{\infty} dk_y  Y_{lm}(\theta_k,\phi_k)\frac{e^{ik_xx+ik_yy \pm i\sqrt{k^2-k_x^2-k_y^2}z}}{k\sqrt{k^2-k_x^2-k_y^2}},\quad z\gtrless 0.
\end{split}
\end{equation}
Here $k\cos\theta_k=k_z=\pm\sqrt{k^2-k_x^2-k_y^2}$ for $z\gtrless 0$. The integral representations \eqref{eq5_9_1} and \eqref{eq4_2_3} will be very useful later. In fact \eqref{eq5_9_1} is equivalent to the well-known formula
\begin{equation}\label{eq4_3_3}
e^{i\mathbf{k}\cdot\mathbf{r}}=4\pi\sum_{l=0}^{\infty}\sum_{m=-l}^l i^l j_l(kr)Y_{lm}(\theta,\phi)Y_{lm}^*(\theta_k,\phi_k).
\end{equation}

Now, consider the expansions which define the components of the translation matrices:
\begin{align}\label{eq4_26_1}
\varphi_{l'm'}^{\text{reg}}(\mathbf{x}-\mathbf{L},k)=&\sum_{l=0}^{\infty}\sum_{m=-l}^{l} V_{lm,l'm'}(-\mathbf{L})\varphi_{lm}^{\text{reg}}(\mathbf{x},k),
\\
\varphi_{l'm'}^{\text{out}}(\mathbf{x}-\mathbf{L},k)=&\sum_{l=0}^{\infty}\sum_{m=-l}^{l} U_{lm,l'm'}(-\mathbf{L})\varphi_{lm}^{\text{reg}}(\mathbf{x},k),\label{eq4_27_5}\\
\varphi_{lm}^{\text{out}}(\mathbf{x}'+\mathbf{L},k)=&\sum_{l'=0}^{\infty}\sum_{m'=-l'}^{l'} U_{l'm',lm}(\mathbf{L})\varphi_{l'm'}^{\text{reg}}(\mathbf{x}',k),\label{eq4_27_6}\\
\varphi_{lm}^{\text{out}}(\mathbf{x}'+\mathbf{L},k)=&\sum_{l'=0}^{\infty}\sum_{m'=-l'}^{l'} W_{l'm',lm}(\mathbf{L})\varphi_{l'm'}^{\text{out}}(\mathbf{x}',k).\label{eq4_26_2}
\end{align}Since
\begin{equation}\label{eq4_9_6}
\begin{split}
\left(P_{l''m''}\varphi_{lm}^{\text{reg}}\right)(\mathbf{0})=&\frac{\mathcal{C}_{l}^{\text{reg}}}{4\pi i^{l}}\int_0^{2\pi}d\phi_k\int_0^{\pi}d\theta_k \sin\theta_k
Y_{lm}(\theta_k,\phi_k)
Y_{l''m''}(\theta_k,\phi_k)=\frac{(-1)^{m}}{4\pi i^{l}}\delta_{l,l''}\delta_{m'',-m}\mathcal{C}_{l}^{\text{reg}},
\end{split}\end{equation}
apply  the operator $\mathcal{P}_{l,-m}$ to both sides of \eqref{eq4_26_1} and set $\mathbf{x}=\mathbf{0}$, we find that
\begin{equation*}
\begin{split}
V_{lm,l'm'}(-\mathbf{L})=&\frac{4\pi i^{l}}{\mathcal{C}_{l}^{\text{reg}}}(-1)^{m}\left(\mathcal{P}_{l,-m}\varphi_{l'm'}^{\text{reg}}\right)( -\mathbf{L})\\
=&  (-1)^{m}\int_0^{2\pi}d\phi_k\int_0^{\pi}d\theta_k \sin\theta_k Y_{l,-m}(\theta_k,\phi_k)
Y_{l'm'}(\theta_k,\phi_k)e^{i\mathbf{k}\cdot\mathbf{L}}.
\end{split}
\end{equation*}Here we have used the fact that $\left(\mathcal{P}_{l,-m}\varphi_{l'm'}^{\text{reg}}\right)( -\mathbf{L})=
(-1)^{l+l'}\left(\mathcal{P}_{l,-m}\varphi_{l'm'}^{\text{reg}}\right)( \mathbf{L})$.
Now using the identities \cite{8}:
\begin{equation}\label{eq4_3_5}
\begin{split}
&Y_{l,-m}(\theta_k,\phi_k)
Y_{l'm'}(\theta_k,\phi_k)\\=&\sum_{l''=0}^{\infty}\sum_{m''=-l''}^{l''}(-1)^{m''} \sqrt{\frac{(2l+1)(2l'+1)(2l''+1)}{4\pi}}\begin{pmatrix}
l& l' &l''\\-m &m' &-m''\end{pmatrix}\begin{pmatrix} l & l' & l''\\0 & 0 & 0\end{pmatrix}Y_{l''m''}(\theta_k,\phi_k),
\end{split}\end{equation}
\begin{equation*}
\begin{pmatrix}
l& l' &l''\\-m &m' &-m''\end{pmatrix}=(-1)^{l+l'+l''}\begin{pmatrix}
l& l' &l''\\m &-m' &m''\end{pmatrix},
\end{equation*}and the fact that
\begin{equation*}
\begin{pmatrix}
l& l' &l''\\m &-m' &m''\end{pmatrix}\begin{pmatrix}
l& l' &l''\\0 &0 &0\end{pmatrix}
\end{equation*}is nonzero only if $l+l'+l''$ is even,  $|l-l'|\leq l''\leq l+l'$, and $m-m'+m''=0$,
we find that
\begin{equation*}\begin{split}
V_{lm,l'm'}(-\mathbf{L})=& (-1)^{m} \sum_{l''=|l-l'|}^{l+l'}\sum_{m''=-l''}^{l''}(-1)^{m''}\delta_{m'',m'-m}\frac{4\pi i^{l''}}{\mathcal{C}_{l''}^{\text{reg}}}\sqrt{\frac{(2l+1)(2l'+1)(2l''+1)}{4\pi}}
\\
&\hspace{4cm}\times \begin{pmatrix}
l& l' &l''\\m &-m' &m''\end{pmatrix}\begin{pmatrix} l & l' & l''\\0 & 0 & 0\end{pmatrix} \varphi_{l''m''}^{\text{reg}}(\mathbf{L},k).\end{split}
\end{equation*}
In the case $\mathbf{L}=L\mathbf{e}_z$, we have $$\varphi_{l''m''}^{\text{reg}}(\mathbf{L},k)
= \mathcal{C}_{l''}^{\text{reg}}\sqrt{\frac{2l''+1}{4\pi}}j_{l''}(kL)\delta_{m'',0}.$$
Passing to the imaginary frequency, we find that
\begin{equation*}
V_{lm,l'm'}(-L\mathbf{e}_z,i\xi)=(-1)^m \delta_{m,m'}\sqrt{\frac{\pi }{2\kappa L}}\sum_{l''=|l-l'|}^{l+l'}(-1)^{l''}H_{ll';m}^{l''}I_{l''+\frac{1}{2}}(\kappa L),
\end{equation*}where
\begin{equation}\label{eq4_3_6}
H_{ll';m}^{l''}=\sqrt{ (2l+1)(2l'+1)}(2l''+1)\begin{pmatrix} l & l' & l''\\0 & 0 & 0\end{pmatrix}\begin{pmatrix} l & l' & l''\\m & -m& 0\end{pmatrix}.
\end{equation}
In the same way, \eqref{eq4_9_6} and \eqref{eq4_27_5} imply that
\begin{equation*}
\begin{split}
U_{lm,l'm'}(-\mathbf{L})=&\frac{4\pi i^{l}}{\mathcal{C}_{l}^{\text{reg}}}(-1)^{m}\left(\mathcal{P}_{l,-m}\varphi_{l'm'}^{\text{out}}\right)( -\mathbf{L})\\
=& \pi (-1)^{l'+1}(-1)^{m}\int_{-\infty}^{\infty} dk_x\int_{-\infty}^{\infty} dk_y  Y_{l,-m}(\theta_k,\phi_k)Y_{l'm'}(\theta_k,\phi_k)\frac{e^{ik_xx+ik_yy +
i\sqrt{k^2-k_x^2-k_y^2}z}}{k\sqrt{k^2-k_x^2-k_y^2}}\Biggr|_{\mathbf{r}=\mathbf{L}}\\
=& (-1)^{l'+1}(-1)^{m}\sum_{l''=|l-l'|}^{l+l'}\sum_{m''=-l''}^{l''}(-1)^{m''}\delta_{m'',m'-m}\frac{2\pi^2 i^{l''}}{\mathcal{C}_{l''}^{\text{out}}}\sqrt{\frac{(2l+1)(2l'+1)(2l''+1)}{4\pi}}\\
&\hspace{4cm}\times \begin{pmatrix}
l& l' &l''\\m &-m' &m''\end{pmatrix}\begin{pmatrix} l & l' & l''\\0 & 0 & 0\end{pmatrix} \varphi_{l''m''}^{\text{out}}(\mathbf{L},k).
\end{split}
\end{equation*}
Specializing to $\mathbf{L}=L\mathbf{e}_z$ and passing to imaginary frequency, we have
\begin{equation*}
U_{lm,l'm'}(-L\mathbf{e}_z,i\xi)=(-1)^{l'+m}\delta_{m,m'}\sqrt{\frac{\pi }{2\kappa L}}\sum_{l''=|l-l'|}^{l+l'}H_{ll';m}^{l''}K_{l''+\frac{1}{2}}(\kappa L).
\end{equation*}Analogously, \eqref{eq4_9_6} and \eqref{eq4_27_6} imply that
\begin{equation*}
U_{l'm',lm}(L\mathbf{e}_z,i\xi)=(-1)^{l'+m}\delta_{m,m'}\sqrt{\frac{\pi }{2\kappa L}}\sum_{l''=|l-l'|}^{l+l'}H_{ll';m}^{l''}K_{l''+\frac{1}{2}}(\kappa L).
\end{equation*}
Finally, notice that for general $\mathbf{L}$,
\begin{equation*}
\begin{split}
\varphi_{lm}^{\text{out}}(\mathbf{x}'+\mathbf{L})=&\sum_{l'=0}^{\infty}\sum_{m'=-l'}^{l'} U_{l'm',lm}(\mathbf{L})\varphi_{l'm'}^{\text{reg}}(\mathbf{x}')\\
=&\sum_{l'=0}^{\infty}\sum_{m'=-l'}^{l'}(-1)^{l'+m'}\sum_{l''=0}^{\infty}\sum_{m''=-l''}^{l''}(-1)^{m''}4\pi\sqrt{\frac{(2l+1)(2l'+1)(2l''+1)}{4\pi}}\\
&\hspace{4cm}\times \begin{pmatrix}
l& l' &l''\\m &-m' &-m''\end{pmatrix}\begin{pmatrix} l & l' & l''\\0 & 0 & 0\end{pmatrix} \varphi_{l''m''}^{\text{out}}
(\mathbf{L},k)\varphi_{l'm'}^{\text{reg}}(\mathbf{x}',k).
\end{split}
\end{equation*}Interchanging $\mathbf{x}'$ and $\mathbf{L}$, we have
\begin{equation*}
\begin{split}
\varphi_{lm}^{\text{out}}(\mathbf{x}'+\mathbf{L})
=& \sum_{l'=0}^{\infty}\sum_{m'=-l'}^{l'}(-1)^{m'} \sum_{l''=0}^{\infty}\sum_{m''=-l''}^{l''}(-1)^{l''+m''} 4\pi \sqrt{\frac{(2l+1)(2l'+1)(2l''+1)}{4\pi}}\begin{pmatrix}
l& l' &l''\\m &-m' &-m''\end{pmatrix}\begin{pmatrix} l & l' & l''\\0 & 0 & 0\end{pmatrix}\\
&\hspace{4cm}\times \varphi_{l''m''}^{\text{reg}}(\mathbf{L},k)\varphi_{l'm'}^{\text{out}}(\mathbf{x}',k).
\end{split}
\end{equation*}Compare to \eqref{eq4_26_2} and specialize to $\mathbf{L}=L\mathbf{e}_z$, we find that
\begin{equation*}
W_{l'm',lm}(L\mathbf{e}_z,i\xi)=(-1)^{m}\delta_{m,m'}\sqrt{\frac{\pi }{2\kappa L}}\sum_{l''=|l-l'|}^{l+l'}(-1)^{l''}H_{ll';m}^{l''}I_{l''+\frac{1}{2}}(\kappa L).
\end{equation*}
Hence, for a scalar interaction between two spheres, the Casimir interaction is given by
\begin{equation*}
E_{\text{Cas}}=\frac{\hbar}{2\pi}\int_0^{\infty}d\xi \text{Tr}\,\ln\left(1-\mathbb{M}(i\xi)\right),
\end{equation*}where the trace  Tr is
\begin{equation}\label{eq4_27_4}
\text{Tr}=\sum_{m=-\infty}^{\infty}\sum_{l=|m|}^{\infty},
\end{equation}and $\mathbb{M}$ is a matrix diagonal in $m$. When the two spheres are outside each other,
\begin{equation*}
\begin{split}
M_{lm,l'm'}(i\xi)=&\delta_{m,m'}T_1^{lm}\sum_{\tilde{l}=|m|}^{\infty}\sqrt{\frac{\pi }{2\kappa L}}\sum_{l''=|l-\tilde{l}|}^{l+\tilde{l}}H_{l\tilde{l};m}^{l''}
K_{l''+\frac{1}{2}}(\kappa L) T^{\tilde{l}m}_2\sqrt{\frac{\pi }{2\kappa L}}\sum_{\tilde{l}''=|l'-\tilde{l}|}^{l'+\tilde{l}}H_{l'\tilde{l};m}^{\tilde{l}''}
K_{\tilde{l}''+\frac{1}{2}}(\kappa L).
\end{split}
\end{equation*}When sphere 1 is inside sphere 2,
\begin{equation*}
\begin{split}
M_{lm,l'm'}(i\xi)=&\delta_{m,m'}T_1^{lm}\sum_{\tilde{l}=|m|}^{\infty}\sqrt{\frac{\pi }{2\kappa L}}\sum_{l''=|l-\tilde{l}|}^{l+\tilde{l}}H_{l\tilde{l};m}^{l''}
I_{l''+\frac{1}{2}}(\kappa L)\widetilde{T}^{\tilde{l}m}_2\sqrt{\frac{\pi }{2\kappa L}}\sum_{\tilde{l}''=|l'-\tilde{l}|}^{l'+\tilde{l}}H_{l'\tilde{l};m}^{\tilde{l}''}
I_{\tilde{l}''+\frac{1}{2}}(\kappa L).
\end{split}
\end{equation*}Here   the fact that $H_{l\tilde{l};m}^{l''}$ and $H_{l'\tilde{l};m}^{\tilde{l}''}$ are nonzero only if
$l+\tilde{l}+l''$ and $l'+\tilde{l}+\tilde{l}''$ are even have been used to write $(-1)^{l''+\tilde{l}''}$ as $(-1)^{l+l'}$, and the cyclic property
of the trace has been used to get rid of this sign factor. $\kappa$ is related to $\xi$ by \eqref{eq3_22_6}.
 $T_1^{lm}$ is the $lm$-diagonal component of the $\mathbb{T}$-matrix for sphere 1. $T_2^{lm}$ and $ \widetilde{T}_2^{lm}$ are components of the $\mathbb{T}$ and $\widetilde{\mathbb{T}}$ matrices  for sphere 2. They have been derived under various boundary
conditions  in Section \ref{sca_sphere}.

 The exact Casimir interaction energy for two spheres that are outside each other has been considered in a number of works. For two Dirichlet or Neumann spheres, it has been considered in \cite{10,13,58}. For two Robin spheres, it has been considered in \cite{5,12,58}. For two semitransparent spheres, it has been considered in \cite{14,16}. The Casimir interaction energy for two  spheres with Robin boundary conditions, where one is inside the other, has been considered in \cite{58}. Besides these scenarios, there are other interesting scenarios whose exact Casimir interaction energy can be obtained from above:

\begin{enumerate}
\item[$\bullet$] Two semitransparent spheres, where one is inside the other.
\item[$\bullet$] A Dirichlet/Neumann/Robin sphere  and a semitransparent sphere, where the two spheres are outside each other or one is inside the other.
\end{enumerate}

Next, we consider electromagnetic fields.
Let $\boldsymbol{\mathcal{L}}$ be the operator
$$\boldsymbol{\mathcal{L}}=  -\mathbf{r} \times \nabla=\mathbf{e}_{\theta}\frac{1}{\sin\theta}\pa_{\phi}-\mathbf{e}_{\phi}\pa_{\theta},$$
and define
\begin{equation*}
\mathbf{X}_{lm}(\theta,\phi)=\frac{1}{\sqrt{l(l+1)}}\boldsymbol{\mathcal{L}}Y_{lm}(\theta,\phi)=\frac{1}{\sqrt{l(l+1)}}\left(\frac{im}{\sin\theta}
Y_{lm}(\theta,\phi)\mathbf{e}_{\theta}-\frac{\pa Y_{lm}(\theta,\phi)}{\pa\theta}\mathbf{e}_{\phi}\right),
\end{equation*}
so that
\begin{equation*}
\mathbf{A}_{lm}^{\text{TE},*}(\mathbf{x},k)=\mathcal{C}_l^*f_l^*(kr)\mathbf{X}_{lm}(\theta,\phi).
\end{equation*}
Let
\begin{equation}\label{eq5_3_4}
\begin{split}
\boldsymbol{\mathcal{P}}_{lm}=&\frac{1}{\sqrt{l(l+1)}}\left(\boldsymbol{\mathcal{L}}\mathcal{P}_{lm}-\mathcal{P}_{lm}\boldsymbol{\mathcal{L}}\right)\\
=&\frac{1}{\sqrt{l(l+1)}}\Biggl(\frac{\mathbf{e}_x}{2i}\left[\sqrt{(l-m)(l+m+1)}\mathcal{P}_{l,m+1}+\sqrt{(l+m)(l-m+1)}\mathcal{P}_{l,m-1}\right]\\
&-\frac{\mathbf{e}_y}{2}\left[\sqrt{(l-m)(l+m+1)}\mathcal{P}_{l,m+1}-\sqrt{(l+m)(l-m+1)}\mathcal{P}_{l,m-1}\right]
-im\mathbf{e}_z\mathcal{P}_{lm}\Biggr).
\end{split}\end{equation}
It follows that
\begin{equation}\label{eq5_3_3}
\mathbf{A}^{\text{TE, reg}}_{lm}(\mathbf{x},k)=\mathcal{C}_l^{\text{reg}}i^{-l}\boldsymbol{\mathcal{P}}_{lm}j_0(kr),\quad \mathbf{A}^{\text{TE, out}}_{lm}=
\mathcal{C}_l^{\text{out}}i^{-l}\boldsymbol{\mathcal{P}}_{lm}h_0(kr).
\end{equation}
In \cite{6}, it was shown that
\begin{equation}\label{eq4_3_9}
\mathbf{X}_{lm}(\theta_k,\phi_k)=\boldsymbol{\mathcal{P}}_{lm}e^{i\mathbf{k}\cdot\mathbf{r}}.
\end{equation}
This implies that
\begin{equation}\label{eq4_3_10}
\begin{split}
\mathbf{A}^{\text{TE, reg}}_{lm}(\mathbf{x},k)=&\frac{\mathcal{C}_l^{\text{reg}}}{4\pi i^l}\int_0^{2\pi}d\phi_k\int_0^{\pi}d\theta_k\sin\theta_k
\mathbf{X}_{lm}(\theta_k,\phi_k)e^{i\mathbf{k}\cdot\mathbf{r}},\\
\mathbf{A}^{\text{TM, reg}}_{lm}(\mathbf{x},k)=&\frac{1}{k}\nabla\times\mathbf{A}^{\text{TE, reg}}_{lm}(\mathbf{x},k)= \frac{\mathcal{C}_l^{\text{reg}}}{4\pi i^l}
\int_0^{2\pi}d\phi_k\int_0^{\pi} d\theta_k\sin\theta_k\frac{i\mathbf{k}}{k}\times
\mathbf{X}_{lm}(\theta_k,\phi_k)e^{i\mathbf{k}\cdot\mathbf{r}},
\end{split}
\end{equation}
\begin{equation}\label{eq4_3_11}
\begin{split}
\mathbf{A}^{\text{TE, out}}_{lm}(\mathbf{x},k)=&\frac{\mathcal{C}_l^{\text{out}}}{2\pi i^l}\int_{-\infty}^{\infty} dk_x\int_{-\infty}^{\infty} dk_y
\mathbf{X}_{lm}(\theta_k,\phi_k) \frac{e^{ik_xx+ik_yy \pm i\sqrt{k^2-k_x^2-k_y^2}z}}{k\sqrt{k^2-k_x^2-k_y^2}},\quad z\gtrless 0\\
\mathbf{A}^{\text{TM, out}}_{lm}(\mathbf{x},k)=&\frac{\mathcal{C}_l^{\text{out}}}{2\pi i^l}\int_{-\infty}^{\infty} dk_x\int_{-\infty}^{\infty} dk_y
\frac{i\mathbf{k}}{k}
\times\mathbf{X}_{lm}(\theta_k,\phi_k) \frac{e^{ik_xx+ik_yy \pm i\sqrt{k^2-k_x^2-k_y^2}z}}{k\sqrt{k^2-k_x^2-k_y^2}},\quad z\gtrless 0.
\end{split}
\end{equation}
Now consider the expansions that define the translation matrices $\mathbb{V}$:
\begin{equation}\label{eq4_2_2}
\begin{split}
\mathbf{A}^{\text{TE, reg}}_{l'm'}(\mathbf{x}-\mathbf{L},k)=&\sum_{l=1}^{\infty}\sum_{m=-l}^{l}\left(V_{lm,l'm'}^{\text{TE,TE}}(-\mathbf{L})
\mathbf{A}^{\text{TE, reg}}_{lm}(\mathbf{x},k)+V_{lm,l'm'}^{\text{TM,TE}}(-\mathbf{L})\mathbf{A}^{\text{TM, reg}}_{lm}(\mathbf{x},k)\right),\\
\mathbf{A}^{\text{TM, reg}}_{l'm'}(\mathbf{x}-\mathbf{L},k)=&\sum_{l=1}^{\infty}\sum_{m=-l}^{l}\left(V_{lm,l'm'}^{\text{TE,TM}}(-\mathbf{L})
\mathbf{A}^{\text{TE, reg}}_{lm}(\mathbf{x},k)+V_{lm,l'm'}^{\text{TM,TM}}(-\mathbf{L})\mathbf{A}^{\text{TM, reg}}_{lm}(\mathbf{x},k)\right).
\end{split}
\end{equation}
Using the relation \eqref{eq3_20_2}, it is easy to see that
\begin{equation*}
V_{lm,l'm'}^{\text{TE,TM}}=V_{lm,l'm'}^{\text{TM,TE}},\quad V_{lm,l'm'}^{\text{TM,TM}}=V_{lm,l'm'}^{\text{TE,TE}}.
\end{equation*}
Since
\begin{equation}\label{eq4_3_8}
\begin{split}
\left(\boldsymbol{\mathcal{P}}_{l''m''}\cdot\mathbf{A}_{lm}^{\text{TE,reg}}\right)(\mathbf{0})=&\frac{\mathcal{C}_{l}^{\text{reg}}}{4\pi i^{l}}
\int_0^{2\pi}d\phi_k\int_0^{\pi}
d\theta_k\,\sin\theta_k \mathbf{X}_{l'',m''}(\theta_k,\phi_k)\cdot\mathbf{X}_{lm}(\theta_k,\phi_k)=\frac{(-1)^{m}}{4\pi i^{l}}\delta_{l,l''}
\delta_{m'',-m}\mathcal{C}_{l}^{\text{reg}},\\
\left(\boldsymbol{\mathcal{P}}_{l''m''}\cdot\mathbf{A}_{lm}^{\text{TM,reg}}\right)(\mathbf{0})=&\frac{\mathcal{C}_{l}^{\text{reg}}}{4\pi i^{l}}
\int_0^{2\pi}d\phi_k\int_0^{\pi}
d\theta_k\,\sin\theta_k \mathbf{X}_{l'',m''}(\theta_k,\phi_k)\cdot\left(\frac{i\mathbf{k}}{k}\times\mathbf{X}_{lm}(\theta_k,\phi_k)\right)=0,
\end{split}\end{equation}applying the operator $\boldsymbol{\mathcal{P}}_{l,-m}\cdot$ to both sides of \eqref{eq4_2_2} and setting $\mathbf{x}=\mathbf{0}$, we find that
\begin{equation}\label{eq4_3_16}
\begin{split}
 V_{lm,l'm'}^{\text{TE,TE}}(-\mathbf{L})=&\frac{4\pi i^{l}}{\mathcal{C}_{l}^{\text{reg}}}(-1)^{m}\left(\boldsymbol{\mathcal{P}}_{l,-m}\cdot
 \mathbf{A}_{l'm'}^{\text{TE,reg}}\right)(-\mathbf{L})\\
 =& (-1)^{m}\int_0^{2\pi}d\phi_k\int_0^{\pi}d\theta_k \sin\theta_k \mathbf{X}_{l,-m}(\theta_k,\phi_k)\cdot
\mathbf{X}_{l'm'}(\theta_k,\phi_k)e^{i\mathbf{k}\cdot\mathbf{L}},\\
 V_{lm,l'm'}^{\text{TE,TM}}(-\mathbf{L})=&\frac{4\pi i^{l}}{\mathcal{C}_{l}^{\text{reg}}}(-1)^{m}\left(\boldsymbol{\mathcal{P}}_{l,-m}\cdot
 \mathbf{A}_{l'm'}^{\text{TM,reg}}\right)(-\mathbf{L})\\
 =&(-1)^{m+1}\int_0^{2\pi}d\phi_k\int_0^{\pi}d\theta_k \sin\theta_k \mathbf{X}_{l,-m}(\theta_k,\phi_k)\cdot
\left(\frac{i\mathbf{k}}{k}\times\mathbf{X}_{l'm'}(\theta_k,\phi_k)\right)e^{i\mathbf{k}\cdot\mathbf{L}}.
\end{split}
\end{equation}Here we have used the fact that
$$\left(\boldsymbol{\mathcal{P}}_{l,-m}\cdot
 \mathbf{A}_{l'm'}^{\text{TE,reg}}\right)(-\mathbf{L})=(-1)^{l+l'}\left(\boldsymbol{\mathcal{P}}_{l,-m}\cdot
 \mathbf{A}_{l'm'}^{\text{TE,reg}}\right)(\mathbf{L})$$ and $$\left(\boldsymbol{\mathcal{P}}_{l,-m}\cdot
 \mathbf{A}_{l'm'}^{\text{TM,reg}}\right)(-\mathbf{L})=(-1)^{l+l'+1}\left(\boldsymbol{\mathcal{P}}_{l,-m}\cdot
 \mathbf{A}_{l'm'}^{\text{TM,reg}}\right)(\mathbf{L}).$$
In Appendix \ref{A2}, we verify that
\begin{equation}\label{eq3_28_1}\begin{split}
\mathbf{X}_{l,-m}(\theta_k,\phi_k)\cdot
\mathbf{X}_{l'm'}(\theta_k,\phi_k)=& \sum_{l''=0}^{\infty}\sum_{m''=-l''}^{l''}\frac{l''(l''+1)-l(l+1)-l'(l'+1)}{2\sqrt{l(l+1)l'(l'+1)}}(-1)^{m''+1}\\
&\times\sqrt{\frac{(2l+1)(2l'+1)(2l''+1)}{4\pi}}\begin{pmatrix}
l& l' &l''\\m &-m' &m''\end{pmatrix}\begin{pmatrix} l & l' & l''\\0 & 0 & 0\end{pmatrix}Y_{l'',m''}(\theta_k,\phi_k),\end{split}
\end{equation}and if $\mathbf{L}=L\mathbf{e}_z$,
\begin{equation}\label{eq3_29_1}\begin{split}
&\int_0^{2\pi}d\phi_k\int_0^{\pi}d\theta_k \sin\theta_k \mathbf{X}_{l,-m}(\theta_k,\phi_k)\cdot
\left(\frac{i\mathbf{k}}{k}\times\mathbf{X}_{l'm'}(\theta_k,\phi_k)\right)e^{i\mathbf{k}\cdot\mathbf{L}}\\=&
\frac{ imkL}{\sqrt{l(l+1)l'(l'+1)}}  \int_0^{2\pi}d\phi_k\int_0^{\pi}d\theta_k \sin\theta_k Y_{l'm'}(\theta_k,\phi_k) Y_{l,-m}(\theta_k,\phi_k)  e^{i\mathbf{k}\cdot\mathbf{L}}.
\end{split}\end{equation}Compare to the scalar case, we find that
\begin{equation*}\begin{split}
\mathbb{V}_{lm,l'm'} (-L\mathbf{e}_z,i\xi)=&  (-1)^{m+1}\delta_{m,m'}\sqrt{\frac{\pi }{2\kappa L}}\sum_{l''=|l-l'|}^{l+l'}(-1)^{l''}
\begin{pmatrix}\Lambda_{ll'}^{l''} & -\tilde{\Lambda}_{ll';m}\\
-\tilde{\Lambda}_{ll';m}&\Lambda_{ll'}^{l''}
\end{pmatrix}H_{ll';m}^{l''}I_{l''+\frac{1}{2}}(\kappa L),
\end{split}\end{equation*}
where
\begin{equation}\label{eq5_9_3}
\Lambda_{ll'}^{l''}=\frac{l''(l''+1)-l(l+1)-l'(l'+1)}{2\sqrt{l(l+1)l'(l'+1)}},\hspace{1cm}\tilde{\Lambda}_{ll';m}=\frac{m\kappa L}{\sqrt{l(l+1)l'(l'+1)}}.
\end{equation}
In the same way, we find that
\begin{equation*}
\mathbb{U}_{lm,l'm'}(-L\mathbf{e}_z,i\xi)=(-1)^{l'+m+1}\delta_{m,m'}\sqrt{\frac{\pi }{2\kappa L}}\sum_{l''=|l-l'|}^{l+l'}
\begin{pmatrix}\Lambda_{ll'}^{l''} & -\tilde{\Lambda}_{ll';m}\\
-\tilde{\Lambda}_{ll';m}&\Lambda_{ll'}^{l''}
\end{pmatrix}H_{ll';m}^{l''}K_{l''+\frac{1}{2}}(\kappa L),
\end{equation*}
\begin{equation*}
\mathbb{U}_{l'm',lm}(L\mathbf{e}_z,i\xi)=(-1)^{l'+m+1}\delta_{m,m'}\sqrt{\frac{\pi }{2\kappa L}}\sum_{l''=|l-l'|}^{l+l'}
\begin{pmatrix}\Lambda_{ll'}^{l''} & \tilde{\Lambda}_{ll';m}\\
\tilde{\Lambda}_{ll';m}&\Lambda_{ll'}^{l''}
\end{pmatrix}H_{ll';m}^{l''}K_{l''+\frac{1}{2}}(\kappa L),
\end{equation*}
\begin{equation*}\begin{split}
\mathbb{W}_{l'm',lm} (L\mathbf{e}_z,i\xi)=&  (-1)^{m+1}\delta_{m,m'}\sqrt{\frac{\pi }{2\kappa L}}\sum_{l''=|l-l'|}^{l+l'}(-1)^{l''}
\begin{pmatrix}\Lambda_{ll'}^{l''} & \tilde{\Lambda}_{ll';m}\\
\tilde{\Lambda}_{ll';m}&\Lambda_{ll'}^{l''}
\end{pmatrix}H_{ll';m}^{l''}I_{l''+\frac{1}{2}}(\kappa L).
\end{split}\end{equation*}

Gathering the results, we find that for electromagnetic interaction between two spheres, the Casimir interaction energy is given by
\begin{equation*}
E_{\text{Cas}}=\frac{\hbar}{2\pi}\int_0^{\infty}d\xi \text{Tr}\,\ln\left(1-\mathbb{M}(i\xi)\right),
\end{equation*}where the trace  Tr is
\begin{equation}\label{eq4_3_13}
\text{Tr}=\sum_{m=-\infty}^{\infty}\sum_{l=\max\{1,|m|\}}^{\infty}\text{tr},
\end{equation} and the trace tr is a trace over $2\times 2$ matrices.  $\mathbb{M}$ is a matrix diagonal in $m$. When the two spheres are outside each other,
\begin{equation*}
\begin{split}
\mathbb{M}_{lm,l'm'}(i\xi)=&\delta_{m,m'}\mathbb{T}_1^{lm}\sum_{\tilde{l}=\max\{1,|m|\}}^{\infty}\sqrt{\frac{\pi }{2\kappa L}}\sum_{l''=|l-\tilde{l}|}^{l+\tilde{l}}
\begin{pmatrix}\Lambda_{l\tilde{l}}^{l''} & -\tilde{\Lambda}_{l\tilde{l};m}\\
-\tilde{\Lambda}_{l\tilde{l};m}&\Lambda_{l\tilde{l}}^{l''}
\end{pmatrix}H_{l\tilde{l};m}^{l''}K_{l''+\frac{1}{2}}(\kappa L)\\&\times \mathbb{T}^{\tilde{l}m}_2\sqrt{\frac{\pi }{2\kappa L}}
\sum_{\tilde{l}''=|l'-\tilde{l}|}^{l'+\tilde{l}}\begin{pmatrix}\Lambda_{l'\tilde{l}}^{\tilde{l}''} & \tilde{\Lambda}_{l'\tilde{l};m}\\
\tilde{\Lambda}_{l'\tilde{l};m}&\Lambda_{l'\tilde{l}}^{\tilde{l}''}
\end{pmatrix}H_{l'\tilde{l};m}^{\tilde{l}''}K_{\tilde{l}''+\frac{1}{2}}(\kappa L).
\end{split}
\end{equation*}When sphere 1 is inside sphere 2,
\begin{equation*}
\begin{split}
\mathbb{M}_{lm,l'm'}(i\xi)=&\delta_{m,m'}\mathbb{T}_1^{lm}\sum_{\tilde{l}=\max\{1,|m|\}}^{\infty}\sqrt{\frac{\pi }{2\kappa L}}\sum_{l''=|l-\tilde{l}|}^{l+\tilde{l}}
\begin{pmatrix}\Lambda_{l\tilde{l}}^{l''} & -\tilde{\Lambda}_{l\tilde{l};m}\\
-\tilde{\Lambda}_{l\tilde{l};m}&\Lambda_{l\tilde{l}}^{l''}
\end{pmatrix}H_{l\tilde{l};m}^{l''}I_{l''+\frac{1}{2}}(\kappa L)\\&\times\widetilde{\mathbb{T}}^{\tilde{l}m}_2\sqrt{\frac{\pi }{2\kappa L}}
\sum_{\tilde{l}''=|l'-\tilde{l}|}^{l'+\tilde{l}}\begin{pmatrix}\Lambda_{l'\tilde{l}}^{\tilde{l}''} & \tilde{\Lambda}_{l'\tilde{l};m}\\
\tilde{\Lambda}_{l'\tilde{l};m}&\Lambda_{l'\tilde{l}}^{\tilde{l}''}
\end{pmatrix}H_{l'\tilde{l};m}^{\tilde{l}''}I_{\tilde{l}''+\frac{1}{2}}(\kappa L).
\end{split}
\end{equation*}
Here $\kappa=n_m(i\xi)\xi/c$, where $n_m$ is the refractive index of the medium between the two spheres.
The  matrices $\mathbb{T}_1^{lm}$ for sphere 1 and $\mathbb{T}_2^{lm}, \widetilde{\mathbb{T}}_2^{lm}$ for sphere 2  have been derived in Section \ref{sca_sphere}.

As in the scalar case, the electromagnetic Casimir interaction   between two spheres has been considered in a number of works. For two perfectly conducting spheres that are outside each other, it has been considered in \cite{12,13,3, 58,61}.  For two magnetodielectric spheres that are outside each other, it has been considered in \cite{12,3,61}. For two   perfectly conducting spheres, where one is inside the other, it has been considered in \cite{58, 59,60}. For two magnetodielectric spheres, where one is inside the other, it has been considered in \cite{60}.  

\section{Casimir effect between a sphere and a plane}\label{sphere_plane}
In this section, we consider the Casimir effect between a sphere and a plane. This scenario has been extensively studied for its experimental value. It is more appropriately regarded as one object (the sphere) is inside
the other (the plane). Assume that the sphere is given by $x^2+y^2+z^2=R^2$, with  center at $O=(0,0,0)$ and radius $R$; the plane is given by $-H/2\leq x,y\leq H/2$, $z=L$, with center  at $O'=(0,0,L)$.

Since the $\mathbb{T}, \widetilde{\mathbb{T}}$ matrices for a sphere and for a plane have been considered respectively in Section \ref{sca_sphere} and Section \ref{sca_plane},
 let us consider now  the translation matrices $\mathbb{V}$ and $\mathbb{W}$.
 In the scalar case, they are defined by
 \begin{equation}\label{eq4_3_4}
\varphi_{\mathbf{k}_{\perp}}^{\text{reg}}(\mathbf{x}-\mathbf{L},k)=\sum_{l=0}^{\infty}\sum_{m=-l}^{l} V_{lm,\mathbf{k}_{\perp}}(-\mathbf{L})
\varphi_{lm}^{\text{reg}}(\mathbf{x},k),
\end{equation}\begin{equation}\label{eq4_27_3}\varphi_{lm}^{\text{out}}(\mathbf{x}'+\mathbf{L},k)=H^2\int_{-\infty}^{\infty}\frac{dk_x}{2\pi}\int_{-\infty}^{\infty}
\frac{dk_y}{2\pi} W_{\mathbf{k}_{\perp},lm}(\mathbf{L})\varphi_{\mathbf{k}_{\perp}}^{\text{out}}(\mathbf{x}',k).
\end{equation}
For $V_{lm,\mathbf{k}_{\perp}}(-\mathbf{L})$, changing $z$ to $-z$ in \eqref{eq4_3_3}, and multiplying by $e^{i\sqrt{k^2-k_{\perp}^2}L}$, we find that
\begin{equation*}
e^{ik_xx+ik_yy-i\sqrt{k^2-k_{ \perp}^2}(z-L)}=4\pi\sum_{l=0}^{\infty}\sum_{m=-l}^l i^l (-1)^{l+m} j_l(kr)Y_{lm}(\theta,\phi)Y_{lm}^*(\theta_k,\phi_k)
e^{i\sqrt{k^2-k_{\perp}^2}L}.
\end{equation*}
Compare to \eqref{eq4_3_4}, we have
\begin{equation*}
\begin{split}
V_{lm,\mathbf{k}_{\perp}}(-L\mathbf{e}_z)=&4\pi(-1)^mY_{lm}^*(\theta_k,\phi_k)e^{i\sqrt{k^2-k_{\perp}^2}L}\\
=&4\pi (-1)^m\sqrt{\frac{2l+1}{4\pi}\frac{(l-m)!}{(l+m)!}}P_l^m\left(\frac{\sqrt{k^2-k_{\perp}^2}}{k}\right)\left(\frac{k_x-ik_y}{k_{\perp}}\right)^{m}
e^{i\sqrt{k^2-k_{\perp}^2}L}.
\end{split}\end{equation*}
For $W_{\mathbf{k}_{\perp},lm}(\mathbf{L})$, compare \eqref{eq4_27_3} to  \eqref{eq4_2_3}, we obtain immediately   that
\begin{equation*}
\begin{split}
W_{\mathbf{k}_{\perp},lm}(L\mathbf{e}_z)=&-  \frac{\pi^2}{H^2k\sqrt{k^2-k_{\perp}^2}}Y_{lm}(\theta_k,\phi_k)e^{i\sqrt{k^2-k_{\perp}^2}L}\\
=&-  \frac{\pi^2}{H^2k\sqrt{k^2-k_{\perp}^2}}\sqrt{\frac{2l+1}{4\pi}\frac{(l-m)!}{(l+m)!}}P_l^m\left(\frac{\sqrt{k^2-k_{\perp}^2}}{k}\right)
\left(\frac{k_x+ik_y}{k_{\perp}}\right)^me^{i\sqrt{k^2-k_{\perp}^2}L}.
\end{split}\end{equation*}
Consequently, we find that the scalar Casimir interaction energy between a sphere and a plane is
\begin{equation*}
E_{\text{Cas}}=\frac{\hbar}{2\pi}\int_0^{\infty} d\xi\text{Tr}\ln\left(1-\mathbb{M}(i\xi)\right),
\end{equation*}
where the trace is given by \eqref{eq4_27_4}, and
\begin{equation*}
\begin{split}
M_{lm,l'm'}(\omega)=&-\pi T^{lm}\int_{-\infty}^{\infty}dk_x\int_{-\infty}^{\infty}dk_y\frac{1}{k\sqrt{k^2-k_{\perp}^2}}
Y_{l,-m}(\theta_k,\phi_k)Y_{l'm'}(\theta_k,\phi_k)\widetilde{T}^{\mathbf{k}_{\perp}}e^{2i\sqrt{k^2-k_{\perp}^2}L}\\
=&-\pi \delta_{m,m'}T^{lm}\sum_{l''=|l-l'|}^{l+l'}\frac{H_{ll';m}^{l''}}{\sqrt{4\pi(2l''+1)}}\int_{-\infty}^{\infty}dk_x
\int_{-\infty}^{\infty}dk_y\frac{1}{k\sqrt{k^2-k_{\perp}^2}}
Y_{l'',0}(\theta_k,\phi_k)\widetilde{T}^{\mathbf{k}_{\perp}}e^{2i\sqrt{k^2-k_{\perp}^2}L}.
\end{split}\end{equation*} Here \eqref{eq4_3_5} and \eqref{eq4_3_6} have been applied. The $\mathbb{T}$-matrix component $T^{lm}$ for the sphere has been derived in Section
\ref{sca_sphere} and the $\widetilde{\mathbb{T}}$-matrix component $\widetilde{T}^{\mathbf{k}_{\perp}}$ for the plane has been derived in Section \ref{sca_plane}.
If the plane is subjected to Dirichlet or Neumann boundary conditions,
$\widetilde{T}^{\mathbf{k}_{\perp}}=1$ or $-1$, in this case,
 we obtain from \eqref{eq4_2_3} that
\begin{equation*}
M_{lm,l'm'}(i\xi)=\pm\delta_{m,m'}T^{lm}\sqrt{\frac{\pi}{4\kappa L}}\sum_{l''=|l-l'|}^{l+l'}H_{ll';m}^{l''}K_{l''+\frac{1}{2}}(2\kappa L).
\end{equation*}
In the general case where the plane is semitransparent or is subjected to Robin boundary conditions,
\begin{equation*}
\begin{split}
M_{lm,l'm'}(i\xi)=& \frac{\pi}{2} \delta_{m,m'}T^{lm}\sum_{l''=|l-l'|}^{l+l'} H_{ll';m}^{l''}  \int_{0}^{\infty}dk_{\perp}
\frac{k_{\perp}}{\kappa\sqrt{\kappa^2+k_{\perp}^2}}
P_{l''}\left(\frac{\sqrt{\kappa^2+k_{\perp}^2}}{\kappa}\right)\widetilde{T}^{\mathbf{k}_{\perp}}e^{-2\sqrt{\kappa^2+k_{\perp}^2}L}.
\end{split}\end{equation*}Making a change of variables
$$  k_{\perp}=\kappa\sinh\theta,$$ we find that
\begin{equation*}
\begin{split}
M_{lm,l'm'}(i\xi)=&\frac{\pi}{2} \delta_{m,m'}T^{lm}\sum_{l''=|l-l'|}^{l+l'} H_{ll';m}^{l''}  \int_{0}^{\infty}d\theta
\sinh\theta
P_{l''}\left(\cosh\theta\right)\widetilde{T}^{\mathbf{k}_{\perp}}e^{-2\kappa L\cosh\theta}\\
=&\frac{\pi}{2} \delta_{m,m'}T^{lm}\sum_{l''=|l-l'|}^{l+l'} H_{ll';m}^{l''}  \int_{1}^{\infty}dz
P_{l''}\left(z\right)\widetilde{T}^{\mathbf{k}_{\perp}}e^{-2\kappa Lz}.
\end{split}
\end{equation*}
For a Robin plane,
\begin{equation*}
\widetilde{T}^{\mathbf{k}_{\perp}}(i\xi)= \frac{u-\kappa\cosh\theta}{u+\kappa\cosh\theta}=\frac{u-\kappa z}{u+\kappa z};
\end{equation*}whereas for a semitransparent plane,
\begin{equation*}
\widetilde{T}^{\mathbf{k}_{\perp}}(i\xi)= \frac{\lambda}{\lambda+2\kappa\cosh\theta}= \frac{\lambda}{\lambda+2\kappa z}.
\end{equation*}In the formulas above, $\kappa$ is related to $\xi$ by \eqref{eq3_22_6}.

The exact Casimir interaction between a Dirichlet sphere and a Dirichlet plane has been considered in \cite{7,10,9,62,65}. The interaction between a Robin sphere and a Dirichlet plane has been considered in \cite{64}. Besides these scenarios, there are other scenarios whose exact Casimir interaction energy can be obtained from above:

\begin{enumerate}
\item[$\bullet$] A Dirichlet/Neumann/Robin sphere in front of a Robin plane.

\item[$\bullet$] A semitransparent sphere in front of a semitransparent plane.

\item[$\bullet$] A Dirichlet/Neumann/Robin sphere in front of a semitransparent plane.

\item[$\bullet$] A semitransparent sphere in front of a Dirichlet/Neumann/Robin plane.
\end{enumerate}
Note that the formula for the Casimir interaction energy between a Robin sphere and a Robin plane is considerably more complicated than the corresponding formula for a Robin sphere in front of a Dirichlet/Neumann plane.

For the electromagnetic case, the translation matrices $\mathbb{V}$ and $\mathbb{W}$ are defined by
\begin{equation}\label{eq4_3_6_2}
\begin{split}
\mathbf{A}^{\text{TE, reg}}_{\mathbf{k}_{\perp}}(\mathbf{x}-\mathbf{L},k)=&\sum_{l=1}^{\infty}\sum_{m=-l}^{l}\left(V_{lm,\mathbf{k}_{\perp}}^{\text{TE,TE}}(-\mathbf{L})
\mathbf{A}^{\text{TE, reg}}_{lm}(\mathbf{x},k)+V_{lm,\mathbf{k}_{\perp}}^{\text{TM,TE}}(-\mathbf{L})\mathbf{A}^{\text{TM, reg}}_{lm}(\mathbf{x},k)\right),\\
\mathbf{A}^{\text{TM, reg}}_{\mathbf{k}_{\perp}}(\mathbf{x}-\mathbf{L},k)=&\sum_{l=1}^{\infty}\sum_{m=-l}^{l}\left(V_{lm,\mathbf{k}_{\perp}}^{\text{TE,TM}}(-\mathbf{L})
\mathbf{A}^{\text{TE, reg}}_{lm}(\mathbf{x},k)+V_{lm,\mathbf{k}_{\perp}}^{\text{TM,TM}}(-\mathbf{L})\mathbf{A}^{\text{TM, reg}}_{lm}(\mathbf{x},k)\right),
\end{split}
\end{equation}
\begin{equation}\label{eq4_3_7}
\begin{split}
\mathbf{A}^{\text{TE, out}}_{lm}(\mathbf{x}'+\mathbf{L},k)=&H^2\int_{-\infty}^{\infty}\frac{dk_x}{2\pi}\int_{-\infty}^{\infty}
\frac{dk_y}{2\pi} \left(W_{\mathbf{k}_{\perp},lm}^{\text{TE,TE}}(\mathbf{L})
\mathbf{A}^{\text{TE, out}}_{\mathbf{k}_{\perp}}(\mathbf{x}',k)+W_{\mathbf{k}_{\perp},lm}^{\text{TM,TE}}(\mathbf{L})
\mathbf{A}^{\text{TM, out}}_{\mathbf{k}_{\perp}}(\mathbf{x}',k)\right),\\
\mathbf{A}^{\text{TM, out}}_{lm}(\mathbf{x}'+\mathbf{L},k)=&H^2\int_{-\infty}^{\infty}\frac{dk_x}{2\pi}\int_{-\infty}^{\infty}
\frac{dk_y}{2\pi} \left(W_{\mathbf{k}_{\perp},lm}^{\text{TE,TM}}(\mathbf{L})
\mathbf{A}^{\text{TE, out}}_{\mathbf{k}_{\perp}}(\mathbf{x}',k)+W_{\mathbf{k}_{\perp},lm}^{\text{TM,TM}}(\mathbf{L})
\mathbf{A}^{\text{TM, out}}_{\mathbf{k}_{\perp}}(\mathbf{x}',k)\right).
\end{split}
\end{equation}
As in Section \ref{sphere_sphere}, the relation \eqref{eq3_20_2} implies that for $Z=  V$ or $W$,
\begin{equation*}
Z^{\text{TE,TM}}=Z^{\text{TM,TE}},\quad Z^{\text{TM,TM}}=Z^{\text{TE,TE}}.
\end{equation*}
For $\mathbb{V}$, \eqref{eq4_3_6_2} and
\eqref{eq4_3_8} imply that
\begin{equation*}
\begin{split}
 V_{lm,\mathbf{k}_{\perp}}^{\text{TE,TE}}(-\mathbf{L})=&\frac{4\pi i^{l}}{\mathcal{C}_{l}^{\text{reg}}}(-1)^{m}
 \left(\boldsymbol{\mathcal{P}}_{l,-m}\cdot\mathbf{A}_{\mathbf{k}_{\perp}}^{\text{TE,reg}}\right)(-\mathbf{L})\\
 =&4\pi(-1)^{l+m} \boldsymbol{\mathcal{P}}_{l,-m}\cdot e^{ik_xx+ik_yy- i\sqrt{k^2-k_{\perp}^2}z}
\left(\frac{ik_y}{k_{\perp}}\mathbf{e}_x-\frac{ik_x}{k_{\perp}}\mathbf{e}_y\right)\Biggr|_{\mathbf{r}=-\mathbf{L}},\\
 V_{lm,\mathbf{k}_{\perp}}^{\text{TE,TM}}(-\mathbf{L})=&\frac{4\pi i^{l}}{\mathcal{C}_{l}^{\text{reg}}}(-1)^{m}
 \left(\boldsymbol{\mathcal{P}}_{l,-m}\cdot\mathbf{A}_{\mathbf{k}_{\perp}}^{\text{TM,reg}}\right)(-\mathbf{L})\\
 =&4\pi(-1)^{l+m} \boldsymbol{\mathcal{P}}_{l,-m}\cdot e^{ik_xx+ik_yy- i\sqrt{k^2-k_{\perp}^2}z}
\left(\frac{ k_x\sqrt{k^2-k_{\perp}^2}}{kk_{\perp}}\mathbf{e}_x +\frac{ k_y\sqrt{k^2-k_{\perp}^2}}{kk_{\perp}}\mathbf{e}_y+
\frac{k_{\perp}}{k}\mathbf{e}_z\right)\Biggr|_{\mathbf{r}=-\mathbf{L}}.
\end{split}
\end{equation*}
Using the fact that
\begin{equation*}
\begin{split}
&\frac{ik_y}{k_{\perp}}\mathbf{e}_x-\frac{ik_x}{k_{\perp}}\mathbf{e}_y=-i\mathbf{e}_{\phi_k},\\
&\frac{ k_x\sqrt{k^2-k_{\perp}^2}}{kk_{\perp}}\mathbf{e}_x +\frac{ k_y\sqrt{k^2-k_{\perp}^2}}{kk_{\perp}}\mathbf{e}_y+
\frac{k_{\perp}}{k}\mathbf{e}_z=-\mathbf{e}_{\tilde{\theta}_k},\\
&\text{where}\quad\cos\tilde{\theta}_k=-\frac{\sqrt{k^2-k_{\perp}^2}}{k},
\end{split}
\end{equation*}and the relation \eqref{eq4_3_9} with $k_z=-\sqrt{k^2-k_{\perp}^2}$, we find that
\begin{equation*}
\begin{split}
 V_{lm,\mathbf{k}_{\perp}}^{\text{TE,TE}}(-L\mathbf{e}_z)=&4\pi(-1)^{l+m}\mathbf{X}_{l,-m}(\tilde{\theta}_k,\phi_k)\cdot
 \left(-i\mathbf{e}_{\phi_k}\right)e^{i\sqrt{k^2-k_{\perp}^2}L}\\
 =&-\frac{4\pi i }{\sqrt{l(l+1)}}  \frac{\pa Y_{l,-m}(\theta_k,\phi_k)}{\pa\theta_k}e^{i\sqrt{k^2-k_{\perp}^2}L},\\
 V_{lm,\mathbf{k}_{\perp}}^{\text{TE,TM}}(-L\mathbf{e}_z)=&4\pi(-1)^{l+m}\mathbf{X}_{l,-m}(\tilde{\theta}_k,\phi_k)\cdot
 \left(- \mathbf{e}_{\tilde{\theta}_k}\right)e^{i\sqrt{k^2-k_{\perp}^2}L}\\=&\frac{4\pi i  }{\sqrt{l(l+1)}} \frac{m}{\sin\theta_k} Y_{l,-m}(\theta_k,\phi_k) e^{i\sqrt{k^2-k_{\perp}^2}L}.
\end{split}
\end{equation*}
Here
$$\cos\theta_k=\frac{\sqrt{k^2-k_{\perp}^2}}{k},$$ and we have used the fact that $P_l^m(\cos\tilde{\theta}_k)=(-1)^{l+m}P_l^m(\cos\theta_k)$.

For the matrix $\mathbb{W}$, we obtain from \eqref{eq4_3_11} that
\begin{equation}\label{eq4_3_12}
\begin{split}
\mathbf{A}^{\text{TE, out}}_{lm}(\mathbf{x}'+\mathbf{L},k)=&-\frac{1}{4\sqrt{l(l+1)}}
\int_{-\infty}^{\infty} dk_x\int_{-\infty}^{\infty} dk_y\left(\frac{im}{\sin\theta_k}Y_{lm}(\theta_k,\phi_k)\mathbf{e}_{\theta_k}
-\frac{\pa Y_{lm}(\theta_k,\phi_k)}{\pa\theta_k}\mathbf{e}_{\phi_k}\right) \\&\hspace{5cm}\times\frac{e^{ik_xx'+ik_yy'
+i\sqrt{k^2-k_{\perp}^2}(z'+L)}}{k\sqrt{k^2-k_{\perp}^2}}.\end{split}
\end{equation}
Since
\begin{equation*}\begin{split}
\mathbf{e}_{\theta_k}=&-\left(-\frac{k_x\sqrt{k^2-k_{\perp}^2}}{kk_{\perp}}\mathbf{e}_x-\frac{ k_y\sqrt{k^2-k_{\perp}^2}}{kk_{\perp}}\mathbf{e}_y+
\frac{k_{\perp}}{k}\mathbf{e}_z\right),\\
\mathbf{e}_{\phi_k}=&i\left(\frac{ik_y}{k_{\perp}}\mathbf{e}_x-\frac{ik_x}{k_{\perp}}\mathbf{e}_y\right),
\end{split}
\end{equation*}compare to \eqref{eq4_3_7}, it follows immediately that
\begin{equation*}
\begin{split}
W_{\mathbf{k}_{\perp},lm}^{\text{TE,TE}}(L\mathbf{e}_z)=&\frac{i }{H^2\sqrt{l(l+1)}} \frac{\pi^2}{k\sqrt{k^2-k_{\perp}^2}}
\frac{\pa Y_{lm}(\theta_k,\phi_k)}{\pa\theta_k}e^{i\sqrt{k^2-k_{\perp}^2}L},\\
W_{\mathbf{k}_{\perp},lm}^{\text{TM,TE}}(L\mathbf{e}_z)=&\frac{i }{H^2\sqrt{l(l+1)}} \frac{\pi^2}{k\sqrt{k^2-k_{\perp}^2}}
\frac{m}{\sin\theta_k} Y_{lm}(\theta_k,\phi_k) e^{i\sqrt{k^2-k_{\perp}^2}L}.
\end{split}
\end{equation*}
Hence, the electromagnetic Casimir interaction energy between a sphere and a plane is
\begin{equation*}
E_{\text{Cas}}=\frac{\hbar}{2\pi}\int_0^{\infty} d\xi\text{Tr}\ln\left(1-\mathbb{M}(i\xi)\right),
\end{equation*}
where the trace Tr is given by \eqref{eq4_3_13}, and
\begin{equation*}
\begin{split}
\mathbb{M}_{lm,l'm'}(\omega)=&\frac{ \pi}{\sqrt{l(l+1)l'(l'+1)}} \mathbb{T}^{lm}\int_{-\infty}^{\infty}dk_x\int_{-\infty}^{\infty}dk_y\frac{1}{k\sqrt{k^2-k_{\perp}^2}}
\left(\begin{aligned}\frac{\pa Y_{l,-m}(\theta_k,\phi_k)}{\pa\theta_k}\hspace{0.7cm}
&\hspace{0.2cm}-\frac{m}{\sin\theta_k}Y_{l,-m}(\theta_k,\phi_k)\\-\frac{m}{\sin\theta_k}Y_{l,-m}(\theta_k,\phi_k)
\hspace{0.4cm}&\hspace{0.7cm}\frac{\pa Y_{l,-m}(\theta_k,\phi_k)}{\pa\theta_k}\end{aligned}\right)
\\&\times\widetilde{\mathbb{T}}^{\mathbf{k}_{\perp}}\left(\begin{aligned}
\frac{\pa Y_{l'm'}(\theta_k,\phi_k)}{\pa\theta_k}\hspace{0.7cm} &\hspace{0.2cm}\frac{m'}{\sin\theta_k}Y_{l'm'}(\theta_k,\phi_k)\\
\frac{m'}{\sin\theta_k}Y_{l'm'}(\theta_k,\phi_k)\hspace{0.4cm}&\hspace{0.7cm}\frac{\pa Y_{l'm'}(\theta_k,\phi_k)}{\pa\theta_k}\end{aligned}\right)
e^{2i\sqrt{k^2-k_{\perp}^2}L}.
\end{split}\end{equation*}Here $k=n_m\omega/c$, where $n_m$ is the refractive index of the medium between the sphere and the plane.
The $\mathbb{T}$-matrix component $\mathbb{T}^{lm}$ for the sphere has been derived in Section
\ref{sca_sphere} and the $\widetilde{\mathbb{T}}$-matrix component $\widetilde{\mathbb{T}}^{\mathbf{k}_{\perp}}$ for the plane has been derived in Section \ref{sca_plane}.
If the plane is perfectly conducting,
we take $\vep_e\rightarrow\infty$ in \eqref{eq4_3_14}, which gives
\begin{equation*}
\widetilde{\mathbb{T}}^{\mathbf{k}_{\perp}}=\begin{pmatrix} 1 & 0 \\ 0 & -1\end{pmatrix}.
\end{equation*}
Then
\begin{equation*}
\mathbb{M}_{lm,l'm'}(\omega)=\pi \mathbb{T}^{lm}\int_{-\infty}^{\infty}dk_x\int_{-\infty}^{\infty}dk_y\frac{1}{k\sqrt{k^2-k_{\perp}^2}}
\begin{pmatrix} \Lambda_1 & \Lambda_2\\ -\Lambda_2 &-\Lambda_1\end{pmatrix}e^{2i\sqrt{k^2-k_{\perp}^2}L},
\end{equation*}
where
\begin{equation*}
\begin{split}
\Lambda_1=&\frac{1}{\sqrt{l(l+1)l'(l'+1)}}\left(\frac{mm'}{\sin^2\theta_k}Y_{l,-m}(\theta_k,\phi_k)Y_{l'm'}(\theta_k,\phi_k)
+\frac{\pa Y_{l,-m}(\theta_k,\phi_k)}{\pa\theta_k}\frac{\pa Y_{l'm'}(\theta_k,\phi_k)}{\pa\theta_k}\right)\\
=&\mathbf{X}_{l,-m}(\theta_k,\phi_k)\cdot\mathbf{X}_{l'm'}(\theta_k,\phi_k),\\
\Lambda_2=&\frac{1}{\sqrt{l(l+1)l'(l'+1)}}\left(\frac{m}{\sin\theta_k}Y_{l,-m}(\theta_k,\phi_k)\frac{\pa Y_{l'm'}(\theta_k,\phi_k)}{\pa\theta_k}+
\frac{m'}{\sin\theta_k}Y_{l'm'}(\theta_k,\phi_k)\frac{\pa Y_{l,-m}(\theta_k,\phi_k)}{\pa\theta_k}\right)\\
=&\mathbf{X}_{l,-m}(\theta_k,\phi_k)\cdot\left(\frac{i\mathbf{k}}{k}\times\mathbf{X}_{l'm'}(\theta_k,\phi_k)\right).
\end{split}
\end{equation*}
Compare to the translation matrix $\mathbb{U}_{lm,l'm'}(-\mathbf{L})$ for the case of two spheres, we deduce that when the plane is perfectly conducting,
\begin{equation*}
\mathbb{M}_{lm,l'm'}(i\xi)=\delta_{m,m'}\mathbb{T}^{lm}\sqrt{\frac{\pi }{4\kappa L}}\sum_{l''=|l-l'|}^{l+l'}\begin{pmatrix}\Lambda_{ll'}^{l''} & \hat{\Lambda}_{ll';m}\\
-\hat{\Lambda}_{ll';m}&-\Lambda_{ll'}^{l''}
\end{pmatrix}H_{ll';m}^{l''}K_{l''+\frac{1}{2}}(2\kappa L),
\end{equation*}where $\Lambda_{ll'}^{l''} $ is given by \eqref{eq5_9_3}, and
\begin{equation*}
 \hat{\Lambda}_{ll';m}=\frac{2m\kappa L}{\sqrt{l(l+1)l'(l'+1)}}.
\end{equation*}
Here $\kappa=\kappa_m=n_m(i\xi)\xi/c$. In the general case, $\mathbb{M}_{lm,l'm'}(\omega)$ is nonzero only if $m=m'$. Passing to imaginary frequency and making the change of variables
$k_{\perp}=\kappa\sinh\theta$,
we have
\begin{equation*}\begin{split}
\mathbb{M}_{lm,l'm'}(i\xi)=&\delta_{m,m'} \frac{(-1)^{m}\pi}{2}\sqrt{\frac{(2l+1)(2l'+1)}{l(l+1)l'(l'+1)}\frac{(l-m)!(l'-m)!}{(l+m)!(l'+m)!}}   \mathbb{T}^{lm}
 \int_{0}^{\infty}d\theta \sinh\theta e^{-2\kappa_m L\cosh\theta}\\&
\times
 \left(\begin{aligned} \sinh\theta P_l^{m\prime}(\cosh\theta)\hspace{0.5cm} &-\frac{m}{\sinh\theta}P_l^m(\cosh\theta)\\ -\frac{m}{\sinh\theta}P_l^m(\cosh\theta)
 \hspace{0.4cm} & \quad\sinh\theta P_l^{m\prime}(\cosh\theta) \end{aligned}\right)\widetilde{\mathbb{T}}^{\mathbf{k}_{\perp}}
 \left(\begin{aligned} \sinh\theta P_{l'}^{m'\prime}(\cosh\theta)\hspace{0.5cm} & \frac{m'}{\sinh\theta}P_{l'}^{m'}(\cosh\theta)\\  \frac{m'}{\sinh\theta}P_{l'}^{m'}(\cosh\theta)
 \hspace{0.4cm} & \quad\sinh\theta P_{l'}^{m'\prime}(\cosh\theta) \end{aligned}\right).
\end{split}\end{equation*}
 The  matrix $\widetilde{\mathbb{T}}^{\mathbf{k}_{\perp}}$ is a diagonal matrix. If $z>L$ is a dielectric half-space with permittivity
$\vep_e$ and permeability $\mu_e$,
 \begin{equation}\label{eq4_3_14_2}
\begin{split}
\widetilde{T}_{\mathbf{k}_{\perp}}^{\text{TE}} =&\frac{\mu_m\sqrt{\frac{n_e^2}{n_m^2}+\sinh^2\theta}-\mu_e\cosh\theta}
{\mu_m\sqrt{\frac{n_e^2}{n_m^2}+\sinh^2\theta}+\mu_e\cosh\theta},\\
\widetilde{T}_{\mathbf{k}_{\perp}}^{\text{TM}} =&\frac{\vep_m\sqrt{\frac{n_e^2}{n_m^2}+\sinh^2\theta}-\vep_e\cosh\theta}
{\vep_m\sqrt{\frac{n_e^2}{n_m^2}+\sinh^2\theta}+\vep_e\cosh\theta}.
\end{split}
\end{equation}

The exact electromagnetic Casimir interaction between a sphere and a plane has been considered in a number of works.
For perfectly conducting sphere in front of a perfectly conducting plane, it has been considered in \cite{7,67,65,64,63}. For a dielectric sphere in front of a perfectly conducting plane, it has been considered in \cite{7}. For a magnetodielectric sphere in front of a magnetodielectric half-space, it has been discussed in \cite{4,69,68}.

\section{Casimir effect between two parallel cylinders}\label{cylinder_cylinder}
In this section, we derive the $TGTG$ formula for the Casimir interaction between two   parallel cylinders of length $H$. Assume that the radius of the cylinders are $R_1$ and $R_2$ respectively, and the centers are at $O=(0,0,0)$ and $O'=(L,0,0)$ respectively. The symmetric axes of the cylinders are parallel to the $z$-axis.

The $\mathbb{T}$ and $\widetilde{\mathbb{T}}$ matrices of a cylinder have been derived in Section \ref{sca_cylinder}.
Now we consider the translation matrices between the cylinders.
These have been derived using different approaches, and they are essentially the addition formula for Bessel functions. Here we mimic the approach for spheres.
For our application, we only consider a translation by a vector $\mathbf{L}$ that is perpendicular to the $z$-axis.
It is obvious that all the translation matrices should then be diagonal in $k_z$.

Consider the well-known formula:
\begin{equation*}
e^{ik_xx+ik_yy}=\sum_{n=-\infty}^{\infty} i^n J_n(k_{\perp}\rho)e^{in\phi}e^{-in\phi_k},
\end{equation*}
where
\begin{equation*}
\begin{split}
&x=\rho\cos\phi,\quad y=\rho\sin\phi,\\
& k_x=k_{\perp}\cos\phi_k,\quad k_y=k_{\perp}\sin\phi_k.
\end{split}
\end{equation*}
From this formula, we can deduce that
\begin{equation*}
 \varphi_{nk_z}^{\text{reg}}(\mathbf{x},k)=\mathcal{C}_n^{\text{reg}}J_n(k_{\perp}\rho)e^{in\phi}e^{ik_zz}=
\frac{\mathcal{C}_n^{\text{reg}}}{2\pi i^n}\int_0^{2\pi}d\phi_k e^{in\phi_k}e^{i\mathbf{k}_{\perp}\cdot\boldsymbol{\rho}+ik_zz}.
\end{equation*}Here $\mathbf{k}_{\perp}=k_x\mathbf{e}_x+k_y\mathbf{e}_y$, $\boldsymbol{\rho}=x\mathbf{e}_x+y\mathbf{e}_y$ and $k_{\perp}=\sqrt{k^2-k_z^2}=\sqrt{k_x^2+k_y^2}$.

Define the operator $\mathcal{Q}_n$ by
\begin{equation*}
\begin{split}
\mathcal{Q}_n=&\left(\frac{\pa_x+i\pa_y}{ik_{\perp}}\right)^n, \\
\mathcal{Q}_{-n}=&\left(\frac{\pa_x-i\pa_y}{ik_{\perp}}\right)^n,
\end{split}
\end{equation*}where $n\geq 0$. It is obvious that for all $n$,
\begin{equation*}
\mathcal{Q}_ne^{i\mathbf{k}_{\perp}\cdot\boldsymbol{\rho}}=e^{in\phi_k}e^{i\mathbf{k}_{\perp}\cdot\boldsymbol{\rho}}.
\end{equation*}
Hence,
\begin{equation}\label{eq4_10_5}
\begin{split}
 \varphi_{nk_z}^{\text{reg}}(\mathbf{x},k)=\frac{\mathcal{C}_n^{\text{reg}}}{2\pi i^n}\mathcal{Q}_n \int_0^{2\pi}d\phi_k
 e^{i\mathbf{k}_{\perp}\cdot\boldsymbol{\rho}+ik_zz}
=\mathcal{C}_n^{\text{reg}}i^{-n}\mathcal{Q}_nJ_0(k_{\perp}\rho)e^{ik_zz}.
\end{split}\end{equation}This identity can also be proved by induction. In fact, using induction, one can also prove  that
\begin{equation*}
i^{-n}\mathcal{Q}_n H_0^{(1)}(k_{\perp}\rho)=H_n^{(1)}(k_{\perp}\rho)e^{in\phi_k}.
\end{equation*}
On the other hand, one can show that
\begin{equation}\label{eq5_3_1}
H_{0}^{(1)}(k_{\perp}\rho)=\frac{1}{\pi}\int_{-\infty}^{\infty}dk_y\frac{e^{\pm i\sqrt{k_{\perp}^2-k_y^2}x+ik_yy}}{\sqrt{k_{\perp}^2-k_y^2}},\quad x\gtrless 0.
\end{equation}
Hence,
\begin{equation}\label{eq4_5_7}
\begin{split}
\varphi_{nk_z}^{\text{out}}(\mathbf{x},k)= \frac{\mathcal{C}_n^{\text{out}}}{ \pi i^n} \int_{-\infty}^{\infty}dk_y e^{in\phi_k}
\frac{e^{\pm i\sqrt{k_{\perp}^2-k_y^2}x+ik_yy+ik_zz}}{\sqrt{k_{\perp}^2-k_y^2}},\quad x\gtrless 0.
\end{split}\end{equation}
Now consider the expansions that define the translation matrices:
\begin{align}
\varphi_{n'k_z }^{\text{reg}}(\mathbf{x}-\mathbf{L},k )=&\sum_{n=-\infty}^{\infty} V_{n,n'}(-\mathbf{L})\varphi_{nk_z}^{\text{reg}}(\mathbf{x},k),\label{eq4_27_7}\\
\varphi_{n'k_z}^{\text{out}}(\mathbf{x}-\mathbf{L},k )=&\sum_{n=-\infty}^{\infty}   U_{n,n'}(-\mathbf{L})\varphi_{nk_z}^{\text{reg}}(\mathbf{x},k),\label{eq4_27_8}\\
\varphi_{nk_z}^{\text{out}}(\mathbf{x}'+\mathbf{L},k )=&\sum_{n'=-\infty}^{\infty}  U_{n',n}(\mathbf{L})\varphi_{n'k_z}^{\text{reg}}(\mathbf{x}',k),\label{eq4_27_9}\\
\varphi_{nk_z}^{\text{out}}(\mathbf{x}'+\mathbf{L},k )=&\sum_{n'=-\infty}^{\infty}  W_{n',n}(\mathbf{L})\varphi_{n'k_z}^{\text{out}}(\mathbf{x}',k),\label{eq4_27_10}
\end{align}  with translation vector $\mathbf{L}=L_x\mathbf{e}_x+L_y\mathbf{e}_y$.
  Since
\begin{equation}\label{eq4_5_5}
 \left(\mathcal{Q}_{n''}\varphi_{nk_z}^{\text{reg}}\right)\left(\mathbf{0}\right)
=\frac{\mathcal{C}_{n}^{\text{reg}}}{2\pi i^n}\int_0^{2\pi}d\phi_k e^{in''\phi_k}e^{in\phi_k}=
\mathcal{C}_n^{\text{reg}}i^{-n}\delta_{n'',-n},
\end{equation}
applying the operator $\mathcal{Q}_{-n}$ to both sides of \eqref{eq4_27_7}, we find that
\begin{equation*}
\begin{split}
V_{n,n'}(-\mathbf{L})=& \frac{i^n}{\mathcal{C}_n^{\text{reg}}}\left(\mathcal{Q}_{-n} \varphi_{n'k_z}^{\text{reg}}\right)( -\mathbf{L} )\\
=& \frac{1}{2\pi}\int_{0}^{2\pi}d\phi_k e^{i(n'-n)\phi_k}e^{i\mathbf{k}\cdot\mathbf{L}}\\
=& (-1)^{n'-n}\varphi_{n'-n,k_z}^{\text{reg}} (\mathbf{L},k).
\end{split}
\end{equation*}Here we have used the fact that $\left(\mathcal{Q}_{-n} \varphi_{n'k_z}^{\text{reg}}\right)( -\mathbf{L} )=(-1)^{n+n'}\left(\mathcal{Q}_{-n} \varphi_{n'k_z}^{\text{reg}}\right)( \mathbf{L} )$. Specialize to $\mathbf{L}=L\mathbf{e}_x$ and passing to imaginary frequency, we find that
\begin{equation*}
V_{n,n'}(-L\mathbf{e}_x,i\xi)=  (-1)^{n'-n}I_{n-n'}( \gamma L),
\end{equation*}
where
\begin{equation*}
\gamma=\sqrt{\kappa^2+k_z^2},
\end{equation*}and $k=i\kappa$. In the same way, we deduce from \eqref{eq4_5_5}, \eqref{eq4_27_8} and \eqref{eq4_27_9} that
\begin{equation*}
\begin{split}
U_{n,n'}(-L\mathbf{e}_x,i\xi)=&  (-1)^{n' }K_{n-n'}( \gamma L),\\
U_{n',n}(L\mathbf{e}_x,i\xi)=  &(-1)^{n' }K_{n-n'}(\gamma L).
\end{split}
\end{equation*}
In fact, for the latter, we have
\begin{equation*}
\varphi_{nk_z}^{\text{out}}(\mathbf{x}'+\mathbf{L},k)=\sum_{n'=-\infty}^{\infty}  (-1)^{n' }
\varphi_{n-n',k_z}^{\text{out}}(\mathbf{L},k)\varphi_{n'k_z}^{\text{reg}}(\mathbf{x}',k).
\end{equation*}Interchanging $\mathbf{x}'$ and $\mathbf{L}$ gives
\begin{equation*}
\begin{split}
\varphi_{nk_z}^{\text{out}}(\mathbf{x}'+\mathbf{L},k)=&\sum_{n'=-\infty}^{\infty}  (-1)^{n' }
\varphi_{n-n',k_z}^{\text{out}}(\mathbf{x}',k)\varphi_{n'k_z}^{\text{reg}}(\mathbf{L},k)\\
=&\sum_{n'=-\infty}^{\infty}  (-1)^{n-n' }
\varphi_{n-n',k_z}^{\text{reg}}(\mathbf{L},k)\varphi_{n'k_z}^{\text{out}}(\mathbf{x}',k).
\end{split}
\end{equation*}Compare to \eqref{eq4_27_10}, we obtain
\begin{equation*}
W_{n',n}(L\mathbf{e}_x,i\xi)=(-1)^{n-n' }
\varphi_{n-n'}^{\text{reg}}(L\mathbf{e}_x, i\kappa)=(-1)^{n-n'}I_{n-n'}(\gamma L).
\end{equation*}
Consequently, for a scalar interaction between two cylinders, the Casimir interaction is given by
\begin{equation}\label{eq4_5_1}
E_{\text{Cas}}=\frac{\hbar H}{2\pi}\int_0^{\infty}d\xi \int_{-\infty}^{\infty}\frac{dk_z}{2\pi}\text{Tr}\,\ln\left(1-\mathbb{M}(i\xi)\right),
\end{equation}where the trace  Tr is
\begin{equation}\label{eq4_5_2}
\text{Tr}=\sum_{n=-\infty}^{\infty}.
\end{equation} We have used the fact that  the $\mathbb{T}, \widetilde{\mathbb{T}}$  and the translation matrices are diagonal in $k_z$ to take out the trace over $k_z$.
When the two cylinders are outside each other,
\begin{equation}\label{eq4_5_3}
\begin{split}
M_{n,n'}(i\xi)=&T_1^{nk_z}\sum_{\tilde{n}=-\infty}^{\infty}K_{\tilde{n}-n}(\gamma L)T_2^{\tilde{n},k_z}K_{\tilde{n}-n'}(\gamma L).
\end{split}
\end{equation}When cylinder 1 is inside cylinder 2,
\begin{equation}\label{eq4_5_4}
\begin{split}
M_{n,n'}(i\xi)=&T_1^{nk_z}\sum_{\tilde{n}=-\infty}^{\infty}I_{ \tilde{n}-n}(\gamma L)\widetilde{T}_2^{\tilde{n},k_z}I_{\tilde{n}-n'}(\gamma L).\end{split}
\end{equation}Here   the cyclic property
of the trace has been used to get rid of the sign factor $(-1)^{n-n'}$; $\gamma =\sqrt{\kappa^2+k_z^2}$ and $\kappa$ is related to $\xi$ by \eqref{eq3_22_6}.
  $T_1^{n}$ is a component of the $\mathbb{T}$-matrix for cylinder 1. $T_2^{n}$ and $ \widetilde{T}_2^{n}$ are components of the $\mathbb{T}$ and $\widetilde{\mathbb{T}}$ matrices for cylinder 2. They have been derived under various boundary
conditions   in Section \ref{sca_cylinder}.

Next, we consider the electromagnetic case.
Notice that
\begin{equation*}
\begin{split}
 \mathbf{A}_{nk_z}^{\text{TE}, *}(\mathbf{x},k)
=& \mathcal{C}_n^*\left(\frac{in}{k_{\perp}\rho }f_n^*(k_{\perp}\rho)\mathbf{e}_{\rho}-f_n^{*\prime}(k_{\perp}\rho)\mathbf{e}_{\phi}\right)e^{in\phi +ik_zz}\\
=&\mathcal{C}_n^*\Biggl(\frac{i\mathbf{e}_x}{2}\left( f_{n+1}^*(k_{\perp}\rho)e^{i(n+1)\phi_k}+f_{n-1}^*(k_{\perp}\rho)e^{i(n-1)\phi_k}\right)
\\&+\frac{ \mathbf{e}_y}{2}\left( f_{n+1}^*(k_{\perp}\rho)e^{i(n+1)\phi_k}-f_{n-1}^*(k_{\perp}\rho)e^{i(n-1)\phi_k}\right)\Biggr)e^{ik_zz},
\end{split}
\end{equation*}where $* = $ reg or out, $f_n^{\text{reg}}(z)=J_n(z)$ and $f_n^{\text{out}}(z)=H_n^{(1)}(z)$. Define
\begin{equation*}
\boldsymbol{\mathcal{Q}}_n=\frac{ \mathbf{e}_x}{2}\left(\mathcal{Q}_{n+1}-\mathcal{Q}_{n-1}\right)+\frac{ \mathbf{e}_y}{2i}\left(\mathcal{Q}_{n+1}+\mathcal{Q}_{n-1}\right).
\end{equation*}
Then
$$\mathbf{A}_{nk_z}^{\text{TE},*}(\mathbf{x},k) =i^{-n}\mathcal{C}_n^*\boldsymbol{\mathcal{Q}}_nf_0^*(k_{\perp}\rho)e^{ik_zz}.$$
Since
\begin{equation*}
\begin{split}
\boldsymbol{\mathcal{Q}}_ne^{i\mathbf{k}_{\perp}\cdot\boldsymbol{\rho}}=&\left[\frac{\mathbf{e}_x}{2}
\left(e^{i(n+1)\phi_k}-e^{i(n-1)\phi_k}\right)+\frac{\mathbf{e}_y}{2i}
\left(e^{i(n+1)\phi_k}+e^{i(n-1)\phi_k}\right)\right]e^{i\mathbf{k}_{\perp}\cdot\boldsymbol{\rho}}\\
=&\left(i\sin\phi_k\mathbf{e}_x-i\cos\phi_k\mathbf{e}_y\right)e^{in\phi_k}e^{i\mathbf{k}_{\perp}\cdot\boldsymbol{\rho}},
\end{split}
\end{equation*}
we have
\begin{equation}\label{eq4_9_4}
\begin{split}
\mathbf{A}_{nk_z}^{\text{TE, reg}}(\mathbf{x},k) =&\mathcal{C}_n^{\text{reg}}i^{-n}\boldsymbol{\mathcal{Q}}_nJ_0(k_{\perp}\rho)e^{ik_zz}\\
=&\frac{\mathcal{C}_n^{\text{reg}}}{2\pi i^n}\boldsymbol{\mathcal{Q}}_n\int_0^{2\pi}d\phi_k  e^{i\mathbf{k}_{\perp}\cdot\boldsymbol{\rho}+ik_zz}\\
=&\frac{\mathcal{C}_n^{\text{reg}}}{2\pi i^n} \int_0^{2\pi}d\phi_k \left(i\sin\phi_k\mathbf{e}_x-i\cos\phi_k\mathbf{e}_y\right)e^{in\phi_k}
e^{i\mathbf{k}_{\perp}\cdot\boldsymbol{\rho}+ik_zz},\\
\mathbf{A}_{nk_z}^{\text{TM, reg}}(\mathbf{x},k) =&\frac{1}{k}\nabla\times \mathbf{A}_{nk_z}^{\text{TE, reg}}(\mathbf{x},k) \\
=&\frac{\mathcal{C}_n^{\text{reg}}}{2\pi i^n} \int_0^{2\pi}d\phi_k \frac{i\mathbf{k}}{k}\times\left(i\sin\phi_k\mathbf{e}_x-i\cos\phi_k\mathbf{e}_y\right)
e^{in\phi_k}e^{i\mathbf{k}_{\perp}\cdot\boldsymbol{\rho}+ik_zz}\\
=&\frac{\mathcal{C}_n^{\text{reg}}}{2\pi i^n} \int_0^{2\pi}d\phi_k  \left( -\frac{k_z\cos\phi_k}{k}\mathbf{e}_x
-\frac{k_z\sin\phi_k}{k}\mathbf{e}_y+\frac{k_{\perp}}{k}\mathbf{e}_z\right)e^{in\phi_k}e^{i\mathbf{k}_{\perp}\cdot\boldsymbol{\rho}+ik_zz}.
\end{split}
\end{equation}
Similarly,
\begin{equation}\label{eq4_9_5}
\begin{split}
\mathbf{A}_{nk_z}^{\text{TE, out}}(\mathbf{x},k)=&\frac{\mathcal{C}_n^{\text{out}}}{ \pi i^n} \int_
{-\infty}^{\infty}dk_y \left(i\sin\phi_k\mathbf{e}_x-i\cos\phi_k\mathbf{e}_y\right)e^{in\phi_k}\frac{e^{\pm i
\sqrt{k_{\perp}^2-k_y^2}x+ik_yy+ik_zz}}{\sqrt{k_{\perp}^2-k_y^2}},\quad x\gtrless 0\\
\mathbf{A}_{nk_z}^{\text{TM, out}}(\mathbf{x},k) =&\frac{\mathcal{C}_n^{\text{out}}}{ \pi i^n} \int_
{-\infty}^{\infty}dk_y \left( -\frac{k_z\cos\phi_k}{k}\mathbf{e}_x
-\frac{k_z\sin\phi_k}{k}\mathbf{e}_y+\frac{k_{\perp}}{k}\mathbf{e}_z\right)e^{in\phi_k}\frac{e^{\pm i
\sqrt{k_{\perp}^2-k_y^2}x+ik_yy+ik_zz}}{\sqrt{k_{\perp}^2-k_y^2}},\quad x\gtrless 0.
\end{split}
\end{equation}
Consider the expansion that defines the translation matrix $\mathbb{V}$:
\begin{equation*}
\begin{split}
\mathbf{A}_{n'k_z}^{\text{TE,reg}}(\mathbf{x}-\mathbf{L},k)=&\sum_{n=-\infty}^{\infty}V_{n,n'}^{\text{TE,TE}}(-\mathbf{L})
\mathbf{A}_{nk_z}^{\text{TE,reg}}(\mathbf{x},k)+V_{n,n'}^{\text{TM,TE}}(-\mathbf{L})
\mathbf{A}_{nk_z}^{\text{TM,reg}}(\mathbf{x},k),\\
\mathbf{A}_{n'k_z}^{\text{TM,reg}}(\mathbf{x}-\mathbf{L},k)=&\sum_{n=-\infty}^{\infty}V_{n,n'}^{\text{TE,TM}}(-\mathbf{L})
\mathbf{A}_{nk_z}^{\text{TE,reg}}(\mathbf{x},k)+V_{n,n'}^{\text{TM,TM}}(-\mathbf{L})
\mathbf{A}_{nk_z}^{\text{TM,reg}}(\mathbf{x},k).
\end{split}
\end{equation*}It follows from \eqref{eq3_20_2} that $V_{n,n'}^{\text{TM,TE}}=V_{n,n'}^{\text{TE,TM}}$ and $V_{n,n'}^{\text{TM,TM}}=V_{n,n'}^{\text{TE,TE}}$. Now, since
\begin{equation}\label{eq4_9_3}
\begin{split}
\left(\boldsymbol{\mathcal{Q}}_{n''}\cdot\mathbf{A}_{nk_z}^{\text{TE,reg}}\right)(\mathbf{0})=
&\frac{\mathcal{C}_n^{\text{reg}}}{2\pi i^n}\int_0^{2\pi}d\phi_k \left(i\sin\phi_k\mathbf{e}_x-i\cos\phi_k\mathbf{e}_y\right)\cdot
\left(i\sin\phi_k\mathbf{e}_x-i\cos\phi_k\mathbf{e}_y\right)
e^{i(n''+n)\phi_k}\\=&-\mathcal{C}_n^{\text{reg}}i^{-n}\delta_{n'',-n},\\
\left(\boldsymbol{\mathcal{Q}}_{n''}\cdot\mathbf{A}_{nk_z}^{\text{TM,reg}}\right)(\mathbf{0})=
&\frac{\mathcal{C}_n^{\text{reg}}}{2\pi i^n}\int_0^{2\pi}d\phi_k \left(i\sin\phi_k\mathbf{e}_x-i\cos\phi_k\mathbf{e}_y\right)\cdot
\left( -\frac{k_z\cos\phi_k}{k}\mathbf{e}_x
-\frac{k_z\sin\phi_k}{k}\mathbf{e}_y+\frac{k_{\perp}}{k}\mathbf{e}_z\right)
e^{i(n''+n)\phi_k}=0,
\end{split}
\end{equation}
we find that
\begin{equation*}
\begin{split}
V_{n,n'}^{\text{TE,TE}}(-L\mathbf{e}_x,i\xi)=&-\frac{i^n}{\mathcal{C}_n^{\text{reg}}}
\left(\boldsymbol{\mathcal{Q}}_{-n}\cdot\mathbf{A}_{n'k_z}^{\text{TE,reg}}\right)(-L\mathbf{e}_x)=(-1)^{n'-n}I_{n'-n}(\kappa_{\perp}L),\\
V_{n,n'}^{\text{TE,TM}}(-L\mathbf{e}_x,i\xi)=&-\frac{i^n}{\mathcal{C}_n^{\text{reg}}}
\left(\boldsymbol{\mathcal{Q}}_{-n}\cdot\mathbf{A}_{n'k_z}^{\text{TM,reg}}\right)(-L\mathbf{e}_x)=0.
\end{split}
\end{equation*}Hence,
\begin{equation*}
\begin{split}
\mathbb{V}_{n,n'}(-L\mathbf{e}_x)=&V_{n,n'}(-L\mathbf{e}_x)\mathbb{I},
\end{split}
\end{equation*}where $\mathbb{I}$ is the $2\times 2$ identity matrix.
In the same way, we find that
\begin{equation*}
\begin{split}
\mathbb{U}_{n,n'}(-L\mathbf{e}_x )=&  U_{n,n'}(-L\mathbf{e}_x )\mathbb{I},\\
\mathbb{U}_{n',n}(L\mathbf{e}_x)=  &U_{n',n}(L\mathbf{e}_x)\mathbb{I},\\
\mathbb{W}_{n',n}(L\mathbf{e}_x)=  &W_{n',n}(L\mathbf{e}_x)\mathbb{I}.
\end{split}
\end{equation*}
Consequently, the electromagnetic Casimir interaction energy is similar to the scalar case given by \eqref{eq4_5_1}, \eqref{eq4_5_3} and \eqref{eq4_5_4},
 except that we have to replace
$T_1^{nk_z}, T_2^{nk_z}$ and $\widetilde{T}_2^{nk_z}$ by $2\times 2$  matrices $\mathbb{T}_1^{nk_z}, \mathbb{T}_2^{nk_z}$ and
$\widetilde{\mathbb{T}}_2^{nk_z}$ derived in Section \ref{sca_cylinder}, and
$\gamma$ is replaced by $\gamma_m=\sqrt{\kappa_m^2+k_z^2}$, with   $\kappa_m=n_m(i\xi)\xi/c$. $n_m$ is the refractive index of the medium between the two cylinders.

The exact Casimir interaction energy between two eccentric Dirichlet/Neumann/perfectly conducting cylinders have been derived in \cite{18,19,50} using the mode summation approach and in \cite{4} using the multiple scattering approach. For two Dirichlet/Neumann/perfectly conducting cylinders that are outside each other, the Casimir interaction energy has been considered in \cite{49,25,4,70,71}. For two magnetodielectric cylinders, the Casimir interaction has been considered in \cite{66}. For two semitransparent cylinders that are outside each other, the Casimir interaction energy has been derived in \cite{14,16}. In addition to these scenarios, the results in this section also provide the exact formula for the Casimir interaction energy in the following scenarios:

\begin{enumerate}
\item[$\bullet$]Two Robin cylinders that are outside each other, or one is inside the other.
\item[$\bullet$] One Dirichlet/Neumann/Robin cylinder and one semi-transparent cylinder that are outside each other, or one is inside the other.
\end{enumerate}

\section{Casimir effect of a cylinder parallel to a plane}\label{cylinder_plane}
In this section, we consider the Casimir interaction between a cylinder  and a plane that are parallel to each other. Assume that  the cylinder is represented by $x^2+y^2=R^2$, $-H/2\leq z\leq H/2$, and the plane is represented by $x=L, -H/2\leq y, z\leq H/2$.
The center  of the cylinder is at $O=(0,0,0)$ and the center of the plane is at $O'=(L,0,0)$.

In this case, we have to choose a different basis for  plane waves. Choosing $\mathbf{e}_x$ as the distinct direction,
plane wave basis are parametrized by $ (k_y, k_z)\in \mathbb{R}^2$, with
\begin{equation*}\begin{split}
\varphi_{k_yk_z}^{\substack{\text{reg}\\\text{out}}}(\mathbf{x},k)=& e^{\mp i\sqrt{k_{\perp}^2-k_y^2}x+ik_yy+ik_zz},
\\
\mathbf{A}_{k_yk_z}^{\text{TE}, \substack{\text{reg}\\\text{out}}}(\mathbf{x},k)=&
\frac{1}{\sqrt{k_y^2+k_z^2}}\nabla\times \varphi_{k_yk_z}^{\substack{\text{reg}\\\text{out}}}(\mathbf{x},k)\mathbf{e}_x\\
=& \frac{1}{\sqrt{k_y^2+k_z^2}}e^{\mp i\sqrt{k_{\perp}^2-k_y^2}x+ik_yy+ik_zz}
\left(ik_z\mathbf{e}_y-ik_y\mathbf{e}_z\right),\\
\mathbf{A}_{k_yk_z}^{\text{TM}, \substack{\text{reg}\\\text{out}}}(\mathbf{x},k)=&\frac{1}{k\sqrt{k_y^2+k_z^2}}
\nabla\times\nabla\times \varphi_{k_yk_z}^{\substack{\text{reg}\\\text{out}}}(\mathbf{x},k)\mathbf{e}_x\\
=& \frac{1}{k\sqrt{k_y^2+k_z^2}}e^{\mp i\sqrt{k_{\perp}^2-k_y^2}x+ik_yy+ik_zz}
\left((k_y^2+k_z^2)\mathbf{e}_x \pm k_y\sqrt{ k_{\perp}^2-k_y^2}\mathbf{e}_y \pm k_z\sqrt{ k_{\perp}^2-k_y^2}\mathbf{e}_z\right).
\end{split}\end{equation*}
Here $k_{\perp}=\sqrt{k^2-k_z^2}$.

Since the $\mathbb{T}$-matrix for a cylinder and the $\widetilde{\mathbb{T}}$-matrix for a plane have been considered respectively in Section \ref{sca_cylinder} and Section \ref{sca_plane},
 let us now consider the translation matrices $\mathbb{V}$ and $\mathbb{W}$. For our application, we only consider a translation by a vector $\mathbf{L}$
 that is perpendicular to the $z$-axis.
  It is obvious that the translation matrices are diagonal in $k_z$.
 In the scalar case, they are defined by
 \begin{equation}
\varphi_{k_yk_z}^{\text{reg}}(\mathbf{x}-\mathbf{L},k)=\sum_{n=-\infty}^{\infty} V_{n,k_y}(-\mathbf{L})\varphi_{nk_z}^{\text{reg}}(\mathbf{x},k),
\end{equation}
\begin{equation}\label{eq4_5_8}
\varphi_{nk_z}^{\text{out}}(\mathbf{x}'+\mathbf{L},k)=H\int_{-\infty}^{\infty}\frac{dk_y}{2\pi}
W_{k_y,n}(\mathbf{L})\varphi_{k_yk_z}^{\text{out}}(\mathbf{x}',k).
\end{equation}
Using \eqref{eq4_5_5}, we find that
\begin{equation*}
\begin{split}
V_{n,k_y}(-\mathbf{L})=&\frac{i^n}{\mathcal{C}_n^{\text{reg}}}\left(\mathcal{Q}_{-n} \varphi_{k_yk_z}^{\text{reg}}\right)( -\mathbf{L} )\\
=&(-1)^n e^{-in\tilde{\phi}_k}e^{- i\sqrt{k_{\perp}^2-k_y^2}x+ik_yy+ik_zz}\Bigr|_{\mathbf{r}=-\mathbf{L}},
\end{split}
\end{equation*}where
\begin{equation*}
\cos\tilde{\phi}_k=-\frac{\sqrt{k_{\perp}^2-k_y^2}}{k_{\perp}}, \quad \sin\tilde{\phi}_k=\frac{k_y}{k}.
\end{equation*}
Hence,
\begin{equation*}
V_{n,k_y}(-L\mathbf{e}_x,i\xi)=\left(\frac{\sqrt{\gamma^2+k_y^2}+k_y}{\gamma}\right)^ne^{- \sqrt{\gamma^2+k_y^2}L},
\end{equation*}where $\gamma=\sqrt{\kappa^2+k_z^2}$.

For $\mathbb{W}$, compare \eqref{eq4_5_7} to \eqref{eq4_5_8}, we find immediately that
\begin{equation*}
W_{k_y,n}(\mathbf{L})=\frac{\pi i}{H\sqrt{k_{\perp}^2-k_y^2}} e^{in\phi_k}e^{i\sqrt{k_{\perp}^2-k_y^2}L},
\end{equation*}
where
\begin{equation*}
\cos\phi_k= \frac{\sqrt{k_{\perp}^2-k_y^2}}{k_{\perp}}, \quad \sin\phi_k=\frac{k_y}{k}.
\end{equation*}
Hence,
\begin{equation*}
W_{k_y,n}(L\mathbf{e}_x,i\xi)=\frac{\pi  }{H\sqrt{\gamma^2+k_y^2}} \left(\frac{\sqrt{\gamma^2+k_y^2}+k_y}{\gamma}\right)^ne^{- \sqrt{\gamma^2+k_y^2}L}.
\end{equation*}
Hence, for a scalar interaction between a cylinder and a plane, the Casimir interaction energy is given by
\begin{equation}\label{eq4_5_1_2}
E_{\text{Cas}}=\frac{\hbar H}{2\pi}\int_0^{\infty}d\xi \int_{-\infty}^{\infty}\frac{dk_z}{2\pi}\text{Tr}\,\ln\left(1-\mathbb{M}(i\xi)\right),
\end{equation}where the trace  Tr is
given by \eqref{eq4_5_2}. We have used the fact that the $\mathbb{T}, \widetilde{\mathbb{T}}$ and the translation matrices are diagonal in $k_z$ to take out the trace over $k_z$.
The matrix $\mathbb{M}$ is given by
\begin{equation*}
M_{n,n'}(\omega)=T^{nk_z}\int_{-\infty}^{\infty}\frac{dk_y}{2\pi}e^{in\phi_k}\widetilde{T}^{k_yk_z}
\frac{\pi i}{\sqrt{k_{\perp}^2-k_y^2}} e^{in'\phi_k}e^{2i\sqrt{k_{\perp}^2-k_y^2}L},
\end{equation*}where $T^{nk_z}$ is a component of the $\mathbb{T}$-matrix  of the cylinder derived in Section \ref{sca_cylinder}, and $\widetilde{T}^{k_yk_z}$ is a component of the $\widetilde{\mathbb{T}}$-matrix of the plane derived in Section \ref{sca_plane}, with $k_{\perp}$ replaced by $\sqrt{k_y^2+k_z^2}$. When the plane is subjected to Dirichlet or
Neumann boundary conditions, $\widetilde{T}^{k_yk_z}=1$ or $-1$. Then
\begin{equation*}
\begin{split}
M_{n,n'}(\omega)=&\pm T^{nk_z}\int_{-\infty}^{\infty}\frac{dk_y}{2\pi}e^{in\phi_k}
\frac{\pi i}{\sqrt{k_{\perp}^2-k_y^2}} e^{in'\phi_k}e^{2i\sqrt{k_{\perp}^2-k_y^2}L}\\
=&\pm T^{nk_z}\varphi_{n+n',k_z}^{\text{out}}(2L\mathbf{e}_x).
\end{split}
\end{equation*}In imaginary frequency,
\begin{equation*}
\begin{split}
M_{n,n'}(L\mathbf{e}_x,i\xi)= \pm T^{nk_z}K_{n+n'}(2\gamma L).
\end{split}
\end{equation*}
If general Robin boundary condition is imposed on the plane or the plane is semitransparent, we have
\begin{equation*}
\begin{split}
M_{n,n'}(i\xi)=\frac{1}{2}T^{nk_z}\int_{-\infty}^{\infty}\frac{dk_y}{\sqrt{\gamma^2+k_y^2}} \widetilde{T}^{k_yk_z}
\left(\frac{\sqrt{\gamma^2+k_y^2}+k_y}{\gamma}\right)^{n+n'}e^{- 2\sqrt{\gamma^2+k_y^2}L}.
\end{split}
\end{equation*}
Let $$k_y=\gamma\sinh\theta.$$ Then
\begin{equation}\label{eq5_9_4}
M_{n,n'}(i\xi)= \frac{1}{2}T^{nk_z}\int_{-\infty}^{\infty}d\theta \widetilde{T}^{k_yk_z}
e^{(n+n')\theta}e^{- 2\gamma L\cosh \theta  }.
\end{equation}
For a Robin plane,
\begin{equation*}
\widetilde{T}^{k_yk_z}=\frac{u-\gamma\cosh\theta}{u+\gamma\cosh\theta}.
\end{equation*}For a semitransparent plane,
\begin{equation*}
\widetilde{T}^{k_yk_z}=\frac{\lambda}{\lambda+2\gamma\cosh\theta}.
\end{equation*}In the formulas above, $\gamma=\sqrt{\kappa^2+k_z^2}$, and $\kappa$ is related to $\xi$ by \eqref{eq3_22_6}.

Next we turn to the electromagnetic case. As in the scalar case, the translation matrices are diagonal in $k_z$. They are defined by
\begin{equation}\label{eq4_9_1}
\begin{split}
\mathbf{A}^{\text{TE, reg}}_{k_yk_z}(\mathbf{x}-\mathbf{L},k)=&\sum_{n=-\infty}^{\infty} \left(V_{n,k_y}^{\text{TE,TE}}(-\mathbf{L})
\mathbf{A}^{\text{TE, reg}}_{nk_z}(\mathbf{x},k)+V_{n,k_y}^{\text{TM,TE}}(-\mathbf{L})\mathbf{A}^{\text{TM, reg}}_{nk_z}(\mathbf{x},k)\right),\\
\mathbf{A}^{\text{TM, reg}}_{k_yk_z}(\mathbf{x}-\mathbf{L},k)=&\sum_{n=-\infty}^{\infty}\left(V_{n,k_y}^{\text{TE,TM}}(-\mathbf{L})
\mathbf{A}^{\text{TE, reg}}_{nk_z}(\mathbf{x},k)+V_{n,k_y}^{\text{TM,TM}}(-\mathbf{L})\mathbf{A}^{\text{TM, reg}}_{nk_z}(\mathbf{x},k)\right),
\end{split}
\end{equation}
\begin{equation}\label{eq4_9_2}
\begin{split}
\mathbf{A}^{\text{TE, out}}_{nk_z}(\mathbf{x}'+\mathbf{L},k)=&H\int_{-\infty}^{\infty}
\frac{dk_y}{2\pi} \left(W_{k_y,n}^{\text{TE,TE}}(\mathbf{L})
\mathbf{A}^{\text{TE, out}}_{k_yk_z}(\mathbf{x}',k)+W_{k_y,n}^{\text{TM,TE}}(\mathbf{L})\mathbf{A}^{\text{TM, out}}_{k_yk_z}(\mathbf{x}',k)\right),\\
\mathbf{A}^{\text{TM, out}}_{nk_z}(\mathbf{x}'+\mathbf{L},k)=&H\int_{-\infty}^{\infty}
\frac{dk_y}{2\pi} \left(W_{k_y,n}^{\text{TE,TM}}(\mathbf{L})
\mathbf{A}^{\text{TE, out}}_{k_yk_z}(\mathbf{x}',k)+W_{k_y,n}^{\text{TM,TM}}(\mathbf{L})\mathbf{A}^{\text{TM, out}}_{k_yk_z}(\mathbf{x}',k)\right).
\end{split}
\end{equation}The relation \eqref{eq3_20_2} implies that for $Z=  V$ or $W$,
\begin{equation*}
Z^{\text{TE,TM}}=Z^{\text{TM,TE}},\quad Z^{\text{TM,TM}}=Z^{\text{TE,TE}}.
\end{equation*}For $\mathbb{V}$, using \eqref{eq4_9_3}, we find that
\begin{equation*}
\begin{split}
&V_{n,k_y}^{\text{TE,TE}}(-L\mathbf{e}_x)\\=&-\frac{i^n}{\mathcal{C}_n^{\text{reg}}}
\left(\boldsymbol{\mathcal{Q}}_{-n}\cdot\mathbf{A}_{k_yk_z}^{\text{TE,reg}}\right)(-L\mathbf{e}_x), \\
=&(-1)^{n+1}\boldsymbol{\mathcal{Q}}_{-n}\cdot\frac{1}{\sqrt{k_y^2+k_z^2}}e^{- i\sqrt{k_{\perp}^2-k_y^2}x+ik_yy+ik_zz}
\left(ik_z\mathbf{e}_y-ik_y\mathbf{e}_z\right)\Biggr|_{\mathbf{r}=-L\mathbf{e}_x},\\
=&(-1)^{n+1}e^{-in\tilde{\phi}_k}\frac{1}{\sqrt{k_y^2+k_z^2}}\left(i\sin\tilde{\phi}_k\mathbf{e}_x-i\cos\tilde{\phi}_k\mathbf{e}_y\right)\cdot
\left(ik_z\mathbf{e}_y-ik_y\mathbf{e}_z\right)e^{i\sqrt{k_{\perp}^2-k_y^2}L},\\
=&(-1)^{n+1}e^{-in\tilde{\phi}_k}\frac{k_z\cos\tilde{\phi}_k}{\sqrt{k_y^2+k_z^2}} e^{i\sqrt{k_{\perp}^2-k_y^2}L},
\\
&V_{n,k_y}^{\text{TE,TM}}(-L\mathbf{e}_x,i\xi)\\=&-\frac{i^n}{\mathcal{C}_n^{\text{reg}}}
\left(\boldsymbol{\mathcal{Q}}_{-n}\cdot\mathbf{A}_{k_yk_z}^{\text{TM,reg}}\right)(-L\mathbf{e}_x) \\
=&(-1)^{n+1}\boldsymbol{\mathcal{Q}}_{-n}\cdot\frac{1}{k\sqrt{k_y^2+k_z^2}}e^{- i\sqrt{k_{\perp}^2-k_y^2}x+ik_yy+ik_zz}
\left((k_y^2+k_z^2)\mathbf{e}_x + k_y\sqrt{ k_{\perp}^2-k_y^2}\mathbf{e}_y + k_z\sqrt{ k_{\perp}^2-k_y^2}\mathbf{e}_z\right)\Biggr|_{\mathbf{r}=-L\mathbf{e}_x}\\
=&(-1)^{n+1}e^{-in\tilde{\phi}_k}\cdot\frac{1}{k\sqrt{k_y^2+k_z^2}}\left(i\sin\tilde{\phi}_k\mathbf{e}_x-i\cos\tilde{\phi}_k\mathbf{e}_y\right)
\cdot\left((k_y^2+k_z^2)\mathbf{e}_x + k_y\sqrt{ k_{\perp}^2-k_y^2}\mathbf{e}_y + k_z\sqrt{ k_{\perp}^2-k_y^2}\mathbf{e}_z\right)e^{i\sqrt{k_{\perp}^2-k_y^2}L}\\
=&(-1)^{n+1}e^{-in\tilde{\phi}_k}\frac{ikk_y}{k_{\perp}\sqrt{k_y^2+k_z^2}} e^{i\sqrt{k_{\perp}^2-k_y^2}L}.
\end{split}
\end{equation*}In imaginary frequency,
\begin{equation*}
\begin{split}
V_{n,k_y}^{\text{TE,TE}}(-L\mathbf{e}_x,i\xi)=&\frac{k_z}{\sqrt{k_y^2+k_z^2}}\frac{\sqrt{\gamma^2+k_y^2}}{\gamma}
\left(\frac{\sqrt{\gamma^2+k_y^2}+k_y}{\gamma}\right)^ne^{- \sqrt{\gamma^2+k_y^2}L},\\
V_{n,k_y}^{\text{TE,TM}}(-L\mathbf{e}_x,i\xi)=&-\frac{k_y}{\sqrt{k_y^2+k_z^2}}\frac{i\kappa}{\gamma}
\left(\frac{\sqrt{\gamma^2+k_y^2}+k_y}{\gamma}\right)^ne^{- \sqrt{\gamma^2+k_y^2}L}.
\end{split}
\end{equation*}

For $\mathbb{W}$, since
\begin{equation*}
\begin{split}
&i\sin\phi_k\mathbf{e}_x-i\cos\phi_k\mathbf{e}_y\\
=&\frac{ik_y}{k_{\perp}}\mathbf{e}_x- \frac{i\sqrt{k_{\perp}^2-k_y^2}}{k_{\perp}}\mathbf{e}_y\\
=&\frac{ik_yk}{k_{\perp}\sqrt{k_y^2+k_z^2}}
\left(\frac{\sqrt{k_y^2+k_z^2}}{k}
\mathbf{e}_x - \frac{k_y\sqrt{ k_{\perp}^2-k_y^2}}{k\sqrt{k_y^2+k_z^2}}\mathbf{e}_y - \frac{k_z\sqrt{ k_{\perp}^2-k_y^2}}{k\sqrt{k_y^2+k_z^2}}\mathbf{e}_z\right)
-\frac{k_z\sqrt{k_{\perp}^2-k_y^2}}{k_{\perp}\sqrt{k_y^2+k_z^2}}\left(\frac{ik_z}{\sqrt{k_y^2+k_z^2}}\mathbf{e}_y-\frac{ik_y}{\sqrt{k_y^2+k_z^2}}\mathbf{e}_z\right),
\end{split}
\end{equation*}
Compare \eqref{eq4_9_5} to \eqref{eq4_9_2}, we find that
\begin{equation*}
\begin{split}
W_{k_y,n}^{\text{TE,TE}}(L\mathbf{e}_x,i\xi)=&-\frac{\pi}{H\sqrt{\gamma^2+k_y^2}}\frac{k_z}{\sqrt{k_y^2+k_z^2}}\frac{\sqrt{\gamma^2+k_y^2}}{\gamma}
\left(\frac{\sqrt{\gamma^2+k_y^2}+k_y}{\gamma}\right)^ne^{- \sqrt{\gamma^2+k_y^2}L},\\
W_{k_y,n}^{\text{TM,TE}}(L\mathbf{e}_x,i\xi)=&\frac{\pi}{H\sqrt{\gamma^2+k_y^2}}\frac{k_y}{\sqrt{k_y^2+k_z^2}}\frac{i\kappa}{\gamma}
\left(\frac{\sqrt{\gamma^2+k_y^2}+k_y}{\gamma}\right)^ne^{- \sqrt{\gamma^2+k_y^2}L}.
\end{split}
\end{equation*}
Hence, for an electromagnetic Casimir interaction between a cylinder and a plane, the Casimir interaction energy is given by \eqref{eq4_5_1_2}, where
\begin{equation*}
\begin{split}
\mathbb{M}_{n,n'}(i\xi)=&-\frac{1}{2}\mathbb{T}^{nk_z}\int_{-\infty}^{\infty}\frac{dk_y}{\sqrt{\gamma_m^2+k_y^2}}\frac{1}{\gamma_m^2(k_y^2+k_z^2)}
\begin{pmatrix}k_z\sqrt{\gamma_m^2+k_y^2} & -i\kappa_m k_y\\-i\kappa_m k_y &k_z\sqrt{\gamma_m^2+k_y^2} \end{pmatrix}\mathbb{T}^{k_yk_z}
\begin{pmatrix}k_z\sqrt{\gamma_m^2+k_y^2} & -i\kappa_m k_y\\-i\kappa_m k_y &k_z\sqrt{\gamma_m^2+k_y^2} \end{pmatrix}\\
&\hspace{2cm}\times \left(\frac{\sqrt{\gamma_m^2+k_y^2}+k_y}{\gamma}\right)^{n+n'}e^{-2 \sqrt{\gamma_m^2+k_y^2}L}.
\end{split}
\end{equation*}
Here $\kappa_m=n_m(i\xi)\xi/c$, $\gamma_m=\sqrt{\kappa_m^2+k_z^2}$ and $n_m$ is the refractive index of the medium between the cylinder and the plane.
Making a change of variables $k_y=\gamma_m\sinh\theta$, we find that
\begin{equation*}
\begin{split}
\mathbb{M}_{n,n'}(i\xi)=&-\frac{1}{2}\mathbb{T}^{nk_z}\int_{-\infty}^{\infty}d\theta\frac{1}{\kappa_m^2\sinh^2\theta+k_z^2\cosh^2\theta}
\begin{pmatrix}k_z\cosh\theta & -i\kappa_m \sinh\theta\\-i\kappa_m\sinh\theta &k_z\cosh\theta\end{pmatrix}\mathbb{T}^{k_yk_z}
\begin{pmatrix}k_z\cosh\theta & -i\kappa_m \sinh\theta
\\-i\kappa_m\sinh\theta &k_z\cosh\theta\end{pmatrix}\\
&\hspace{2cm}\times e^{(n+n')\theta}e^{-2 \gamma_m L\cosh\theta  }.
\end{split}
\end{equation*}
If the plane is perfectly conducting,
\begin{equation*}
\widetilde{\mathbb{T}}^{k_yk_z}=\begin{pmatrix} 1 & 0 \\ 0 & -1\end{pmatrix}.
\end{equation*}In this case,
\begin{equation*}
\begin{split}
\mathbb{M}_{n,n'}(i\xi)=&-\frac{1}{2}\mathbb{T}^{nk_z}\begin{pmatrix} 1 & 0 \\ 0 & -1\end{pmatrix}\int_{0}^{\infty}d\theta \cosh\left([n+n']\theta\right)
e^{-2 \gamma_m L\cosh\theta  }\\
=&-\frac{1}{2}\mathbb{T}^{nk_z}\begin{pmatrix} 1 & 0 \\ 0 & -1\end{pmatrix}K_{n+n'}(2\gamma_m L).
\end{split}
\end{equation*}In general, $\widetilde{\mathbb{T}}^{\mathbf{k}_{\perp}}$ is a diagonal matrix.  If $x>L$ is dielectric with permittivity $\vep_e$ and permeability $\mu_e$,
then
\begin{equation}
\begin{split}
\widetilde{T}_{k_yk_z}^{\text{TE}}(i\xi)=&\frac{\mu_m\sqrt{\gamma_e^2+\gamma_m^2\sinh^2\theta}-\mu_e\gamma_m\cosh\theta}
{\mu_m\sqrt{\gamma_e^2+\gamma_m^2\sinh^2\theta}+\mu_e\gamma_m\cosh\theta},\\
\widetilde{T}_{k_yk_z}^{\text{TM}}(i\xi)=&\frac{\vep_m\sqrt{\gamma_e^2+\gamma_m^2\sinh^2\theta}-\vep_e\gamma_m\cosh\theta}
{\vep_m\sqrt{\gamma_e^2+\gamma_m^2\sinh^2\theta}+\vep_e\gamma_m\cosh\theta},
\end{split}
\end{equation}where $\gamma_e=\sqrt{\kappa_e^2+k_z^2}$, $\kappa_e=n_e(i\xi)\xi/c$. 

The exact Casimir interaction energy for the cylinder-plane configuration has been considered in a number of works. For scalar interaction with Dirichlet or Neumann boundary conditions, or electromagnetic interaction with perfectly conducting boundary conditions, it has been considered in \cite{11,49,4,9,18,19,50,72}. The case where a magnetodielectric cylinder is in front of a magnetodielectric half-space was briefly discussed in \cite{4}.  In \cite{51}, the Casimir interaction  between a cylindrical plasma sheet and a dielectric half-space or a planar plasma sheet were treated as a scalar problem.  It is interesting to note that the formula \eqref{eq5_9_4} for the scalar interaction between a cylinder and a plane is very similar to the formula obtained in \cite{51}.

In this section, we have also obtained the exact formula for the  Casimir interaction energy for:

\begin{enumerate}
\item[$\bullet$] A Robin cylinder in front of a Robin plane.
\item[$\bullet$] A semitransparent cylinder in front of a semitransparent plane.
\item[$\bullet$] A Robin cylinder in front of a semitransparent plane.
\item[$\bullet$] A semitransparent cylinder in front of a Robin plane.

\end{enumerate}
 
\section{Conclusions}

We have reported on the derivation of the Casimir interaction energy between two objects from the point of view of mode summation approach, for both scalar interactions and electromagnetic interactions. We consider the case where the two objects are exterior to each other, and the case where one object is inside the other. A closer scrutiny reveals that this approach actually has some flavor of the multiple scattering approach of \cite{3,5,4}. The advantage over the multiple scattering approach is that we do not have to rely on the path integral quantization  nor the Green's functions. In some sense, this provides a simpler prescription to obtain the Casimir interaction energy, and it is easier to generalize to other fields such as spinor fields and massive vector fields.

In practice, all the difficulties in writing down the Casimir interaction energy lie in the computation of the $\mathbb{T}, \widetilde{\mathbb{T}}$  and the translation matrices, and the latter is usually more difficult.
In the mode summation approach, one need to have a coordinate system for each of the objects where the wave equations are separable. In this paper, we consider planes, spheres and cylinders where the corresponding coordinate systems are readily available. After choosing a convenient basis, it is straightforward to compute the $\mathbb{T}$ and $\widetilde{\mathbb{T}}$ matrices of an object by matching the boundary conditions.  To compute the translation matrices, additional tools are needed, for which we have chosen the operator approach of \cite{6}.

We discuss some situations that are of interest, such as a scalar field with Dirichlet, Neumann or Robin boundary conditions, or with a Dirac delta potential supported on the boundary of an object, and an electromagnetic field propagating in the presence of magnetodielectric object. This is not meant to be exhaustive but they are boundary conditions that have been explored so far. It is obvious that the mode summation approach can be straightforwardly generalized to other boundary conditions, which are only going to affect the $\mathbb{T}$ and $\widetilde{\mathbb{T}}$ matrices of an object.

In this paper, we have dealt exclusively with the zero temperature Casimir energy. Nevertheless, as mentioned at the end of Section \ref{gen}, the extension to finite temperature is straightforward by using the Matsubara formalism. We only have to replace the formula \eqref{eq3_20_7} for zero temperature Casimir energy by the formula \eqref{eq5_7_1} for finite temperature Casimir free energy.

The results reported here do not only contain the known results about the sphere-sphere, cylinder-cylinder, sphere-plane and cylinder-plane configurations, but actually more. In particular, we have included results on scalar interaction between a Dirichlet/Neumann/Robin object and a semi-transparent object. The new results have been listed in each corresponding section.

\begin{acknowledgments}\noindent
 I would like to thank  M. Bordag and H. Gies for giving helpful comments on the first draft of this paper. This work is supported by the Ministry of Higher Education of Malaysia  under the FRGS grant FRGS/2/2010/SG/UNIM/02/2.
\end{acknowledgments}
\appendix
\section{Casimir interaction between multiple objects}\label{A1}
In this section, we explain how the mode summation approach in Section \ref{gen} can be generalized to $N$ objects, where $N\geq 3$. Let the $N$ objects be labeled as O$_1$, O$_2, \ldots,$ O$_N$. As in Section \ref{gen}, we   consider the following two cases separately:

\begin{enumerate}
\item[] \textbf{Case A} The $N$ objects are outside each other.

\item[] \textbf{Case B} The objects  O$_1$, O$_2, \ldots,$ O$_{N-1}$ are outside each other, but they are all inside the object O$_N$.
\end{enumerate}

Besides these two scenarios, there are other scenarios, such as object O$_1$ is inside object O$_2$, object O$_2$ is inside object O$_3$, and so on. The treatment of these scenarios are conceptually the same but technically more complicated. Therefore we only restrict ourselves here to the Case A and Case B mentioned above. For simplicity, we discuss only scalar fields.

Let $O_i$ be the coordinate origin of an appropriate coordinate system for object O$_i$, and let $\mathbf{x}_i$ be the position vector of a point with respect to $O_i$. If $\mathbf{L}_{ij}$ is the vector from O$_i$ to O$_j$, then $\mathbf{x}_i=\mathbf{x}_j+\mathbf{L}_{ij}$. Obviously, $\mathbf{L}_{ji}=-\mathbf{L}_{ij}$. All the other notations are the same as in Section \ref{gen}.

\subsection{Case A}
The scalar field is represented by
\begin{equation}\label{eq4_24_3}
\varphi=\int_{-\infty}^{\infty} d\omega \sum_{\alpha_i} A_i^{\alpha_i}\varphi_{\alpha_i}^{\text{reg}}(\mathbf{x}_i,\omega)e^{-i\omega t}
\end{equation}inside object O$_i$. In the region outside all the objects and close to object O$_i$,
\begin{equation}\label{eq4_24_4}
\varphi=\int_{-\infty}^{\infty} d\omega \sum_{\alpha_i} \left(a_i^{\alpha_i}\varphi_{\alpha_i}^{\text{reg}}(\mathbf{x}_i,\omega)
+b_i^{\alpha_i}\varphi_{\alpha_i}^{\text{out}}(\mathbf{x}_i,\omega)\right)e^{-i\omega t}.
\end{equation}
As in Section \ref{gen}, the boundary conditions on the boundary of O$_i$ give rise to two equations between $A_i^{\alpha_i}$, $a_i^{\alpha_i}$ and $b_i^{\alpha_i}$. Eliminating $A_i^{\alpha_i}$ gives a relation between $a_i^{\alpha_i}$ and $b_i^{\alpha_i}$:
\begin{equation}\label{eq4_24_1}
b_i^{\alpha_i}=-T_i^{\alpha_i}a_i^{\alpha_i},
\end{equation} which defines  $T_i^{\alpha_i}$.
On the other hand, the outgoing waves $\varphi_{\alpha_i}^{\text{out}}$ outside object O$_i$ will propagate to other objects. In a vicinity of object O$_j$, where $j\neq i$, it contributes to $\varphi_{\alpha_j}^{\text{reg}}$ via the translation relation:
\begin{equation}\label{eq4_24_5}
\varphi_{\alpha_i}^{\text{out}}(\mathbf{x}_i,\omega)=\sum_{\alpha_j}U_{\alpha_j,\alpha_i}(\mathbf{L}_{ij})\varphi_{\alpha_j}^{\text{reg}}(\mathbf{x}_j,\omega).
\end{equation}
In fact, $\varphi_{\alpha_i}^{\text{reg}}$ is a superposition of the waves propagated from the outgoing waves close to other objects. Hence,
\begin{equation}\label{eq4_24_2}
a_i^{\alpha_i}=\sum_{j=1}^N\sum_{\alpha_i}U_{\alpha_i,\alpha_j}(\mathbf{L}_{ji})b_j^{\alpha_j}.
\end{equation}
We are using the convention that $U_{\alpha_i,\alpha_i}(\mathbf{L}_{ii})=0$.
Let $\mathbf{a}$ and $\mathbf{b}$ be respectively the column matrices
\begin{equation*}\mathbf{a}=\begin{pmatrix}\mathbf{a}_1\\ \mathbf{a}_2 \\\vdots \\ \mathbf{a}_N\end{pmatrix}, \hspace{1cm}\mathbf{b}=\begin{pmatrix}\mathbf{b}_1\\ \mathbf{b}_2 \\\vdots \\ \mathbf{b}_N\end{pmatrix},\end{equation*}where
$\mathbf{a}_i$ is the infinite column vector with components $a_i^{\alpha_i}$, and $\mathbf{b}_i$ is the infinite column vector with components $b_i^{\alpha_i}$. Let
$\mathbb{T}$ and $\mathbb{U}$ be respectively the matrices given by
\begin{equation*}
\mathbb{T}=\begin{pmatrix} \mathbb{T}_1 & 0 &\ldots & 0\\
0 & \mathbb{T}_2 & \ldots & 0\\
\vdots & \vdots & \ddots &\vdots\\
0 & 0 &\ldots & \mathbb{T}_N\end{pmatrix},\hspace{1cm} \mathbb{U}=\begin{pmatrix} 0 & \mathbb{U}_{12} &\ldots & \mathbb{U}_{1N}\\
\mathbb{U}_{21} & 0 & \ldots & \mathbb{U}_{2N}\\
\vdots & \vdots & \ddots &\vdots\\
\mathbb{U}_{N1} & \mathbb{U}_{N2} &\ldots & 0\end{pmatrix},
\end{equation*}where $\mathbb{T}_i$ is an infinite diagonal matrix with $(\alpha_i,\alpha_i)$-element given by $T_i^{\alpha_i}$, and $\mathbb{U}_{ij}$ is an infinite matrix with $(\alpha_i,\alpha_j)$ given by $U_{\alpha_i,\alpha_j}(\mathbf{L}_{ji})$.
\eqref{eq4_24_1} and \eqref{eq4_24_2} imply that
\begin{equation*}
\mathbf{b}=-\mathbb{T}\mathbf{a}=-\mathbb{T}\mathbb{U}\mathbf{b}.
\end{equation*}As in Section \ref{gen}, this implies that the Casimir interaction energy is given by \eqref{eq3_20_7}, with
\begin{equation*}
\mathbb{M}=-\mathbb{T}\mathbb{U}.
\end{equation*}In particular, when $N=2$,
\begin{equation*}
\mathbb{M}=\begin{pmatrix} 0 & -\mathbb{T}_1\mathbb{U}_{12}\\ -\mathbb{T}_2\mathbb{U}_{21} & 0\end{pmatrix}.
\end{equation*}
Hence,
\begin{equation*}
\det\left(\mathbb{I}-\mathbb{M}\right)=\det\begin{pmatrix} \mathbb{I} & -\mathbb{T}_1\mathbb{U}_{12}\\ -\mathbb{T}_2\mathbb{U}_{21} &\mathbb{I}\end{pmatrix}
=\det\left(\mathbb{I}-\mathbb{T}_1\mathbb{U}_{12}\mathbb{T}_2\mathbb{U}_{21}\right),
\end{equation*}which recovers the result of Section \ref{gen}.

\subsection{Case B}
Inside the object O$_i$, $1\leq i\leq N-1$, the scalar field is still represented by \eqref{eq4_24_3}. Outside the object O$_N$, the scalar field is represented as
\begin{equation}\label{eq4_24_3_2}
\varphi=\int_{-\infty}^{\infty} d\omega \sum_{\alpha_N} B_N^{\alpha_N}\varphi_{\alpha_N}^{\text{out}}(\mathbf{x}_N,\omega)e^{-i\omega t}.
\end{equation}
In the region outside the objects O$_i$, $1\leq i\leq N-1$ and inside O$_N$, the scalar field is represented as \eqref{eq4_24_4} in a vicinity of O$_j$, $1\leq j\leq N$. The  matrix $\mathbb{T}_i$ for the object O$_i$, $1\leq i\leq N-1$ is defined by \eqref{eq4_24_1}. For the object O$_N$, the matrix $\widetilde{\mathbb{T}}_N$ is defined by
\begin{equation}\label{eq4_24_8}
a_N^{\alpha_N}=-\widetilde{T}_N^{\alpha_N}b_N^{\alpha_N},
\end{equation}which is obtained by eliminating $B_N^{\alpha_N}$ from the boundary conditions on O$_N$.
The translation formula \eqref{eq4_24_5} still holds for $1\leq i,j\leq N-1$. Moreover, we have
\begin{equation}\label{eq4_24_6}
\varphi_{\alpha_N}^{\text{reg}}(\mathbf{x}_N,\omega)=\sum_{\alpha_i}V_{\alpha_i,\alpha_N}(\mathbf{L}_{Ni})\varphi_{\alpha_i}^{\text{reg}}(\mathbf{x}_i,\omega),
\end{equation}
\begin{equation}\label{eq4_24_7}
\varphi_{\alpha_i}^{\text{out}}(\mathbf{x}_i,\omega)=\sum_{\alpha_N}W_{\alpha_N,\alpha_i}(\mathbf{L}_{iN})\varphi_{\alpha_N}^{\text{out}}(\mathbf{x}_N,\omega),
\end{equation} for $1\leq i\leq N-1$.
These imply that for $1\leq i\leq N-1$,
\begin{equation}\label{eq4_24_9}
a_i^{\alpha_i}=\sum_{j=1}^{N-1}\sum_{\alpha_j}U_{\alpha_i,\alpha_j}(\mathbf{L}_{ji})b_j^{\alpha_j}+\sum_{\alpha_N}V_{\alpha_i,\alpha_N}(\mathbf{L}_{Ni})a_N^{\alpha_N},
\end{equation}and
\begin{equation}\label{eq4_24_10}
b_N^{\alpha_N}=\sum_{i=1}^{N-1}\sum_{\alpha_i} W_{\alpha_N,\alpha_i}(\mathbf{L}_{iN})b_i^{\alpha_i}.
\end{equation}

Let
\begin{equation*}\widetilde{\mathbf{a}}=\begin{pmatrix}\mathbf{a}_1\\ \mathbf{a}_2 \\\vdots \\\mathbf{a}_{N-1} \\ \mathbf{b}_N\end{pmatrix}, \hspace{1cm}\widetilde{\mathbf{b}}=\begin{pmatrix}\mathbf{b}_1\\ \mathbf{b}_2 \\\vdots \\ \mathbf{b}_{N-1}\\\mathbf{a}_N\end{pmatrix},\end{equation*}
\begin{equation*}
\widetilde{\mathbb{T}}=\begin{pmatrix} \mathbb{T}_1 & 0 &\ldots &0& 0\\
0 & \mathbb{T}_2 & \ldots & 0& 0\\
\vdots & \vdots & \ddots &\vdots &\vdots\\
0 & 0 &\ldots & \mathbb{T}_{N-1}& 0\\
0 & 0 &\ldots &0 & \widetilde{\mathbb{T}}_N\end{pmatrix},\hspace{1cm} \widetilde{\mathbb{U}}=\begin{pmatrix} 0 & \mathbb{U}_{12} &\ldots &\mathbb{U}_{1,N-1} & \mathbb{V}_{1N}\\
\mathbb{U}_{21} & 0 & \ldots &\mathbb{U}_{2,N-1}& \mathbb{V}_{2N}\\
\vdots & \vdots & \ddots &\vdots &\vdots\\
\mathbb{U}_{N-1,1} &\mathbb{U}_{N-1,2} & \ldots & 0 &\mathbb{V}_{N-1,N}\\
\mathbb{W}_{N1} & \mathbb{W}_{N2} &\ldots &\mathbb{W}_{N,N-1} & 0\end{pmatrix}.
\end{equation*}
\eqref{eq4_24_1}, \eqref{eq4_24_8}, \eqref{eq4_24_9} and \eqref{eq4_24_10} imply that
\begin{equation*}
\widetilde{\mathbf{b}}=-\widetilde{\mathbb{T}}\widetilde{\mathbf{a}}=-\widetilde{\mathbb{T}}\widetilde{\mathbb{U}}\widetilde{\mathbf{b}}.
\end{equation*}
Therefore, the Casimir interaction energy is given by \eqref{eq3_20_7}, with
\begin{equation*}
\mathbb{M}=-\widetilde{\mathbb{T}}\widetilde{\mathbb{U}}.
\end{equation*}When $N=2$,
\begin{equation*}
\det\left(\mathbb{I}-\mathbb{M}\right)=\det\begin{pmatrix} \mathbb{I} & -\mathbb{T}_1\mathbb{V}_{12}\\ -\mathbb{T}_2\mathbb{W}_{21} &\mathbb{I}\end{pmatrix}
=\det\left(\mathbb{I}-\mathbb{T}_1\mathbb{V}_{12}\mathbb{T}_2\mathbb{W}_{21}\right),
\end{equation*}which recovers the result of Section \ref{gen}.

\section{Verification of identities}\label{A2}
First we verify the identity \eqref{eq3_28_1}. By definition,
\begin{equation*}
\mathbf{X}_{l,-m}(\theta_k,\phi_k)\cdot
\mathbf{X}_{l'm'}(\theta_k,\phi_k)=\frac{1}{\sqrt{l(l+1)l'(l'+1)}}\left(\frac{mm'}{\sin^2\theta_k}Y_{l,-m}(\theta_k,\phi_k)Y_{l'm'}(\theta_k,\phi_k)
+\frac{\pa Y_{l,-m}(\theta_k,\phi_k)}{\pa\theta_k}\frac{\pa Y_{l'm'}(\theta_k,\phi_k)}{\pa\theta_k}\right).
\end{equation*}
Let
\begin{equation}\label{eq3_29_2}
\mathbf{X}_{l,-m}(\theta_k,\phi_k)\cdot
\mathbf{X}_{l'm'}(\theta_k,\phi_k)=\sum_{l^{\prime\prime} =0}^{\infty}\sum_{m^{\prime\prime} =-l^{\prime\prime} }^{l^{\prime\prime} }
\Xi_{l^{\prime\prime} ,m^{\prime\prime} }Y_{l^{\prime\prime} m^{\prime\prime} }(\theta_k,\phi_k).
\end{equation}Then
\begin{equation*}
\begin{split}
\Xi_{l''m''}=& \frac{1}{\sqrt{l(l+1)l'(l'+1)}}\int_0^{2\pi}d\phi_k\int_0^{\pi}d\theta_k\sin\theta_k \\&\times\left(\frac{mm'}{\sin^2\theta_k}
Y_{l,-m}(\theta_k,\phi_k)Y_{l'm'}(\theta_k,\phi_k)+\frac{\pa Y_{l,-m}(\theta_k,\phi_k)}{\pa\theta_k}\frac{\pa Y_{l'm'}(\theta_k,\phi_k)}{\pa\theta_k}\right)
Y_{l''m''}^*(\theta_k,\phi_k).
\end{split}
\end{equation*}
For the integration with respect to $\theta_k$, integration by parts gives:
\begin{equation*}
\begin{split}
&\int_0^{\pi}d\theta_k\sin\theta_k \frac{\pa Y_{l,-m}(\theta_k,\phi_k)}{\pa\theta_k}\frac{\pa Y_{l'm'}(\theta_k,\phi_k)}{\pa\theta_k} Y_{l''m''}^*(\theta_k,\phi_k)\\
=&-\frac{1}{2}\int_0^{\pi}d\theta_k\sin\theta_k Y_{l,-m}(\theta_k,\phi_k)\frac{1}{\sin\theta_k}\frac{\pa}{\pa\theta_k}\left(\sin\theta_k
\frac{\pa Y_{l'm'}(\theta_k,\phi_k)}{\pa\theta_k}\right)Y_{l''m''}^*(\theta_k,\phi_k)\\
&-\frac{1}{2}\int_0^{\pi}d\theta_k\sin\theta_k \frac{1}{\sin\theta_k}\frac{\pa}{\pa\theta_k}\left(\sin\theta_k \frac{\pa Y_{l,-m}(\theta_k,\phi_k)}{\pa\theta_k}\right)
Y_{l'm'}(\theta_k,\phi_k)Y_{l''m''}^*(\theta_k,\phi_k)\\
&-\frac{1}{2}\int_0^{\pi}d\theta_k\sin\theta_k \left(Y_{l,-m}(\theta_k,\phi_k)\frac{\pa Y_{l'm'}(\theta_k,\phi_k)}{\pa\theta_k}
+\frac{\pa Y_{l,-m}(\theta_k,\phi_k)}{\pa\theta_k}Y_{l'm'}(\theta_k,\phi_k)\right)\frac{\pa Y_{l''m''}^*}{\pa\theta_k}\\
=&\frac{1}{2}\int_0^{\pi}d\theta_k\sin\theta_k \left(l'(l'+1)-\frac{m^{\prime 2}}{\sin^2\theta_k}\right)Y_{l,-m}(\theta_k,\phi_k)Y_{l'm'}(\theta_k,\phi_k)
Y_{l''m''}^*(\theta_k,\phi_k)\\
&+\frac{1}{2}\int_0^{\pi}d\theta_k\sin\theta_k \left(l(l+1)-\frac{m^{2}}{\sin^2\theta_k}\right)Y_{l,-m}(\theta_k,\phi_k)Y_{l'm'}(\theta_k,\phi_k)
Y_{l''m''}^*(\theta_k,\phi_k)\\
&-\frac{1}{2}\int_0^{\pi}d\theta_k\sin\theta_k \left(l''(l''+1)-\frac{m^{\prime\prime 2}}{\sin^2\theta_k}\right)Y_{l,-m}(\theta_k,\phi_k)
Y_{l'm'}(\theta_k,\phi_k)Y_{l''m''}^*(\theta_k,\phi_k).
\end{split}
\end{equation*}
Using the fact that
$$\int_0^{2\pi}d\phi_kY_{l,-m}(\theta_k,\phi_k)Y_{l'm'}(\theta_k,\phi_k)Y_{l''m''}^*(\theta_k,\phi_k)$$ is nonzero only if $m'=m+m''$, we find that
\begin{equation*}
\begin{split}
\Xi_{l''m''}=&\frac{l(l+1)+l'(l'+1)-l''(l''+1)}{2\sqrt{l(l+1)l'(l'+1)}} \int_0^{2\pi}d\phi_k\int_0^{\pi}d\theta_k\sin\theta_k Y_{l,-m}(\theta_k,\phi_k)
Y_{l'm'}(\theta_k,\phi_k)Y_{l''m''}^*(\theta_k,\phi_k)\\
=&\frac{l''(l''+1)-l(l+1)-l'(l'+1)}{2\sqrt{l(l+1)l'(l'+1)}}(-1)^{m''+1}\sqrt{\frac{(2l+1)(2l'+1)(2l''+1)}{4\pi}}\begin{pmatrix}
l& l' &l''\\m &-m' &m''\end{pmatrix}\begin{pmatrix} l & l' & l''\\0 & 0 & 0\end{pmatrix}.\end{split}
\end{equation*}This proves \eqref{eq3_28_1}.

Next, we verify \eqref{eq3_29_1}.
Since
\begin{equation*}
\begin{split}
&\mathbf{X}_{l,-m}(\theta_k,\phi_k)\cdot
\left(\frac{i\mathbf{k}}{k}\times\mathbf{X}_{l'm'}(\theta_k,\phi_k)\right)\\=&
\frac{1}{\sqrt{l(l+1)l'(l'+1)}}\left(\frac{m}{\sin\theta_k}Y_{l,-m}(\theta_k,\phi_k)
\frac{\pa Y_{l'm'}(\theta_k,\phi_k)}{\pa\theta_k}+
\frac{m'}{\sin\theta_k}Y_{l'm'}(\theta_k,\phi_k)\frac{\pa Y_{l,-m}(\theta_k,\phi_k)}{\pa\theta_k}\right)\\
=&\frac{ik^2}{\sqrt{l(l+1)l'(l'+1)}}\Bigl(\nabla Y_{l'm'}(\theta_k,\phi_k)\times\nabla Y_{l,-m}(\theta_k,\phi_k)\Bigr)\cdot\mathbf{e}_k\\
=& \frac{ik^2}{\sqrt{l(l+1)l'(l'+1)}}\left[\nabla \times\Bigl(Y_{l'm'}(\theta_k,\phi_k) \nabla Y_{l,-m}(\theta_k,\phi_k)\Bigr)\right]\cdot\mathbf{e}_k,
\end{split}
\end{equation*}
we find that
\begin{equation}\label{eq3_30_1}
\begin{split}
&\int_0^{2\pi}d\phi_k\int_0^{\pi}d\theta_k \sin\theta_k \mathbf{X}_{l,-m}(\theta_k,\phi_k)\cdot
\left(\frac{i\mathbf{k}}{k}\times\mathbf{X}_{l'm'}(\theta_k,\phi_k)\right)e^{i\mathbf{k}\cdot\mathbf{L}}\\
=&-\int_0^{2\pi}d\phi_k\int_0^{\pi}d\theta_k \sin\theta_k\left( i\mathbf{L}\times \frac{ik^2}{\sqrt{l(l+1)l'(l'+1)}} \Bigl(Y_{l'm'}(\theta_k,\phi_k)
\nabla Y_{l,-m}(\theta_k,\phi_k)\Bigr)\right)\cdot\mathbf{e}_ke^{i\mathbf{k}\cdot\mathbf{L}}\\
=& \frac{ k}{\sqrt{l(l+1)l'(l'+1)}} \mathbf{L}\cdot\int_0^{2\pi}d\phi_k\int_0^{\pi}d\theta_k \sin\theta_k Y_{l'm'}(\theta_k,\phi_k) \Bigl( -k\mathbf{e}_k\times
\nabla Y_{l,-m}(\theta_k,\phi_k)\Bigr)  e^{i\mathbf{k}\cdot\mathbf{L}}.
\end{split}\end{equation}
Using the fact that
\begin{equation*}
\begin{split}
&-k\mathbf{e}_k\times  \nabla Y_{l,-m}(\theta_k,\phi_k)=\sqrt{l(l+1)}\mathbf{X}_{l,-m}(\theta_k,\phi_k)\\
=& \frac{\mathbf{e}_x}{2i}\left[\sqrt{(l+m)(l-m+1)}Y_{l,-m+1}(\theta_k,\phi_k)+\sqrt{(l-m)(l+m+1)}Y_{l,-m-1}(\theta_k,\phi_k)\right]\\
&-\frac{\mathbf{e}_y}{2}\left[\sqrt{(l+m)(l-m+1)}Y_{l,-m+1}(\theta_k,\phi_k)-\sqrt{(l-m)(l+m+1)}Y_{l,-m-1}(\theta_k,\phi_k)\right]
\\&+i\mathbf{e}_z mY_{l,-m}(\theta_k,\phi_k),
\end{split}
\end{equation*}
if $\mathbf{L}=L\mathbf{e}_z$, we find that \eqref{eq3_30_1} is equal to
\begin{equation*}
\frac{ imkL}{\sqrt{l(l+1)l'(l'+1)}}  \int_0^{2\pi}d\phi_k\int_0^{\pi}d\theta_k \sin\theta_k Y_{l'm'}(\theta_k,\phi_k) Y_{l,-m}(\theta_k,\phi_k)  e^{i\mathbf{k}\cdot\mathbf{L}}.
\end{equation*}

\end{document}